\documentclass[aps,prd,amsmath,amsfonts,nofootinbib,preprintnumbers, superscriptaddress,eqsecnum]{revtex4-2}
\pdfoutput=1 
\usepackage{color,graphicx,slashed,multirow,hyperref,amssymb,amsmath,cleveref,multirow}
\usepackage{braket}
\usepackage{graphicx}
\usepackage[utf8]{inputenc}
\usepackage[section]{placeins}
\usepackage{xcolor,colortbl}
\usepackage{subcaption}
\usepackage{cleveref}
\usepackage[shortlabels]{enumitem}
\AtBeginDocument{\usepackage{booktabs}}
\renewcommand{\arraystretch}{1.2}

\newcommand{\GeV}{{\rm\ GeV}}
\newcommand{\TeV}{{\rm\ TeV}}
\newcommand{\fb}{{\rm\ fb}}
\newcommand{\Lag}{{\mathcal L}}

\newcommand{\UYD}{{U(1)_{\rm YD}}}
\newcommand{\SUD}{{SU(2)_{\rm D}}}
\newcommand{\UD}{{U(1)_{\rm D}}}
\newcommand{\Z}{{\mathbb Z}}

\newcommand{\be}{\begin{equation}}
\newcommand{\ee}{\end{equation}}
\def\bsp#1\esp{\begin{split}#1\end{split}}
\def\bpm{\begin{pmatrix}}
\def\epm{\end{pmatrix}}

\usepackage[normalem]{ulem}

\begin{document}
\preprint{}

\title{A Fermionic Portal to Vector Dark Matter from a New Gauge Sector}

\author{Alexander Belyaev}
\email{a.belyaev@soton.ac.uk}
\affiliation{School of Physics and Astronomy, University of Southampton, Highfield, Southampton SO17 1BJ, UK}
\affiliation{Particle Physics Department, Rutherford Appleton Laboratory, Chilton, Didcot, Oxon OX11 0QX, UK}

\author{Aldo Deandrea}
\email{deandrea@ip2i.in2p3.fr}
\affiliation{Universit{\' e} de Lyon, Universit{\' e} Claude Bernard Lyon 1, CNRS/IN2P3, IP2I UMR5822, F-69622, Villeurbanne, France}
\affiliation{Department of Physics, University of Johannesburg, PO Box 524, Auckland Park 2006, South Africa}

\author{Stefano Moretti}
\email{s.moretti@soton.ac.uk; stefano.moretti@physics.uu.se}
\affiliation{School of Physics and Astronomy, University of Southampton, Highfield, Southampton SO17 1BJ, UK}
\affiliation{Department of Physics and Astronomy, Uppsala University, Box 516, SE-751 20 Uppsala, Sweden}

\author{Luca Panizzi}
\email{luca.panizzi@physics.uu.se}
\affiliation{Department of Physics and Astronomy, Uppsala University, Box 516, SE-751 20 Uppsala, Sweden}
\affiliation{School of Physics and Astronomy, University of Southampton, Highfield, Southampton SO17 1BJ, UK}

\author{Douglas A. Ross}
\email{d.a.ross@soton.ac.uk}
\affiliation{School of Physics and Astronomy, University of Southampton, Highfield, Southampton SO17 1BJ, UK}

\author{Nakorn Thongyoi}
\email{nakorn.thongyoi@gmail.com}
\affiliation{School of Physics and Astronomy, University of Southampton, Highfield, Southampton SO17 1BJ, UK}

\begin{abstract}
We present a new class of Dark Matter (DM) models wherein the Standard Model (SM) is extended with a new $\SUD$ dark gauge sector. In this framework the stability of DM
is provided by the conservation of a $U(1)$ global symmetry which, upon appropriate charge assignments for the $\SUD$ multiplets, effectively leads to a $\Z_2$ symmetry subgroup. The origin of the global $U(1)$ symmetry which ensures the stability of DM can be justified in the form of a dark EW sector or through an underlying composite structure. 
The key ingredient of the model is
a Vector-Like (VL) fermion doublet of $\SUD$, the members of which are singlets of the SM Electro-Weak (EW) gauge group, which mediate the interactions between the dark sector and the SM, via new Yukawa interactions.
This class of models, labelled as Fermion Portal Vector DM (FPVDM), allows multiple realisations, depending on the properties of the the VL partner and the scalar potential. 
After spontaneous breaking of the $\SUD$ symmetry via a new scalar doublet, the ensuing massive vector bosons with non-zero dark-isospin are DM candidates. 
The new class of FPVDM models  suggested here
has numerous phenomenological implications  for collider and non-collider studies.
As a practical example, we discuss here in detail a realisation involving a VL top partner assuming no mixing between the two physical scalars of the theory, the SM Higgs boson and its counterpart in the dark sector. We thus provide bounds on this setup from both collider and astroparticle observables. 
\end{abstract}

\maketitle
\newpage
\tableofcontents
\newpage

\section{Introduction}\label{sec:intro}
The Standard Model (SM) of particle physics describes fundamental particle fields and their interactions under strong, Electro-Magnetic (EM) and weak forces using the symmetry principle of gauge invariance. Furthermore, through the so-called Higgs mechanism, triggering Electro-Weak Symmetry Breaking (EWSB), the last two forces are actually unified into a single EW force. 
Given the particle content and charges under the gauge group of the SM, $SU(3)_C \times SU(2)_L \times U(1)_Y$, 
some of the particles in it are stable either due to the (unbroken) gauge symmetries themselves (such as the gluons and photon) or due to the fact that they are the lightest ones obeying a conservation law (charge or number conservation) such as the electron and its neutrino. The latter is of some importance here, as the analysis of the gravitational interactions at different scales in the Universe implies the existence of matter without EM interactions, called Dark Matter (DM), for which a particle interpretation is a natural possibility in the framework of the SM. So far, the only viable candidate is the aforementioned neutrino, alas, it is not compliant with corresponding experimental observations. Hence, leaving aside other shortcomings of it, there is an obvious need to surpass the SM.

We consider here DM as a vector (spin-1) gauge particle. Such a theoretical construction is extremely well motivated whilst being constrained in the possible model building choices.
The Higgs portal is the simplest and most favoured mechanism to connect a dark sector where the DM is represented by a new gauge boson which gets its mass through a new scalar, that breaks the gauge symmetry through the Higgs mechanism. In this mechanism the quartic interaction involving two new scalars and two Higgs bosons, $|S|^2 |H|^2$, is not protected by any symmetry, and is the minimal way of connecting the visible with the invisible sector. 
The Higgs portal, however, might not be the dominant connection between the two sectors. It induces a mixing in the scalar sector modifying the Higgs couplings to the SM particles and generating Higgs-DM interactions, which are strongly constrained~\cite{Arcadi:2020jqf}. The size of the dimensionless coupling of the quartic interaction, which in principle can have any value, is thus constrained to be small to respect the size of the scalar mixing. 
This makes the detection of signatures from the dark sector extremely challenging. For the non-Abelian case it is also possible to construct kinetic-mixing terms, which are however non-renormalisable and hence suppressed by the scale of new physics. All these scenarios have been extensively studied in literature~\cite{Hubisz:2004ft,Hambye:2008bq,Chen:2009ab,DiazCruz:2010dc,Bhattacharya:2011tr,Lebedev:2011iq,Farzan:2012hh,Baek:2012se,Koorambas:2013una,Fraser:2014yga,Ko:2014gha,Huang:2015wts,Gross:2015cwa,DiFranzo:2015nli,Ko:2016fcd,Barman:2017yzr,Huang:2017bto,Barman:2018esi,Barman:2019lvm,Buttazzo:2019mvl,Abe:2020mph,Gross:2020zam,Chowdhury:2021tnm,Baouche:2021wwa,Hu:2021pln,Babu:2021hef}. 

Other mediation mechanisms can however be present in case of vector DM, noticeably involving the fermionic sector~\cite{Hisano:2020qkq,Babu:2021hef}. 
The fermionic mediator which was studied in the context of scalar DM is well motivated theoretically~\cite{Servant:2002aq,Cacciapaglia:2009pa} and provides interesting phenomenology with well-defined parameter space~\cite{Giacchino:2015hvk,Garny:2018icg,Arina:2020udz,Arina:2020tuw}.
The interaction of vector DM with SM fermions is also well motivated from the phenomenological point of view: most of the current anomalies observed in SM measurements are associated with the fermion sector (especially with the lepton one)~\cite{Crivellin:2021sff}.
Also, the new fermions might also play a role in the radiative shift of the $W$ boson mass, for which a sizeable discrepancy with respect to the SM expectation has been recently reported by~\cite{CDF:2022hxs}.
Scenarios with Vector-Like (VL) fermion portals, but for scalar DM candidates, have also been explored in the literature~\cite{Baek:2017ykw,Colucci:2018vxz}.
Some version of a non-Abelian vector DM scenario connected to the SM through the Higgs portal and the fermionic sector was suggested in~\cite{Hisano:2020qkq}, to explore EM multipole interactions of DM candidates, where the authors introduced two new fermionic multiplets and assumed a negligibly small Higgs portal, so that the main connection to the SM is at one-loop level via the new fermions. In that paper the authors also assumed vanishing mixing between new and SM fermions.
 
In this paper we propose a new minimal framework for Fermion Portal Vector DM (FPVDM) (albeit closely related to that of \cite{Hisano:2020qkq}) which incorporates just one dark doublet of VL fermions.
The FPVDM scenario relies crucially on the mixing of one of the fermions from the dark doublet with one or more SM fermions sharing the same electric charge, and this mixing provides the tree-level portal connecting 
dark and SM sectors. In addition we have formulated the complete Lagrangian for this FPVDM framework, together with the necessary conditions and dark charge assignments which guarantee the stability of vector DM, ensuring the consistency of the new framework suggested in our approach. In our setup the elements of doublet  VL fermions have different charges under a new ``dark" $SU(2)$ group and are singlets under the $SU(2)_L$ group of the SM. The elements of the fermionic doublet have opposite $\Z_2$ parity. This parity emerges as a subgroup of a new global $U(1)$ symmetry, which has to be imposed to ensure the stability of the dark sector, and for which different members of $\SUD$ multiplets transform differently depending on the third component of their dark-isospin (D-isospin). 
The $U(1)$ global symmetry can in principle be promoted to a local symmetry and gauged, generating a new massless gauge boson besides the DM candidate.
 
The plan of our paper is as follows. In section~\ref{sec:Z2SUD} we give a detailed description of
the class of models we propose.
In the following section~\ref{sec:theory}
we further discuss the possibility of gauging the $U(1)$ global symmetry of the model
which would provide a natural symmetry behind the stability of DM.
In section~\ref{sec:onlyTnoHHD}
we discuss the case of a particular realisation of our model, in connection with new interesting collider features. In this scenario we invoke a top-quark portal and eliminate any mixing between SM and dark Higgs bosons. We discuss various aspects of phenomenological implications of this specific top-portal scenario (a selection of such results is presented in Ref. \cite{Belyaev:2022zjx}).
Finally, in section~\ref{sec:conclusion}
we summarise our findings on the new FPVDM framework and our particular realisation of it.

\section{The dark sector and its interactions with the SM}\label{sec:Z2SUD}

We start by considering a new dark $SU(2)$ group -- the simplest non-Abelian group in terms of number of generators -- which we  label  as $\SUD$. The gauge bosons associated with the  $\SUD$ breaking are labelled as $V_\mu^D=\left(V^0_{D+\mu}~V^0_{D0\mu}~V^0_{D-\mu}\right)$, where, here and in the following, the superscript identifies the electric charge and the subscript the isospin under $\SUD$ (D-isospin). The full covariant derivative, including the SM terms, is
\begin{equation}
\label{eq:covdev}
 D_\mu = \partial_\mu 
 - \left( i \frac{g}{\sqrt 2} W^{\pm}_\mu T^\pm + i g W^3_\mu T_3 + i g^\prime Y B_\mu \right)
 - \left( i \frac{g_D}{\sqrt 2} V^0_{D\pm\mu} T^\pm_D + i g_{D} V^0_{D0\mu} T_{3D} \right)\;,
\end{equation}
where $g$ and $g^\prime$ are, respectively, the weak and hypercharge coupling constants, $g_D$ is the $\SUD$ coupling constant, $T_3$ and $Y$ are the weak-isospin and weak-hypercharge, respectively, while $T_{3D}$ is the dark-isospin third component of $\SUD$. The indices of the $T_D$ matrices act only on the $\SUD$ elements and are diagonal with respect to the $SU(2)_L$ ones while the indices of the $T$ matrices act only on the $SU(2)$ elements and are diagonal with respect to $\SUD$.
The $\SUD$ symmetry needs to be spontaneously broken to generate a mass for its gauge bosons. Two complex scalar doublets are thus needed for the breaking of $SU(2)_L$ and $\SUD$, respectively:
\begin{eqnarray}
\begin{array}{rclcrcll}
 \Phi_H   &=& \left(\begin{array}{c}\phi^+ \\ \phi^0 \end{array}\right) &\longrightarrow& 
\langle \Phi_H \rangle &=& {\frac{1}{\sqrt{2}}} \left(\begin{array}{c} 0 \\ v \end{array}\right) & \quad
\text{(SM-like Higgs doublet  breaking $SU(2)_L\times U(1)_Y$)}\;, \label{eq:Higgsdoublet} \\
 \Phi_D &=& \left(\begin{array}{c} \varphi^0_{D+1/2} \\ \varphi^0_{D-1/2} \end{array}\right) &\longrightarrow&
\langle \Phi_D \rangle &=& {\frac{1}{\sqrt{2}}} \left(\begin{array}{c} 0 \\ v_D \end{array}\right) & \quad\text{(new ``dark" scalar doublet breaking $\SUD$)}\;.\label{eq:HDdoublet}
\end{array}
\end{eqnarray}
The full scalar potential has the following form:
\begin{eqnarray}
 V(\Phi_H,\Phi_D) &=& - \mu^2 \Phi_H^\dagger \Phi_H - \mu_D^2 \Phi_D^\dagger \Phi_D + \lambda (\Phi_H^\dagger \Phi_H)^2 + \lambda_D (\Phi_D^\dagger \Phi_D)^2 + \lambda_{\Phi_H\Phi_D} (\Phi_H^\dagger \Phi_H)(\Phi_D^\dagger \Phi_D)\;,
 \label{eq:scalarpotential}
\end{eqnarray}
where the last  term provides the interaction between $\Phi_H$ and $\Phi_D$ (the Higgs portal).
In the unbroken phase the Lagrangian of $\Phi_D$ is invariant under a $SO(4)\sim SU(2)\times SU(2)$ global symmetry. One of the two $SU(2)$ is gauged to be $\SUD$. The Vacuum Expectation Value (VEV) of $\Phi_D$ selects a direction in the scalar field space keeping three unbroken generators and leaving an unbroken global symmetry, the custodial symmetry associated with the diagonal $SU(2)$, $SO(4) \to SO(3) \sim SU(2)_{\rm{diag}}$. In the absence of new fermions, this custodial symmetry ensures the stability of the new (dark) gauge bosons~\cite{Hambye:2008bq}.

We stress here that the quartic term $\Phi_H^\dagger \Phi_H \Phi_D^\dagger \Phi_D$ is in general not protected by any symmetry and therefore cannot be removed altogether from the Lagrangian. A key point of the model, however, is that this portal does not need to play an important role and can indeed be negligible with respect to the other operators of the potential. The connection between the dark sector and the SM is realised via two new VL fermions, singlets of $SU(2)_L$ but with a $U(1)_Y$ hypercharge identical to one of the corresponding right-handed SM fermions. These VL fermions form a doublet under $\SUD$, labelled as $\Psi=(\psi_D, \;\psi )$. 
The respective mass terms and Yukawa interactions  of the new fermion sector have the following form:
\begin{equation}
 -\Lag_f = M_\Psi \bar \Psi \Psi + (
 y^\prime \bar \Psi_L \Phi_D f^{\rm SM}_R 
 + h.c)\;,
\end{equation}
where $f^{\rm SM}_{R}$ generically denotes a SM right-handed singlet and $y^\prime$ is a new Yukawa coupling connecting the SM fermion with $\Psi$ through the $\Phi_D$ doublet.
The absence of an additional Yukawa term $y^{\prime\prime} \bar \Psi_L \Phi_D^c f^{\rm SM}_R$, which would violate the stability of DM, is protected by the presence of the unbroken global $U(1)_D$.
Without this symmetry such a term would be compulsory since the scalar doublet, $\Phi_D$, is in the pseudo-real representation.
Under this global  $U(1)_D=e^{i \Lambda Y_D}$, the new fields transform non trivially, whilst the SM fields transform into themselves. 

 In analogy with the SM, where the $SU(2)_L\times U(1)_Y$ symmetry breaks down to the EM $U(1)$, the vacuum state of $\Phi_D$ is invariant under a residual $U(1)$, which in this case is global. The invariance of the VEV under the transformation
$e^{i g_D \vec\alpha \cdot \vec\tau} e^{i\Lambda Y_D}$, is ensured if the relations $g_D \alpha_3=\Lambda$ and $(T_D^3+Y_D)_{\langle\Phi_D\rangle}=0$ are satisfied, leading to  the assignment $Y_D=1/2$ for $\Phi_D$. 
The breaking pattern in the dark sector is therefore $SU(2)_D \times U(1)_D \to U(1)^d_D$ associated with the diagonal generator $SU(2)_D\times U(1)_D$ with a conserved quantum number $Q_D=T^3_D+Y_D$, the dark charge of the new particles. For this reason, different elements of $SU(2)_D$ multiplets  have different transformation properties under the residual $U(1)_D^d$, and with the assignment $Y_D=1/2$ for doublets and $Y_D=0$ for triplets, a $\Z_2$ subgroup can be defined as
\begin{equation}
 \Z_2: (-1)^{Q_D}\;,
\end{equation}
under which different members of $SU(2)_D$ multiplets transform differently, guaranteeing the stability of the lightest $\Z_2$ odd state. Specifically, $\SUD$ doublets always contain a $\Z_2$-odd and $\Z_2$-even component, while $\SUD$ triplets have a $(-~+~-)$ transformation structure. 
Clearly, the analogies with the SM EM $U(1)$ can be exploited further by promoting the global $U(1)_D$ to a local symmetry and gauging it. This leads to the presence of renormalisable kinetic mixing between the SM and dark $U(1)_D$ groups in the unbroken phase. This aspect will be addressed in \cref{sec:theory}, but such a construction and its phenomenological consequences is not part of the FPVDM scenario suggested here, and therefore will not be explored in detail.

The particle content of the model is summarised in Table~\ref{tab:particlesQN}.
{
\setlength{\tabcolsep}{3pt}
\setlength{\arraycolsep}{0pt}
\begin{table}[h]
\centering
\begin{minipage}{.48\textwidth}
\begin{tabular}{c|cc|c||c}
\toprule
Scalars & $SU(2)_L$ & $U(1)_Y$ & $\SUD$ & $\Z_2$ \\
\midrule
$\Phi_H=\left(\begin{array}{c} \phi^+ \\ \phi^0 \end{array}\right)$ & $\mathbf{2}$ & $1/2$ & $\mathbf{1}$ & $+$ \\
\midrule
$\Phi_D=\left(\begin{array}{c} \varphi^0_{D+{\frac{1}{2}}} \\ \varphi^0_{D-{\frac{1}{2}}} \end{array}\right)$ & $\mathbf{1}$ & $0$ & $\mathbf{2}$ & $\begin{array}{c} - \\ + \end{array}$ \\
\midrule
\bottomrule
\end{tabular}
\vskip 10pt
\begin{tabular}{c|cc|c||c}
\toprule
\midrule
Vectors & $SU(2)_L$ & $U(1)_Y$ & $\SUD$ & $\Z_2$ \\
\midrule
$W_\mu=\left(\begin{array}{c} W^+_\mu \\ W^3_\mu \\ W^-_\mu \end{array}\right)$ & $\mathbf{3}$ & $0$ & $\mathbf{1}$ & $\begin{array}{c} + \\ + \\ + \end{array}$ \\
\midrule
$B_\mu$ & $\mathbf{1}$ & $0$ & $\mathbf{1}$ & $+$ \\
\midrule
$V^D_{\mu}=\left(\begin{array}{c} V^0_{D+\mu} \\ V^0_{D0\mu} \\ V^0_{D-\mu} \end{array}\right)$ & $\mathbf{1}$ & $0$ & $\mathbf{3}$ & $\begin{array}{c} - \\ + \\ - \end{array}$ \\
\midrule
\bottomrule
\end{tabular}
\end{minipage}
\begin{minipage}{.48\textwidth}
\begin{tabular}{c|cc|c||c}
\toprule
\midrule
Fermions & $SU(2)_L$ & $U(1)_Y$ & $\SUD$ & $\Z_2$ \\
\midrule
$f_{L}^{\rm SM}=\left(\begin{array}{c} f_{u,\nu}^{\rm SM} \\ f_{d,\ell}^{\rm SM} \end{array}\right)_{\!\!\!L}$ & $\mathbf{2}$ & ${\frac{1}{6}},-{\frac{1}{2}}$ & $\mathbf{1}$ & $+$ \\
$u_{R}^{\rm SM},\nu_{R}^{\rm SM}$  & $\mathbf{1}$ & ${\frac{2}{3}},0$ & $\mathbf{1}$ & $+$ \\
$d_{R}^{\rm SM},\ell_{R}^{\rm SM}$ & $\mathbf{1}$ & $-{\frac{1}{3}},-1$ & $\mathbf{1}$ & $+$ \\
\midrule
$\Psi=\left(\begin{array}{c} \psi_D \\ \psi \end{array}\right)$ & $\mathbf{1}$ & $Q$ & $\mathbf{2}$ & $\begin{array}{c} - \\ + \end{array}$ \\
\midrule
\bottomrule
\end{tabular}
\end{minipage}
\caption{\label{tab:particlesQN}The quantum numbers under the EW and dark gauge group $\SUD$ of the particles of the model, and their $\Z_2$ parity.}
\end{table}
}

After imposing the dark charge conservation,
ensuring the stability of the lightest particle in the dark sector which is odd under $\Z_2$, the most general Lagrangian for this scenario,   which is composed of field strength tensors for the vectors (SM and dark), the kinetic and mass terms for the fermions and the scalars,  the Yukawa terms and the potential for $\Phi_H$ and $\Phi_D$, takes the following form:
\begin{eqnarray}
\Lag_D &\supset& - {\frac{1}{4}} (V^{i}_{\mu\nu})^2|_{B,W^i,V^0_{Di}} + \bar f^{\rm SM} i \slashed D f^{\rm SM} + \bar \Psi i \slashed D \Psi + | D_\mu \Phi_H |^2 + | D_\mu \Phi_D |^2 - V(\Phi_H,\Phi_D) \nonumber\\
&-& (y \bar f^{\rm SM}_L \Phi_H f^{\rm SM}_R + y^\prime \bar \Psi_L \Phi_D f^{\rm SM}_R + h.c) - M_\Psi \bar \Psi \Psi \;,
\end{eqnarray}
with the covariant derivative and  scalar potential
given in  eq.(\ref{eq:covdev}) and eq.(\ref{eq:scalarpotential}),
respectively.

The lightest $\Z_2$-odd particles can be either the $V^0_{D\pm}$ dark gauge bosons, or $\psi_D$. If it is $\psi_D$, it can be either a partner of a) SM quarks, b) charged leptons or c) neutrinos. In case a) the DM candidate would form a stable bound state with SM quarks, in case b) the model would be excluded because the DM would be electrically charge, while in case c) the DM would be a neutrino partner.
Conversely, if the lightest $\Z_2$-odd particle is $V^0_{D\pm}$, the DM is a massive dark gauge boson. It is this this scenario, labelled as Fermion Portal Vector Dark Matter (FPVDM), which we discuss in the rest of this paper.

\subsection{Kinetic mixing in the unbroken EW and dark phases}
We discuss here in more detail the origin of the kinetic mixing at loop level. The two scalar doublets are secluded with respect to one another in the sense that the SM one has no dark quantum numbers (singlet with respect to $\SUD$) and the $\SUD$ one has no SM quantum numbers (transforming as a singlet with respect to the  SM). The operators giving rise to kinetic mixing in the effective Lagrangian are of dimension-six for $U(1)_Y$ and dimension-eight for $SU(2)_L$ and, in our case, have the form
\begin{equation}
\mathcal{V}_D^{\mu\nu a} \Phi^\dagger_{D k} (\sigma^a)_{kl} \Phi_{Dl} \left( \frac{\kappa_W}{\Lambda^4} W_{\mu\nu}^b \Phi_{Hi}^\dagger (\sigma^b)_{ij}\Phi_{Hj} + \frac{\kappa_B}{\Lambda^2} B_{\mu\nu} \right)\;,
\label{eq:dim8}
\end{equation}
where $\sigma^a$ is a Pauli matrix generator of $\SUD$ and $\sigma^b$ is a generator of $SU(2)_L$. Here, $\mathcal{V}_D^{\mu\nu a}$ is the field strength tensor of $\SUD$ and $W_{\mu\nu}^b$ and $B_{\mu\nu}$ are, respectively, the field strength tensors of $SU(2)_L$ and $U(1)_Y$. The kinetic mixing term is obtained upon inserting the VEVs of the Higgs doublets but, as already indicated, the operator is suppressed through the fourth power of the large scale $\Lambda$. Concerning the origin of this effective operator in our model, the suppression can be estimated with a one-loop two-point function mixing the two types of gauge bosons, $SU(2)_L \times U(1)$  and $\SUD$. 
\begin{figure}[htb]
\includegraphics[width=0.4\linewidth]{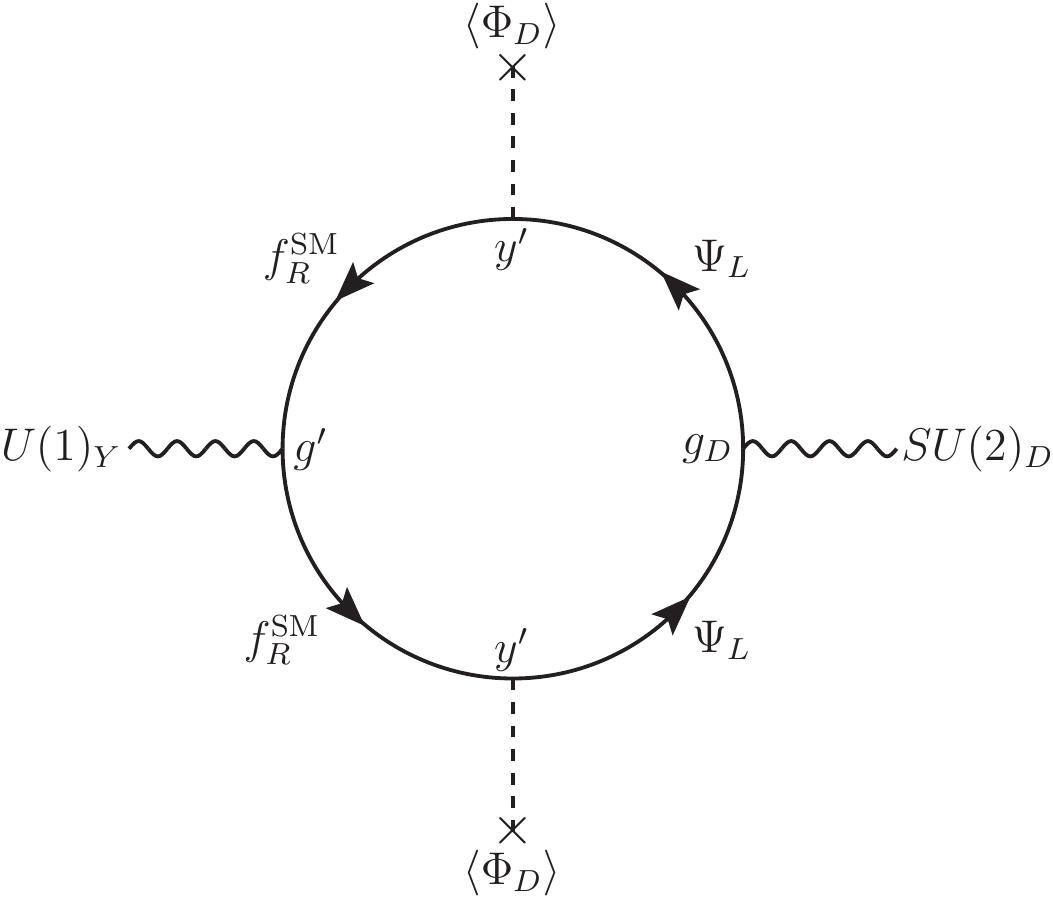}\hskip 20pt
\includegraphics[width=0.4\linewidth]{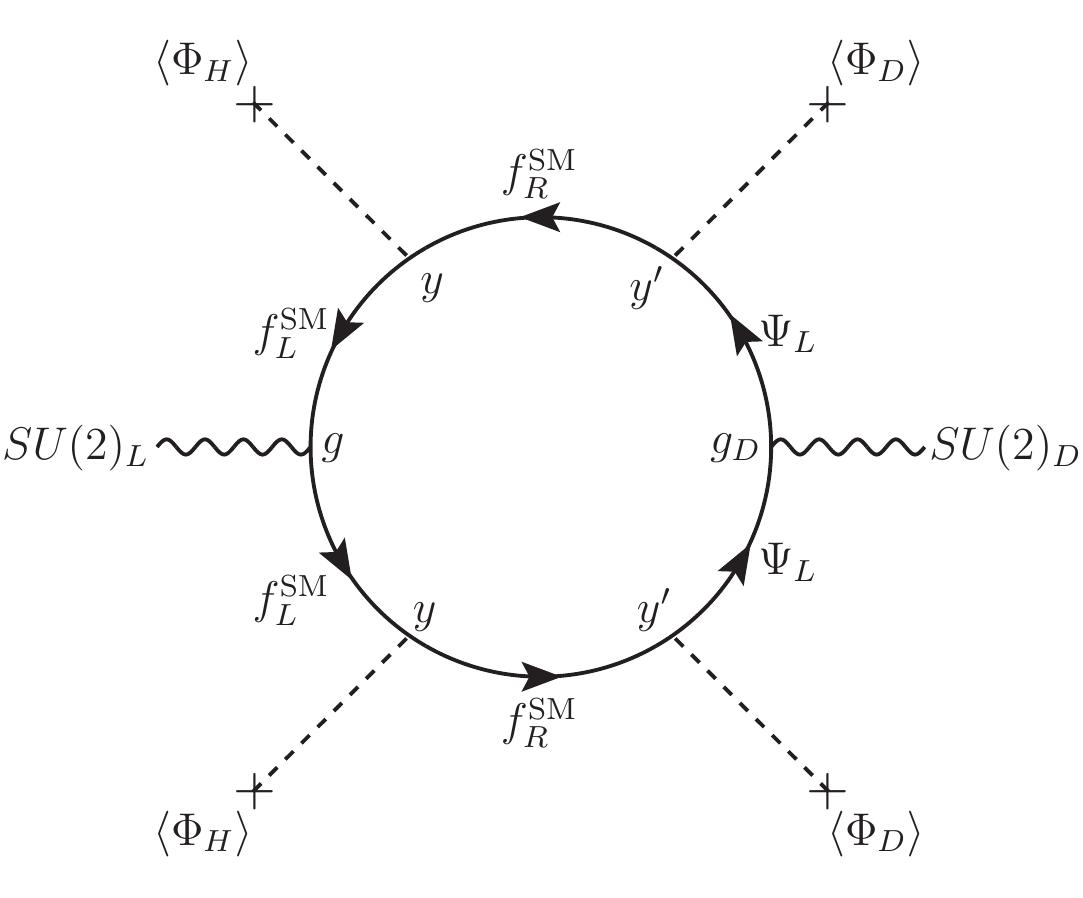}
\caption{Loop realisation of the kinetic mixing operators for $U(1)_Y$ and $SU(2)_L$ in the unbroken EW and dark symmetry phases.}
\label{fig:KMloops}
\end{figure} 
The fermion loops with VEV insertions  allows  the two types of gauge bosons to connect, as shown in \cref{fig:KMloops}, and the interactions are expected to be of order
\begin{equation}
\frac{1}{16 \pi^2 M_\Psi^2 m_f^2} y'^2 g^\prime\; g_D v_D^2 \quad\text{(for $U(1)_Y-SU(2)_D$ mixing)}
\label{eq:suppression1}
\end{equation}
and
\begin{equation}	
\frac{1}{16 \pi^2 M_\Psi^2 m_f^4} y^2y'^2 g\; g_D v^2 v_D^2 \quad\text{(for $SU(2)_L-SU(2)_D$ mixing)}\;,\label{eq:suppression2}
\end{equation}
where $M_\Psi$ is the mass of the VL fermion $\Psi$ with both weak hypercharge and $\SUD$ quantum numbers coupling with a Yukawa type term $y'$ to the Higgs sector.\footnote{
Notice that the Yukawa parameters determine the masses of both $\Z_2$-even fermions, and their expression is a function of all fermion masses. Therefore, \cref{eq:suppression1,eq:suppression2} are finite in the limit $m_f\to 0$: this can be verified by substituting the explicit expressions of the Yukawa couplings (see \cref{eq:yukawas}) and consider that, in the the same limit, the two elements of the VL fermion doublet become degenerate. 
}
A gauge mixing term is also possible using the quartic term in the scalar potential $ \lambda_{\Phi_H\Phi_D} \Phi_H^\dagger \Phi_H \; \Phi_D^\dagger \Phi_D$, but its contribution is more suppressed as it arises at two-loop level. In the broken phase, a kinetic mixing arises between the electrically neutral mass eigenstates~\cite{Holdom:1985ag,Rizzo:2018vlb,Hisano:2020qkq,Rueter:2020qhf}. This is described in more detail in \cref{sec:Dgaugemasses} and has important phenomenological consequences.

\subsection{Electroweak and dark symmetry breaking}\label{sec:EWSB}

\noindent The minimum of the potential reads  as
\begin{equation}
 V(\Phi_H,\Phi_D)_{\rm min} = - {\frac{\mu^2}{2}} v^2 - {\frac{\mu_D^2}{2}} v_D^2 + {\frac{\lambda}{4}} v^4 + {\frac{\lambda_D}{4}} v_D^4 + {\frac{\lambda_{\Phi_H\Phi_D}}{4}} v^2 v_D^2 \;
\end{equation}
and the minimisation conditions are
\begin{equation}
v(- \mu^2 +\lambda v^2 + {\frac{1}{2}} \lambda_{\Phi_H\Phi_D} v_D^2) = 0 \quad\text{and}\quad
v_D(- \mu_D^2 + \lambda_D v_D^2 + {\frac{1}{2}}\lambda_{\Phi_H\Phi_D} v^2) = 0
\end{equation}
whilst the two non-trivial stationary points are
\begin{equation}
\label{eq:vev1}
 v=\sqrt{\frac{4\lambda_D \mu^2-2\lambda_{\Phi_H\Phi_D}\mu_D^2} {4\lambda\lambda_D-\lambda_{\Phi_H\Phi_D}^2}}
 \quad\text{and}\quad 
 v_D=\sqrt{\frac{4\lambda \mu_D^2 - 2\lambda_{\Phi_H\Phi_D}\mu^2}{4\lambda\lambda_D-\lambda_{\Phi_H\Phi_D}^2}}\;,
\end{equation}
where the VEVs are taken to be positive without loss of generality. They are minima if the corresponding Hessian matrix is positive definite ({i.e.}, if its eigenvalues are both positive, being a symmetric matrix),
\begin{equation}
\label{eq:Hess matrix}
\mathcal H|_{v_{\rm min},v_{D\rm min}}=\left(
\begin{array}{cc}
  3\lambda v^2 - \mu^2 + {\frac{\lambda_{\Phi_H\Phi_D}}{2}} \,v_D^2 & \lambda_{\Phi_H\Phi_D} v v_D \\
  \lambda_{\Phi_H\Phi_D} v v_D & 3\lambda_D v_D^2 - \mu_D^2 +  \frac{\lambda_{\Phi_H\Phi_D}}{2}\, v^2
\end{array}
\right)\;,
\end{equation}
which leads to the following conditions for the Lagrangian parameters:
\begin{eqnarray}
\label{eq:minVconditions}
\mu \neq 0 ~\text{and}~ \mu_D \neq 0 ~\text{and}~
\left\{
\begin{array}{l}
\lambda_{\Phi_H\Phi_D}<0 ~\text{and}~ \lambda>0 ~\text{and}~ \lambda_D>0 ~\text{and}~ \lambda_{\Phi_H\Phi_D}^2<4 \lambda \lambda_D\\
\text{or}\\
\lambda_{\Phi_H\Phi_D}>0 ~\text{and}~ 2 \lambda \mu_D^2 > \lambda_{\Phi_H\Phi_D} \mu^2 ~\text{and}~ 2 \lambda_D \mu^2 > \lambda_{\Phi_H\Phi_D} \mu_D^2
\end{array}
\right.\;.
\end{eqnarray}

Finally, if the Higgs  quartic coupling vanishes, $\lambda_{\Phi_H\Phi_D}=0$, the system simply reduces to two independent potentials, $V(\Phi_H,\Phi_D)=V(\Phi_H)+V(\Phi_D)$, where the two terms have identical structure, corresponding to the SM one, and where the minima are simply defined as:
\begin{equation}
 v=\pm\sqrt{\frac{\mu^2}{\lambda}} \quad\text{and}\quad v_D=\pm\sqrt{\frac{\mu_D^2}{\lambda_D}}\;.
\end{equation}

\subsection{Particle spectrum of the model}\label{sec:spectrum}

The model contains new scalar, fermion and vector states. The scalar and fermion ones can mix with SM objects, while the vectors undergo kinetic and mass mixing in the broken EW and dark phases, potentially affecting observables primarily sensitive to the SM itself. In this section, the structure of each particle sector is thus carefully described.

\subsubsection{Fermions}
\label{sec:fermionmasses}

The fermion component with $T_{3D}=+1/2$ gets only the VL mass,  therefore 
\begin{equation}
m_{\psi_D}=M_\Psi\;,
\end{equation}
whereas the other fermion masses are generated after both scalars acquire a VEV. The fermionic mass matrix reads as follows:
\begin{equation}
 \Lag_m^f = (\bar f^{\rm SM}_L \psi_L) \mathcal M_F \left(\begin{array}{c} f^{\rm SM}_R \\ \psi_R \end{array}\right),\quad\text{with}\quad \mathcal M_F = \left(\begin{array}{cc} y \frac{v}{\sqrt 2} & 0 \\ y^\prime \frac{v_D}{\sqrt 2} & M_\Psi \end{array}\right)\;.
\end{equation}
This mass matrix describes the mixing of a VL fermion with a SM fermion but, unlike in well-known VL scenarios where the new states mix with SM fermions via the Higgs boson, in this case the mixing is driven by $\Phi_D$ and the non-zero off-diagonal element is proportional to $v_D$. The mass matrix can be diagonalised by two unitary matrices, $V_{L,R}$, leading to the mass eigenstates $f$ and $F$, where $f$ identifies the SM fermion and $F$ its heavier partner:
\begin{equation}
 \Lag_m^f = (\bar f_L F_L) \mathcal M_F^d \left(\begin{array}{c} f_R \\ F_R \end{array}\right) = (\bar f_L F_L) V_{fL}^\dagger \mathcal M_F V_{fR} \left(\begin{array}{c} f_R \\ F_R \end{array}\right)\;.
\end{equation}
The two rotation matrices $V_{fL}=\left(\begin{array}{cc} \cos\theta_{fL} & \sin\theta_{fL} \\ -\sin\theta_{fL} & \cos\theta_{fL} \end{array} \right)$ and $V_{fR}=\left(\begin{array}{cc} \cos\theta_{fR} & \sin\theta_{fR} \\ -\sin\theta_{fR} & \cos\theta_{fR} \end{array} \right)$ diagonalise the products $\mathcal M_F^d \mathcal M_F^{d\dagger}$ and $\mathcal M_F^{d\dagger} \mathcal M_F^d$, respectively, and the mass eigenvalues are:
\begin{equation}
 m_{f,F}^2=\frac{1}{4} \left[y^2 v^2 + y^{\prime2} v_D^2 + 2 M_\Psi^2 \mp \sqrt{(y^2 v^2 + y^{\prime2} v_D^2 + 2 M_\Psi^2)^2-8y^2v^2M_\Psi^2}\right]\;.
\end{equation}
The fermion sector  therefore contains the SM fermion with mass $m_f$, a $\Z_2$-even partner with mass $m_F$ and a $\Z_2$-odd partner with mass $m_{\psi_D}$. The mass hierarchy is $m_f<m_{\psi_D}\leq m_F$.

It is possible to trade the Yukawa parameters for the masses of the physical fermions $\{m_f,m_{\psi_D},m_F\}$ as:
\begin{equation}
\label{eq:yukawas}
y = \sqrt{2} \frac{m_f m_F}{m_{\psi_D} v},\quad y^\prime = \sqrt2 \frac{\sqrt{(m_F^2 - m_{\psi_D}^2)(m_{\psi_D}^2 - m_f^2)}}{m_{\psi_D} v_D}\;.
\end{equation}
The mixing angles can also be expressed as function of the masses as:
\begin{equation}
\sin^2\theta_{fL} = \frac{m_f^2}{m_{\psi_D}^2} \frac{m_F^2 - m_{\psi_D}^2} {m_F^2 - m_f^2},\quad \sin^2\theta_{fR} = \frac{m_F^2 - m_{\psi_D}^2}{m_F^2 - m_f^2}\;.
\end{equation}
The left-handed mixing angle is suppressed by the ${m_f^2/m_{\psi_{D}}^2}$ ratio. This feature is different from the usual scenarios where a $SU(2)_L$-singlet VL fermion is added to the SM and allowed to mix with SM fermions and where the right-handed mixing angle is suppressed~\cite{Buchkremer:2013bha}. In this case, despite the fact that $\psi$ is a singlet under the SM gauge group, the mixing is driven by the $\SUD$ fermion doublet $\Psi$ and the $\SUD$ scalar doublet $\Phi_D$, the elements of which are also singlets under the EW gauge group and hence involves a right-handed SM fermion.

Finally, the new fermion sector is completely decoupled in the limit $m_F=m_{\psi_D}$, for which $y=y_{\rm SM}=\sqrt2 \frac{m_f}{v}$, $y^\prime=0$, $\sin\theta_{fL}=\sin\theta_{fR}=0$, so that the pure SM scenario is restored.

\subsubsection{Gauge bosons}
\label{sec:Dgaugemasses} 
The kinetic Lagrangian of $\Phi_H$ and $\Phi_D$ evaluated at the minimum of the scalar potential reads as follows:
\begin{equation}
\label{eq:gaugeLag}
 \Lag_{S}^{\rm kin}|_{v,v_D} \supset (\mathcal V^0_{\rm SM})^T \mathcal M_{\mathcal V^0_{\rm SM}}^2 \mathcal V^0_{\rm SM} + \frac{1}{4} g^2 v^2 W^+ W^- + \frac{1}{8} g_D^2 v_D^2 (V^0_{D0})^2 + \frac{g_D^2}{4} v_D^2 V^0_{D+} V^0_{D-} \;,
\end{equation}
where $\mathcal V^0_{{\rm SM}\mu}=(B_\mu~W_\mu^3)^T$. At tree level, the SM gauge bosons are not affected by the new $\Phi_D$ scalar, and therefore their masses correspond to the SM values, while the gauge bosons of $\SUD$ are all degenerate and their masses are
\begin{equation}
m_V\equiv m_{V^0_{D\pm}} = m_{V^0_{D0}} = \frac{g_D}{2} v_D \label{eq:VPmass} \;.
\end{equation}
The only electrically neutral  massive $\Z_2$-odd states of FPVDM scenarios are the $\SUD$ gauge bosons $V^0_{D\pm}$, which are thus identified as DM candidates.

The degeneracy in mass is broken at loop level by different effects. In the following, for making the notation more compact, we will label the two gauge bosons as:
\begin{eqnarray*}
 \left\{\begin{array}{ll}
 V^0_{D\pm} \equiv V_D & \text{ with mass } m_{V_D} \\
 V^0_{D0} \equiv V^\prime     & \text{ with mass } m_{V^\prime}
 \end{array}\right.\;.
\end{eqnarray*}
First of all, in the broken EW and dark gauge symmetry phases, a kinetic mixing arises between $V^\prime$ and both photon and $Z$ boson~\cite{Holdom:1985ag,Rizzo:2018vlb,Hisano:2020qkq,Rueter:2020qhf}. 
\begin{figure}[htb]
\hspace*{-0.5\linewidth}\includegraphics[width=0.35\linewidth]{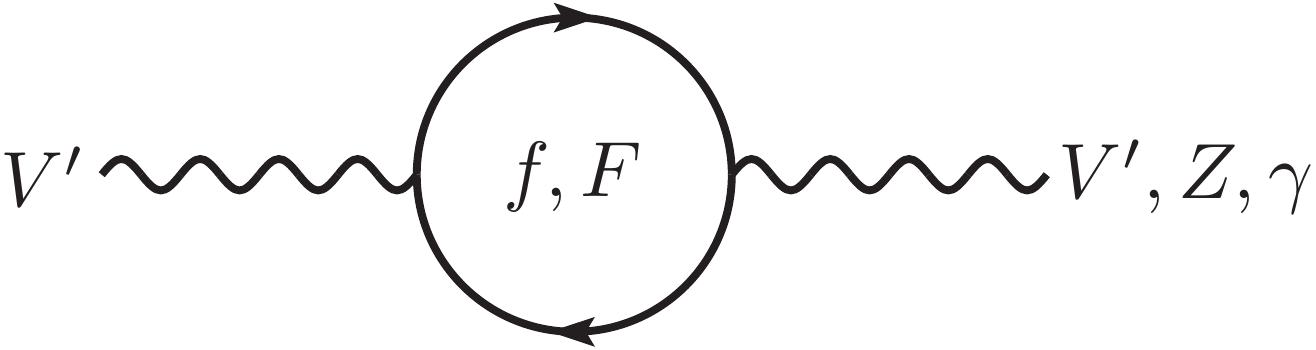}\\
\vspace*{0.5cm}
\hspace*{-0.5\linewidth}\includegraphics[width=0.35\linewidth]{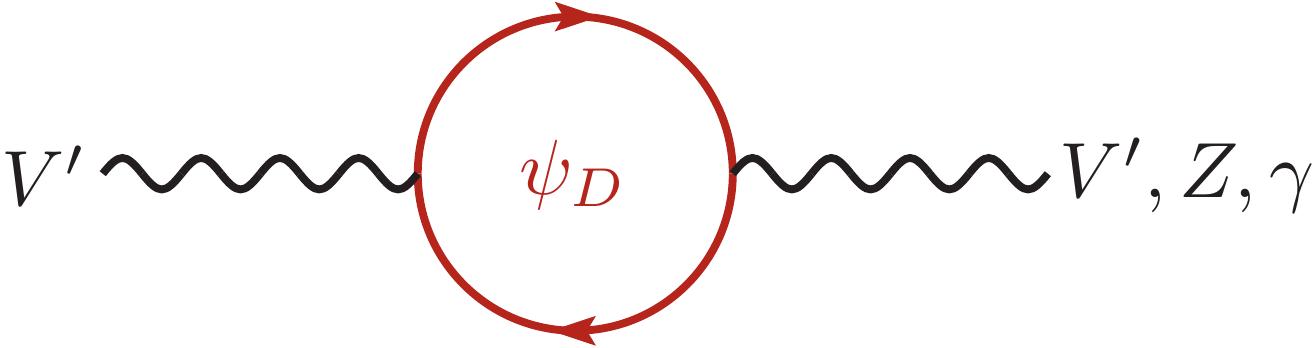}\\
\vspace*{-3cm}
\hspace*{0.5\linewidth}\includegraphics[width=0.35\linewidth]{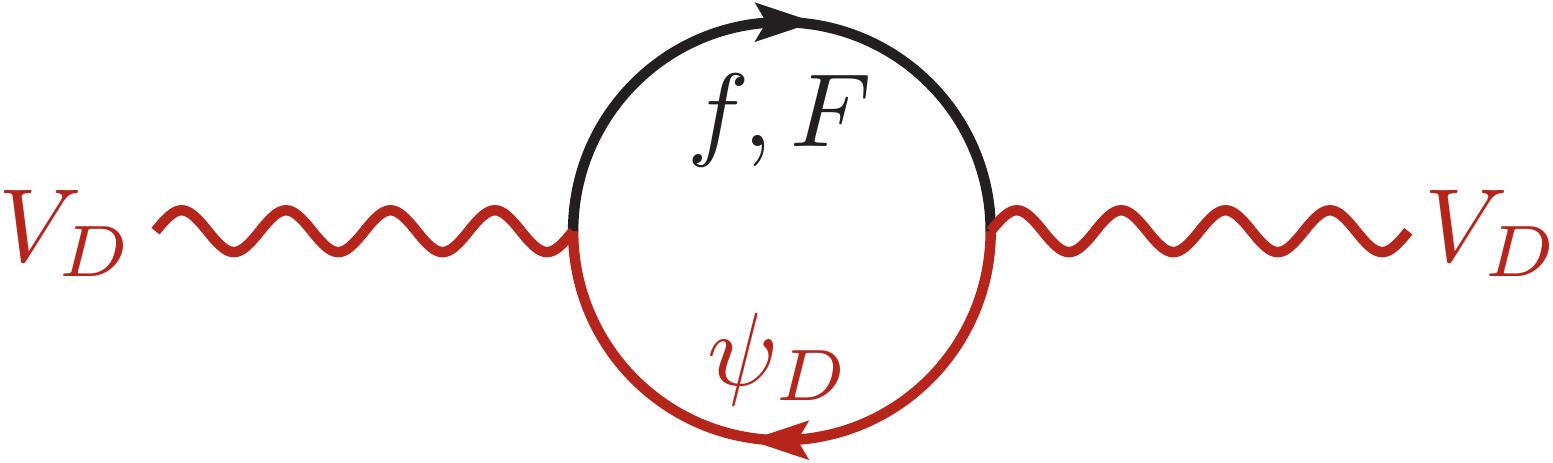}
\vspace*{1.0cm}
	\caption{The Feynman diagrams contributing to mass corrections and mixing of $SU(2)_D$ vector bosons, $V^{\prime}, Z,\gamma$ (left) and $V_D$ (right) at one loop level. $\Z_2$-odd particles are highlighted in red.}
	\label{fig:vector_SE}
\end{figure} 
Using analogous notation to  \cite{Rueter:2020qhf}, and assuming only one VL fermion doublet under $\SUD$ exists, the kinetic mixing parameters $\epsilon_{AV}$ and $\epsilon_{ZV}$ entering the kinetic mixing matrix
\begin{equation}
V^{\rm KM}=\left(
\begin{array}{ccc}
 1 & 0 & -{\epsilon_{AV}\over\sqrt{1-\epsilon_{AV}^2-\epsilon_{ZV}^2}} \\ 
 0 & 1 & -{\epsilon_{ZV}\over\sqrt{1-\epsilon_{AV}^2-\epsilon_{ZV}^2}} \\ 
 0 & 0 & {1\over\sqrt{1-\epsilon_{AV}^2-\epsilon_{ZV}^2}}\\ 
\end{array}
\right)\;,
\end{equation}
which rotates the $(A_\mu\;Z_\mu\;V^\prime_\mu)$ vector of gauge eigenstates, are determined by loops involving the only three fermions charged under the SM and dark gauge groups, $f$, $F$ and $\psi_D$, as shown in \cref{fig:vector_SE}. The scalar fields do not contribute due to the fact that neither $\Phi_H$ nor $\Phi_D$ transform under the SM and dark gauge groups at the same time. 
These loops can be evaluated separately for the $AV$ and $ZV$ mixings using the general expression of the gauge boson vacuum polarisation tensor provided in~\cite{Djouadi:1993ss}. For the $AV$ mixing the tensor is purely transverse and in the $q^2\to0$ limit reads $\Pi^{AV}_T\sim q^2 \epsilon_{AV}$, where 
\begin{eqnarray}
\epsilon_{AV}&=&{g_D e Q_f \over 8 \pi^2}\sum_{i=f,F,\psi_D} (V_{Li}^2+V_{Ri}^2) T^3_{D_i} \ln {m_i^2\over\mu^2}\nonumber\\
&=&{g_D e Q_f \over 8 \pi^2}\left[-{1\over2}(\sin\theta_{fL}^2+\sin\theta_{fR}^2)\ln{m_f^2\over\mu^2} - {1\over2}(\cos\theta_{fL}^2+\cos\theta_{fR}^2)\ln{m_F^2\over\mu^2} + \ln{m_{\psi_D}^2\over\mu^2} \right]\nonumber\\
&=&{g_D e Q_f \over 16 \pi^2}\left[ {m_{\psi_D}^4-m_f^2 m_F^2  \over (m_F^2 - m_f^2) m_{\psi_D}^2} \ln{m_f^2\over m_F^2} + 2\ln{m_{\psi_D}^2\over m_f m_F} \right]
\equiv {g_D e Q_f \over 16 \pi^2}\mathcal F^{AV}(r_f,r_{\psi_D})\;,
\end{eqnarray}
with $\{c,s,t\}_W\equiv\{\cos,\sin,\tan\}\theta_W$, $r_f=m_f/m_{\psi_D}$ and $r_{\psi_D}=m_{\psi_D}/m_F$.
The loop function
\begin{equation}
\mathcal F^{AV}(r_f,r_{\psi_D})={r_{\psi_D}^2-r_f^2 \over 1 - r_f^2 r_{\psi_D}^2} \ln (r_f^2 r_{\psi_D}^2) + \ln{r_{\psi_D}^2 \over r_f^2}
\end{equation}
does not depend on the specific fermion flavour but only on the ratios between fermion masses, and its numerical values are shown in~\cref{fig:KMplot}, where it is possible to see that the contribution of kinetic mixing completely cancels when $r_f=r_{\psi_D}$.
\begin{figure}[htbp!]
\centering
\includegraphics[width=0.45\linewidth]{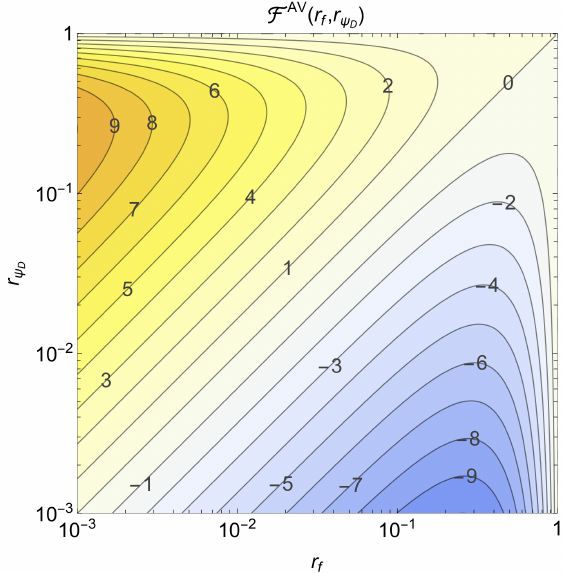}\\
\caption{Numerical values of the loop function $\mathcal F^{AV}(r_f,r_{\psi_D})$, with $r_f=m_f/m_{\psi_D}$ and $r_{\psi_D}=m_{\psi_D}/m_F$.}
\label{fig:KMplot}
\end{figure}

\begin{figure}[htbp!]
\centering
\includegraphics[width=0.3\linewidth]{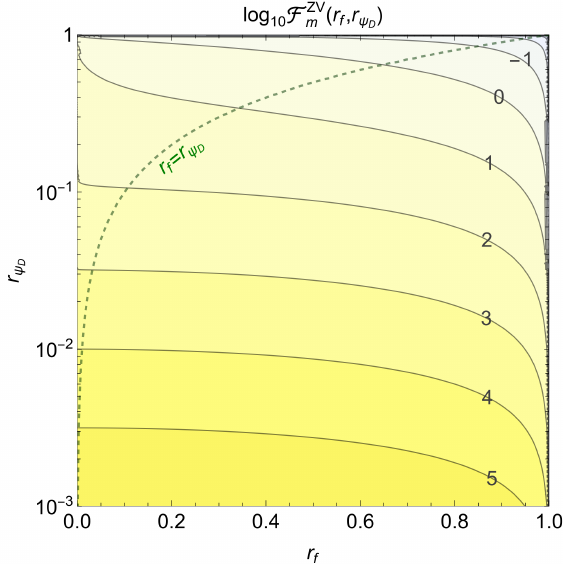}
\includegraphics[width=0.3\linewidth]{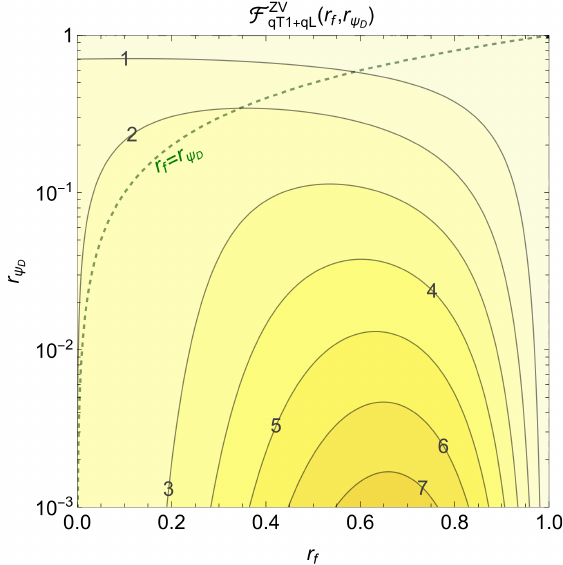}
\includegraphics[width=0.3\linewidth]{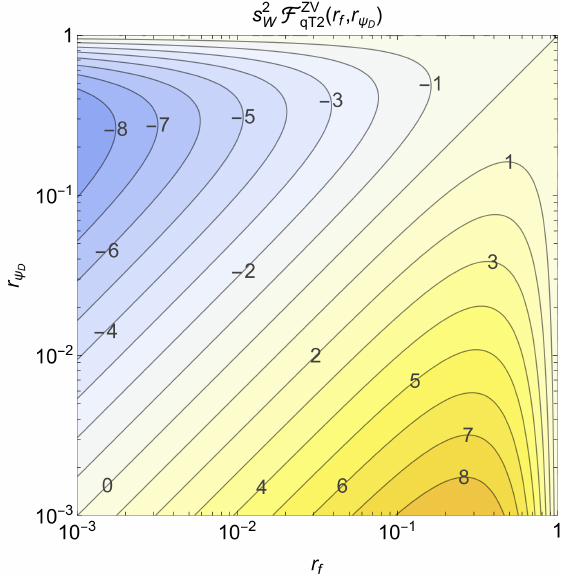}\\
\caption{Numerical values of the loop function $\mathcal F_{m,qT1+qL,qT2}^{ZV}(r_f,r_{\psi_D})$, with $r_f=m_f/m_{\psi_D}$ and $r_{\psi_D}=m_{\psi_D}/m_F$.}
\label{fig:KMZVplot}
\end{figure}

The vacuum polarisation tensor for the $ZV$ mixing, in contrast, is more involved due to the non-vector nature of the couplings on both sides of the loop. Its transverse and longitudinal components in the $q^2\to0$ limit read
\begin{eqnarray}
 \Pi^{ZV}_T(q^2\to0)&\sim& {g g_D \over 64\pi^2 c_w} \bigg[3 m_f^2 \mathcal F_m^{ZV}(r_f,r_{\psi_D}) + q^2 \bigg(\mathcal F_{qT1}^{ZV}(r_f,r_{\psi_D}) + Q_f s_W^2 \mathcal F_{qT2}^{ZV}(r_f,r_{\psi_D})\bigg) \bigg]\;,\\
 \Pi^{ZV}_L(q^2\to0)&\sim& {g g_D \over 64\pi^2 c_w} \bigg[3 m_f^2 \mathcal F_m^{ZV}(r_f,r_{\psi_D}) + q^2 \mathcal F_{qL}^{ZV}(r_f,r_{\psi_D})\bigg]\;,
\end{eqnarray}
such that the total contribution is
\begin{equation}
 \Pi^{ZV}(q^2\to0)\sim {g g_D \over 64\pi^2 c_w} \bigg[6 m_f^2 \mathcal F_m^{ZV}(r_f,r_{\psi_D}) + q^2 \bigg(\mathcal F_{qT1+qL}^{ZV}(r_f,r_{\psi_D}) + Q_f s_W^2 \mathcal F_{qT2}^{ZV}(r_f,r_{\psi_D})\bigg) \bigg]\;,
 \label{eq:ZVKM}
\end{equation}
where the functions $\mathcal F_{m,qT1+qL,qT2}^{ZV}(r_f,r_{\psi_D})$ are provided in \cref{app:KMfunctions} and their numerical values are shown in \cref{fig:KMZVplot}.

Besides the kinetic mixing, a mass mixing is thus induced between the SM $Z$ boson and $V^\prime$. The coefficients of the $ZV$ kinetic and mass mixing read:
\begin{eqnarray}
 \epsilon_{ZV}&=&{g g_D \over 64\pi^2 c_w} \bigg(\mathcal F_{qT1+qL}^{ZV}(r_f,r_{\psi_D}) + Q_f s_W^2 \mathcal F_{qT2}^{ZV}(r_f,r_{\psi_D})\bigg)\;,\\
 \Delta m^2_{ZV}&=&{3 g g_D \over 32\pi^2 c_w} m_f^2 \mathcal F_m^{ZV}(r_f,r_{\psi_D})\;.
\end{eqnarray}
The adimensional function $F_m^{ZV}(r_f,r_{\psi_D})$ appearing in the expression of the mass shift $\Delta m^2_{ZV}$ is small for $r_{\psi_D}\simeq1$ ({i.e.}, in the decoupling limit) and rapidly grows as $r_{\psi_D}$ decreases.
The function $F_{qT1+qL}^{ZV}(r_f,r_{\psi_D})$ has a similar behaviour but with a milder dependence on $r_{\psi_D}$. The function $F_{qT2}^{ZV}(r_f,r_{\psi_D})$ has a similar behaviour as $\mathcal F^{AV}(r_f,r_{\psi_D})$. 

The mass matrix of the $(Z_\mu\;V^\prime_\mu)$ system receives a shift proportional to the mass term in the vacuum polarisation tensor:
\begin{equation}
\tilde M^2_{ZV}=\left(
\begin{array}{cc}
{1\over4}(g^2+g^{\prime2})v^2 & 
{1\over2}\Delta m^2_{ZV} \\
{1\over2}\Delta m^2_{ZV} & 
{1\over4} g_D^2 v_D^2
\end{array}
\right)
=
\left(
\begin{array}{cc}
\tilde m_Z^2 & 
{1\over2} m_f^2 \epsilon^m_{ZV} \\
{1\over2} m_f^2 \epsilon^m_{ZV} & 
\tilde m_{V^\prime}^2
\end{array}
\right)\;,
\end{equation}
where the adimensional parameter $\epsilon^m_{ZV}=\Delta m^2_{ZV}/m_f^2$ has been introduced, and where loop contributions to the diagonal terms have been neglected because of the non-zero tree-level values. This matrix is rotated by $V^{\rm KM}$ into
\begin{equation}
M^2_{ZV}=\left(V^{\rm KM}\right)^T \tilde M^2_{ZV} V^{\rm KM}={1\over4}\left(
\begin{array}{cc}
(g^2+g^{\prime2})v^2 & 
-{(g^2 + g^{\prime2}) v^2 \epsilon_{ZV}-2m_f^2 \epsilon^m_{ZV}\over\sqrt{1-\epsilon_{AV}^2-\epsilon_{ZV}^2}} \\
-{(g^2 + g^{\prime2}) v^2 \epsilon_{ZV}-2m_f^2 \epsilon^m_{ZV}\over\sqrt{1-\epsilon_{AV}^2-\epsilon_{ZV}^2}} & 
{g_D^2 v_D^2+(g^2 + g^{\prime2}) v^2 \epsilon_{ZV}^2 - 4 m_f^2 \epsilon_{ZV}\epsilon^m_{ZV} \over1-\epsilon_{AV}^2-\epsilon_{ZV}^2}
\end{array}
\right)
\end{equation}
and diagonalised through a rotation with angle
\begin{equation}
 \tan 2\theta_{ZV} = \pm {2\left((g^2 + g^{\prime2}) v^2 \epsilon_{ZV} - 2m_f^2 \epsilon^m_{ZV}\right)\sqrt{1-\epsilon_{AV}^2-\epsilon_{ZV}^2} \over (1-\epsilon_{AV}^2-\epsilon_{ZV}^2)(g^2 + g^{\prime2}) v^2 - g_D^2 v_D^2 + 4 \epsilon_{ZV} \epsilon^m_{ZV}m_f^2}\;,
\end{equation}
which is positive for $m_{V^\prime}>m_Z$ and negative otherwise, and in the limit of small $\epsilon_{AV},\epsilon_{ZV}$ and $\epsilon^m_{ZV}$ becomes:
\begin{equation}
\tan 2\theta_{ZV}\simeq 2 \theta_{ZV}\simeq\pm2{2m_f^2 \epsilon^m_{ZV} - (g^2 + g^{\prime2}) v^2 \epsilon_{ZV} \over g_D^2 v_D^2 - (g^2 + g^{\prime2}) v^2}\;. 
\end{equation}
In the same limit the masses of the $Z$ and $V^\prime$ bosons read:
\begin{eqnarray}
m_Z^2&\simeq&{1\over4}(g^2+g^{\prime2})v^2\left[1 + \theta_{ZV}^2\left(1-{g_D^2 v_D^2 \over (g^2 + g^{\prime2}) v^2}\right)\right]\;,\\
m_{V^\prime}^2&\simeq& {1\over4} g_D^2 v_D^2 \left[ 1 + \epsilon_{AV}^2 + (\theta_{ZV} - \epsilon_{ZV})^2\left(1-{(g^2 + g^{\prime2}) v^2 \over g_D^2 v_D^2 }\right)\right]
\end{eqnarray}

The induced modification to the $Z$ boson mass (and an analogous modification to the $W$ boson mass induced by loops involving $F$ and a SM particle, potentially contributing to the $W$ mass anomaly observed by \cite{CDF:2022hxs}) are constrained by EW precision data and depend on specific realisations of the model. 
Another source of $V_D$ and $V^\prime$ mass split are the different fermionic loop corrections from $f, F$ and $\psi_D$ corresponding to the different $\Z_2$ parities of the $\SUD$ gauge bosons, as shown in \cref{fig:vector_SE}.
A detailed discussion of the 1-loop calculations is provided in \cref{app:masssplitting}. The mass splitting $\Delta m_V =m_{V_{D}}-m_{V^{\prime}}$ can be written in a compact form in terms of the parameters
\begin{equation}
	\epsilon_1=\frac{m_F^2-m_{\psi_D}^2}{m_F^2},\quad \epsilon_2=\frac{m_f^2}{m_F^2},\quad \epsilon_3=\frac{m_{V_D}^2}{m_F^2}\;.
\end{equation}
In the approximation of $\epsilon_1,\epsilon_2,\epsilon_3  \ll 1$ one has
\begin{equation}
	\Delta m_V^{\prime}
	\equiv
	\Delta m_V\big|_{\epsilon,\epsilon_2,\epsilon_3  \ll 1}=\frac{1}{640 \pi^2 m_{V_D}}\epsilon_1^2 g_D^2 m_F^2\left[(20+3\epsilon_3-15\epsilon_2+20\epsilon_2\epsilon_3)+10(3\epsilon_2-\epsilon_3-2\epsilon_2\epsilon_3)\log\epsilon_3\right] + o(\epsilon_1^2,\epsilon_2,\epsilon_3)\;.
	\label{eq:simple_mass_splitting_1}
\end{equation}

For practical purposes, the expression for $\Delta m_V$ can be further simplified by neglecting $\epsilon_2$ and $\epsilon_3$ and keeping the leading term in $\epsilon_1$, which leads to the following simple form:
\begin{equation}
	\Delta m_{V}^{\prime\prime}\equiv\Delta m_V^\prime\big|_{\epsilon_2,\epsilon_3=0}=\frac{g_D^2m_F^2}{32 \pi^2 m_{V_D}}
	\epsilon_1^2=
	\frac{g_D^2m_F^2}{32 \pi^2 m_{V_D}}\left(\frac{m_F^2-m_{\psi_D}^2}{m_F^2}\right)^2.
	\label{eq:simple_mass_splitting_2}
\end{equation}

The radiative mass splitting between the $V_D$ and $V^{\prime}$ bosons plays a very important role in the determination of relic density and DM Indirect Detection (ID) rates.
The range of validity of the approximations for $\Delta m_{V}$ presented above depends on the specific realisation of the FPVDM model and its parameter space. A detailed discussion of the respective numerical results for $\Delta m_{V}$ is given in \cref{sec:onlyTnoHHD} for a specific case study.

Finally, it is important to mention that the covariant derivative is modified by  the kinetic mixing as follows:
\begin{equation}
D_\mu \simeq \partial_\mu - i e Q A_\mu - i \left[{g\over c_w}(T_3-Q s_W^2) - g_D T^3_D \theta_{ZV}\right] Z_\mu - i \left[ g_D T^3_D - e Q \epsilon_{AV}  + {g\over c_w}(T_3-Q s_W^2) (\theta_{ZV}-\epsilon_{ZV}) \right] V^\prime_\mu\; \quad ,
\label{eq:modified covariant derivative}
\end{equation}
where we have included only leading terms in $\theta_{ZV}$ and $\epsilon_{ZV}$.

This modification has certain phenomenological consequences. Among the most relevant ones, the interaction of $V^\prime$ with all charged SM particles via the mixing parameter $\epsilon_{AV}$ allows the direct production of $V^\prime$ at the LHC via Drell-Yan topologies, and is therefore constrained by direct searches of heavy resonances. Also, the DM candidate $V_D$ can interact through EM multipoles with atomic matter, contributing to direct detection observables~\cite{Hisano:2020qkq}.
In the case where only one VL representation is present, the constraints coming from these processes depend only on the fermion charge $Q$ and on the mass ratios $r_f$ and $r_{\psi_D}$, but not on the specific flavour of the fermion.

\subsubsection{Scalars}
\label{sec:scalarsector}

The scalar potential of eq. \eqref{eq:scalarpotential} is constructed starting from the 8 degrees of freedom of all the scalar fields of the theory: 4 for $\Phi_H$ and 4 for $\Phi_D$. The theory contains 6 massive gauge bosons: $Z$, $W^\pm$, $V^\prime$ and $V_D$ (with two opposite D-isospin values). Therefore 6 Goldstone bosons are needed to give the corresponding longitudinal components. Thus, 2 degrees of freedom are left, which correspond to physical massive scalars: the SM Higgs boson, $h$, and a further CP-even scalar, $H$. Upon expressing the neutral scalars in the interaction eigenstates in terms of their components in the unitary gauge as
\begin{align}
\phi^0 &= {1\over\sqrt{2}} (v + h_1) \;,\\
\varphi^0_{D-1/2} &= {1\over\sqrt{2}} (v_D + \varphi_1) \;,
\end{align}
the Lagrangian terms for scalar masses can be written as:
\begin{equation}
\Lag_m^{\mathcal S} = 
-(h_1 \; \varphi_1) 
\left(
\begin{array}{cc}
\lambda v^2 & {\lambda_{\Phi_H\Phi_D}\over2} v v_D \\ {\lambda_{\Phi_H\Phi_D}\over2} v v_D & \lambda_D v_D^2
\end{array}
\right)
\left( \begin{array}{c} h_1 \\ \varphi_1 \end{array} \right) \;.
\end{equation}
The mass eigenvalues are obtained by diagonalising the mass matrix via a rotation matrix $V_S=\left(\begin{array}{cc} \cos\theta_S & \sin\theta_S \\ -\sin\theta_S & \cos\theta_S \end{array} \right)$ and are 
\begin{equation}
 m_{h,H}^2=\lambda v^2+\lambda_D v_D^2\mp\sqrt{(\lambda v^2-\lambda_D v_D^2)^2+\lambda_{\Phi_H\Phi_D}^2v^2 v_D^2} 
\end{equation}
whilst the mixing angle is 
\begin{equation}
\sin\theta_S = \sqrt{2{m_H^2 v^2 \lambda - m_h^2 v_D^2 \lambda_D \over m_H^4 - m_h^4}}\;.
\end{equation}
Even in the absence of explicit mixing induced by the quadratic term, {i.e.}, even if $\lambda_{\Phi_H\Phi_D}=0$, $h_1$ and $\varphi_1$ can mix at one-loop via the their interactions with fermions. The consequences of this mixing, which can also affect Higgs-related observables, go beyond the scopes of this analysis, and will be treated in a future work.

\subsection{Flavour structure and Cabibbo-Kobayashi-Maskawa (CKM) matrix}\label{sec:flavourstructure}

The previous treatment assumed the presence of one VL $\SUD$ doublet interacting with one SM fermion, without specifying the flavour structure involved. If the full flavour structure of the SM is considered, different possibilities might arise. A VL fermion can interact with one or more SM flavours and there can be multiple VL fermions.

The most general Lagrangian, accounting for the above-mentioned possibilities, is
\begin{eqnarray}
\Lag_m &=& M_U^I \bar U_I U_I + M_D^J \bar D_J D_J + M_E^K \bar E_K E_K \nonumber\\ 
&+& y_u^i \bar Q_{iL}^{\rm SM} \tilde \Phi_H u^{\rm SM}_{iR} + y_d^{i} \tilde V_{\rm CKM}^{ij} \bar Q^{\rm SM}_{iL} \Phi_H d^{\rm SM}_{jR} + y_l^i \bar L^{\rm SM}_{iL} \Phi_H l^{\rm SM}_{iR} + h.c. \nonumber\\
&+& (y^\prime_u)^{Ij} \bar U_{IL} \Phi_D u^{\rm SM}_{jR} + (y^\prime_d)^{Jj} \bar D_{JL} \Phi_D d^{\rm SM}_{jR} + (y^\prime_l)^{Kj} \bar E_{KL} \Phi_D l^{\rm SM}_{jR} + h.c.\;,
\end{eqnarray}
where $\tilde \Phi_H=i\tau_2\Phi_H^*$, $i,j=1,2,3$ are SM flavour indices and $I,J,K$ run over the flavours of the VL partners. The SM Yukawa couplings have been diagonalised exploiting the flavour symmetries and the SM CKM matrix ({i.e.}, the CKM matrix if no VL states were introduced) and $\tilde V_{\rm CKM}$ has been introduced to parametrise the misalignment between the flavour and mass eigenstates in the down sector. 

The most generic mass matrices read as follows: 
\begin{equation}
\mathcal M_U = 
\left(
\begin{array}{c|c}
y_u^i{v\over\sqrt2} & 0^{iI} \\
\hline
(y_u^\prime)^{Ii}{v_D\over\sqrt2} & M_U^I
\end{array}
\right)\;,\quad
\mathcal M_D = 
\left(
\begin{array}{c|c}
y_d^i \tilde V_{\rm CKM}^{ij} {v\over\sqrt2} & 0^{iJ} \\
\hline
(y_d^\prime)^{Ji}{v_D\over\sqrt2} & M_D^J
\end{array}
\right)\;,\quad
\mathcal M_E = 
\left(
\begin{array}{c|c}
y_l^i{v\over\sqrt2} & 0^{iK} \\
\hline
(y_l^\prime)^{Ki}{v_D\over\sqrt2} & M_E^K
\end{array}
\right)\;.
\end{equation}
The mass matrices can be diagonalised by two unitary matrices $V_{L}$ and $V_{R}$, with dimension $3+\{I,J,K\}$ depending on the fermion type.
If the same VL fermion interacts with multiple flavours of SM fermions, the most constraining effects are represented by modifications to SM observables, induced by Flavour Changing Neutral Currents (FCNCs) \cite{Cacciapaglia:2011fx,Okada:2012gy}. If for each SM fermion there is a VL partner, the matrix proportional to $y^\prime$ is diagonal as well and no mixing is induced between different SM and VL flavours, thus fermions from the dark sector only interact with the corresponding SM flavour. In the following we will limit the analysis to this simpler scenario.

An important property of this construction is that the CKM matrix of the SM receives contributions from new physics. In fact, the SM charged current is 
\begin{eqnarray}
J_{W^+}^\mu &=& {g\over\sqrt2}(\bar u^{\rm SM\;i}_L\; \bar U_L^I) \gamma^\mu \left(\begin{array}{c|c}1_{3\times3} & 0^{3J} \\ \hline 0^{I3} & 0^{IJ} \end{array}\right) \left(\begin{array}{c}d_L^{\rm SM\;i} \\ D^J_L \end{array}\right) \nonumber\\
&=& {g\over\sqrt2}(\bar u_L^i\; \bar u_L^{\prime I}) \gamma^\mu V_{uL}^\dagger\left(\begin{array}{c|c}\tilde V_{\rm CKM} & 0^{3J} \\ \hline 0^{I3} & 0^{IJ} \end{array}\right)V_{dL} \left(\begin{array}{c}d^i_L \\ d^{\prime J}_L \end{array}\right)\;,
\end{eqnarray}
such that the entries of the measured CKM matrix are given by
\begin{equation}
V_{CKM}^{ij}=(V_{uL}^\dagger)^{ik}\tilde V_{\rm CKM}^{kl}V_{dL}^{kj}\;.
\end{equation}

\subsection{FPVDM parameter space}\label{sec:inputparameters}

The Lagrangian parameters of the model are the following:
\begin{itemize}
\item gauge couplings: $g,g^\prime,g_D$;
\item Scalar potential parameters: $\mu,\lambda,\mu_D,\lambda_D,\lambda_{\Phi_H\Phi_D}$;
\item Yukawa couplings and VL quark mass: $y,y^\prime,m_{\psi_D}$;
\item $\tilde V_{\rm CKM}$ parameters.
\end{itemize}
Assuming that the new VL fermion interacts only with one SM flavour, these parameters can be traded for the masses of all the physical states, the weak coupling constant $g$ (or equivalently, the fine structure constant $\alpha_{\rm EM}$), the new gauge coupling $g_D$, the mixing angle between the scalar fields $\theta_S$ and the measured CKM parameters. A complete set of parameters is therefore:
\begin{eqnarray}
\{g,m_W,m_Z\},~\{g_D,m_{V_D}\},~\{m_h,m_H,\sin\theta_S\},~\{m_f,m_F,m_{\psi_D}\}~\text{and}~V_{\rm CKM}\;,
\end{eqnarray}
but, since ${g,m_W,m_Z}$, $m_h$,  $m_f$ and $V_{\rm CKM}$ are precisely measured SM parameters, we are left with the following six independent new physics parameters, namely:
\begin{eqnarray}
g_D,m_{V_D}, m_H, \sin\theta_S, m_F, m_{\psi_D}\;.
\label{eq:pars}
\end{eqnarray}

Approximating the CKM as a diagonal matrix for simplicity, the relations between the Lagrangian parameters connected  to the new physics components and the input parameters take a very simple form:
\begin{eqnarray}
v &=& {2 m_W \over g}, \quad v_D = {2 m_{V_D}\over g_D}\;,\\
\lambda   &=& {g^2\over8m_W^2} (m_h^2 \cos^2\theta_S + m_H^2 \sin^2\theta_S)\;, \\
\lambda_D &=& {g_D^2\over8m_{V_D}^2} (m_h^2 \sin^2\theta_S + m_H^2 \cos^2\theta_S)\;, \\
\lambda_{\Phi_H\Phi_D}&=&{g\;g_D\over 8 m_W m_{V_D}} (m_H^2 - m_h^2) \sin2\theta_S\;,\label{eq:lam4}\\
\mu^2 &=& {1\over2}\left(m_h^2 \cos^2\theta_S + m_H^2 \sin^2\theta_S + {1\over2}{g\over g_D}{m_{V_D}\over m_W} (m_H^2 - m_h^2) \sin2\theta_S\right)\;,\label{eq:mu2}\\
\mu_D^2 &=& {1\over2}\left(m_h^2 \sin^2\theta_S + m_H^2 \cos^2\theta_S + {1\over2}{g_D\over g}{m_W\over m_{V_D}} (m_H^2 - m_h^2) \sin2\theta_S\right)\;,\label{eq:mud2}\\
y &=& {g\;m_f\;m_F \over \sqrt 2 m_{\psi_D} m_W}\;,\\
y^\prime &=& {g_D\sqrt{(m_F^2 - m_{\psi_D}^2)(m_{\psi_D}^2 - m_f^2)}\over \sqrt2 m_{\psi_D} m_{V_D}}\;. \label{eq:yp-pert}
\end{eqnarray}
 
The minimisation conditions of the scalar potential in eq. \eqref{eq:minVconditions} are automatically satisfied. 
If $\lambda_{\Phi_H\Phi_D}<0$, which corresponds to $m_h>m_H$, the condition $\lambda_{\Phi_H\Phi_D}^2<4 \lambda \lambda_D$ translates into ${1\over16} {g^2g_D^2\over m_W^2 m_{V_D}^2} m_h^2 m_H^2>0$, which is always true, whilst, if $\lambda_{\Phi_H\Phi_D}>0$, the conditions $2 \lambda \mu_D^2 > \lambda_{\Phi_H\Phi_D} \mu^2$ and  $2 \lambda_D \mu^2 > \lambda_{\Phi_H\Phi_D} \mu_D^2$ translate  into ${1\over8} {g^2\over m_W^2} m_h^2 m_H^2>0$ and ${1\over8} {g_D^2\over m_{V_D}^2} m_h^2 m_H^2>0$, respectively, again automatically satisfied.

\begin{figure}[h!]
     \centering
     \includegraphics[width=.48\textwidth]{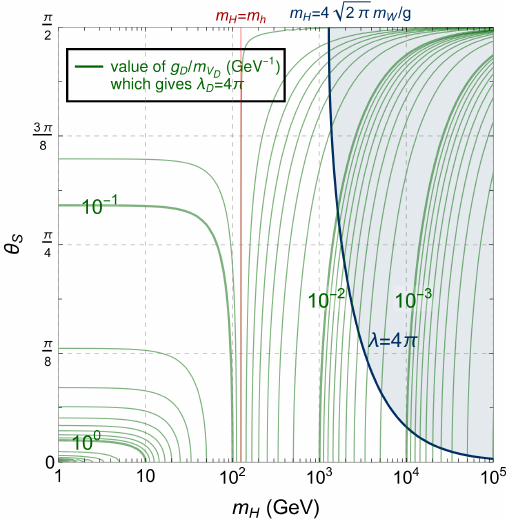}
     \includegraphics[width=.48\textwidth]{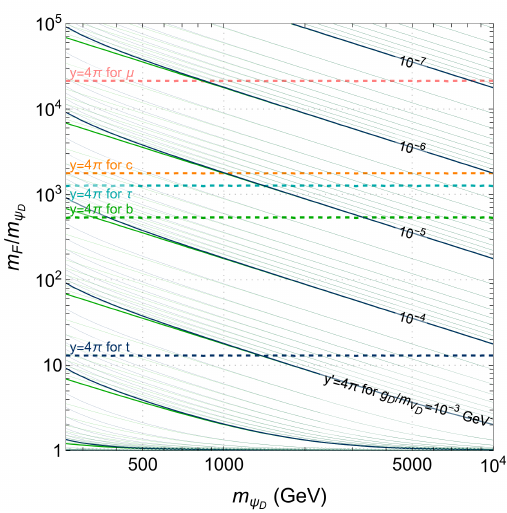}
     \caption{{Left:} the maximum value of $m_H$ and minimum value of $\theta_S$ for $\lambda<4\pi$ and $\lambda_D<4\pi$ as function of $g_D\over m_{V_D}$. The regions corresponding to $\lambda_D<4\pi$ are to the left of the green lines. {Right:} the maximum value of the $m_F/m_{\psi_D}$ ratio as function of $m_{\psi_D}$ and $g_D\over m_{V_D}$, and under different hypotheses about which SM fermion interacts with the $\SUD$ doublet $\Psi$, to satisfy the perturbativity conditions $\{y,y^\prime\} < 4\pi$.}
     \label{fig:maxmH}
 \end{figure}

For a perturbative analysis of the parameter space we need to identify the regions where coupling parameters do not become too large, in order to make sure that all predictions on the model are reliable. A complete loop description of all the sectors of the model is beyond the scope of this analysis and therefore we assume that perturbativity is achieved by the requirement for all couplings of the FPVDM model to be (optimistically) below $4\pi$.
For example, the requirement $\lambda < 4\pi$ defines the maximal value of $m_H$ for a given value of the scalar mixing angle, $\theta_S$, as shown by the blue contour in the left panel of \cref{fig:maxmH}.
 The same figure presents contours for the $g_D/m_{V_D}$ ratio in the $\{m_H,\theta_S\}$ plane corresponding to $\lambda_D = 4\pi$, which indicates the perturbativity limit on the respective parameters.

The perturbative constraints on the Yukawa couplings $y$ and $y^\prime$  imply that  the ratio between the masses of the new fermions $F$ and $\psi_D$ cannot be too large.
The condition for $y$ reads as ${m_F\over m_{\psi_D}}<4\pi {\sqrt2 m_W \over g m_f}$.  At the same time,  the $y^\prime<4\pi$ condition is defined also by  the $g_D/m_{V_D}$ ratio, as one can see from eq.~\eqref{eq:yp-pert}. Both constraints from $y$ and $y'$ perturbativity requirements are presented in the right panel of \cref{fig:maxmH} in the  $(m_{\psi_D},{m_F\over m_{\psi_D}})$ plane.
In our analysis of the parameter space we indicate the respective regions where perturbativity constraints are violated.
%

\section{On the origin of the global $U(1)$ symmetry}\label{sec:theory}

One of the main open questions of the construction presented in this analysis is the origin of the global $U(1)$ symmetry (with its $\Z_2$ parity subset) which has to be imposed to avoid the contemporary presence of two Yukawa interactions involving $\Phi_D$ and $\Phi_D^*$ which would explicitly break $\SUD$, and therefore spoil the stability of the DM candidate. 
A theoretical origin of the symmetry would provide a robust ground for the consistency of the model. In this section we explore two options for explaining such origin. 
The first involves promoting the global $U(1)$ to a local gauge symmetry, $U(1)_{\rm D}$, in the dark sector, which would generate a mirror version of the SM EW sector in the dark sector, the two of which can be connected by the mixed $(\Phi^\dagger \Phi)(\Phi_D^\dagger \Phi_D)$ quartic term in the full potential and by the gauge  kinetic mixing between $U(1)_Y$ and $U(1)_D$. In this scenario the $U(1)_{\rm D}$ local symmetry would be associated to a conserved dark-charge completely analogous to the EM charge of QED, thus giving literal meaning to the notation $V^0_{D0}$ and $V^0_{D\pm}$ for the $\SUD$ gauge bosons in the dark sector. 

The second involves the existence of a strongly-coupled sector whose condensates form the particle in the low energy regime, in particular, a residual parity for the composite sector is present due to the specific vacuum alignment present in this kind of models (which would typically also imply an extended Higgs sector). A detailed discussion is given in \cite{Ma:2015gra} and further used in \cite{Wu:2017iji} for the case of a scalar DM candidate. 

\subsection{A dark electroweak sector}\label{sec:darkEW}
In this scenario the SM is augmented with a dark sector constructed starting from a dark gauge group $\mathcal G_D$ with same structure as the EW gauge group of the SM. The gauge group is spontaneously broken as:
\begin{equation}
 \mathcal G = \mathcal G_{\rm SM} \times \mathcal G_D = SU(2)_L \times U(1)_Y \times \SUD \times \UYD \longrightarrow U(1)_{\rm EM} \times \UD \;.
\end{equation}
The gauge boson associated to $\UYD$ is labelled as $B^0_{D0\mu}$. The full covariant derivative is
\begin{equation}
\label{eq:covdev2}
 D_\mu = \partial_\mu 
 - \left( i {g \over \sqrt 2} W^{\pm}_\mu T^\pm + i g W^3_\mu T_3 + i g^\prime Y B_\mu \right)
 - \left( i {g_{D} \over \sqrt 2} V^0_{D\pm\mu} T^\pm_D + i g_{D} V^0_{D0\mu} T_{3D} + i g^\prime_D Y_D B^0_{D0\mu}\right)\;,
\end{equation}
where $g$ and $g^\prime$ are, respectively, the weak and hypercharge coupling constants, $g_D$ and $g_D^\prime$ are the $\SUD$ and $\UYD$ coupling constants, $T_3$ and $Y$ are the weak-isospin and weak-hypercharge, $T_{3D}$ and $Y_D$ are the dark-isospin associated with$\SUD$ and the dark-hypercharge associated with  $\UYD$ and where the indices of the $T_D$ matrices act only on the $\SUD$ elements and are diagonal with respect $SU(2)_L$  while  the indices of the $T$ matrices act only on the $SU(2)$ elements and are diagonal with respect to $\SUD$.

The unbroken $U(1)_{\rm D}$ continuous symmetry is associated to a conserved charge, labelled D-charge, defined as:
\begin{equation}
 Q_D = T_{3D} + Y_D\;.
\end{equation} 
Notice that \textit{the D-charge is not associated with the electric charge}: electrically neutral particles can be D-charged and vice versa. The only assumption to be made in this scenario is that {\it all the SM states are neutral under the conserved D-charge $Q_D$}. This however does not necessarily imply that all the states of new physics are charged under $\UD$ or that they must be neutral under the conserved SM charges.

The fields responsible for the breaking of the gauge symmetry are the two scalar doublets $\Phi_H$ and $\Phi_D$ described in section  \ref{sec:Z2SUD}. Since $\Phi_H$ is singlet with respect to the dark gauge group and $\Phi_D$ is singlet with respect to the EW gauge group, given  the absence of gauge kinetic mixing terms, no mixing is induced between the fully neutral gauge bosons $W_\mu^3$, $B_\mu$, $V^0_{D0\mu}$ and $B^0_{D0\mu}$. In complete analogy with the SM, by counting the number of bosonic degrees of freedom, one massless gauge boson is predicted in the dark gauge sector and the other dark gauge bosons receive different masses. We can thus define the mass eigenstates $\gamma_D$, $Z^0_D$ and $W^0_{D\pm}$ with values
\begin{eqnarray}
\label{eq:VPmass2}
M_{\gamma_D} &=& 0\;, \\
M_{Z^0_D}   &=& {1\over2} \sqrt{g_{D}^2 + g^{\prime 2}_D}~ v_D\;, \label{eq:ZPmass2}\\
M_{W^0_{D\pm}} &=& {g_D\over2} v_D\;, 
\end{eqnarray}
such that the masses of the DM vector $V^0_{D\pm}$ and of the D-charge-neutral gauge boson $V^0_{D0}$ receive a splitting proportional to ${1\over2} g^\prime_D v_D$.
The particle content of the model is summarised in Table~\ref{tab:particlesQN2}.
One should note that the presence of the massless dark radiation from the  unbroken $U(1)$ is not necessarily a problem as soon as it does not contribute too much to relativistic degrees of freedom at BBN and allows the formation of structures as small scales. As shown in~\cite{Ackerman:2008kmp}, for example, it can be achieved when at the DM decouples from the dark radiation at  high redshifts.

{
\setlength{\tabcolsep}{3pt}
\setlength{\arraycolsep}{0pt}
\begin{table}[h]
\centering
\begin{tabular}{c|cc|cc||cc}
\toprule
& \multicolumn{2}{c|}{EW} & \multicolumn{2}{c|}{Dark} & \multicolumn{2}{c}{Unbroken} \\
&$SU(2)_L$ & $U(1)_Y$ & $\SUD$ & $U(1)_{\rm YD}$ & $U(1)_{\rm EM}$ & $U(1)_{\rm D}$ \\
\midrule       
\midrule
\multicolumn{7}{c}{Scalar fields} \\
\midrule
\midrule
\multirow{2}{*}{$\Phi_H=\left(\begin{array}{c} \phi^+ \\ \phi^0 \end{array}\right)$} & \multirow{2}{*}{$\mathbf{2}$} & \multirow{2}{*}{$1/2$} & \multirow{2}{*}{$\mathbf{1}$} & \multirow{2}{*}{$0$} & 1 & 0 \\
& & & & & 0 & 0 \\
\midrule
$\Phi_D=\left(\begin{array}{c} \varphi^0_{D+{1\over2}} \\ \varphi^0_{D-{1\over2}} \end{array}\right)$ & $\mathbf{1}$ & $0$ & $\mathbf{2}$ & $1/2$ & $\begin{array}{c} 0 \\ 0 \end{array}$ & $\begin{array}{c} 1 \\ 0 \end{array}$ \\
\midrule       
\midrule
\multicolumn{7}{c}{Fermion fields} \\
\midrule
$f_{L}^{\rm SM}=\left(\begin{array}{c} f_{u,\nu}^{\rm SM} \\ f_{d,\ell}^{\rm SM} \end{array}\right)_{\!\!\!L}$ & $\mathbf{2}$ & $1/6,-1/2$ & $\mathbf{1}$ & 0 & $T_{3f}+Y_f$ & 0 \\
$u_{R}^{\rm SM},\nu_{R}^{\rm SM}$  & $\mathbf{1}$ & $2/3,0$ & $\mathbf{1}$ & 0 & $T_{3f}+Y_f$ & 0 \\
$d_{R}^{\rm SM},\ell_{R}^{\rm SM}$ & $\mathbf{1}$ & $-1/3,-1$ & $\mathbf{1}$ & 0 & $T_{3f}+Y_f$ & 0 \\
\midrule
\multirow{2}{*}{$\Psi=\left(\begin{array}{c} \psi^D \\ \psi \end{array}\right)$} & \multirow{2}{*}{$\mathbf{1}$} & \multirow{2}{*}{$Q_\Psi$} & \multirow{2}{*}{$\mathbf{2}$} & \multirow{2}{*}{$1/2$} & \multirow{2}{*}{$Q_\Psi$} & $1$ \\
& & & & & & $0$ \\
\midrule       
\midrule
\multicolumn{7}{c}{Vector fields} \\
\midrule
\midrule
$W_\mu=\left(\begin{array}{c} W^+_\mu \\ W^3_\mu \\ W^-_\mu \end{array}\right)$ & $\mathbf{3}$ & $0$ & $\mathbf{1}$ & $0$ & $\begin{array}{c} 1 \\ 0 \\ -1 \end{array}$ & $\begin{array}{c} 0 \\ 0 \\ 0 \end{array}$ \\
\midrule
$B_\mu$ & $\mathbf{1}$ & 0 & $\mathbf{1}$ & $0$ & 0 & 0 \\
\midrule
$V_{D\mu}=\left(\begin{array}{c} V^0_{D+\mu} \\ V^0_{D0\mu} \\ V^0_{D-\mu} \end{array}\right)$ & $\mathbf{1}$ & $0$ & $\mathbf{3}$ & $0$ & $\begin{array}{c} 0 \\ 0 \\ 0 \end{array}$ & $\begin{array}{c} 1 \\ 0 \\ -1 \end{array}$ \\
\midrule
$B^0_{D0\mu}$ & $\mathbf{1}$ & 0 & $\mathbf{1}$ & $0$ & 0 & 0 \\
\bottomrule
\end{tabular}
\caption{\label{tab:particlesQN2}The quantum numbers under the EW and dark gauge group $\SUD\times U(1)_{\rm D}$ of the particles of the model. The charges of the unbroken groups $U(1)_{\rm EM}$ and $U(1)_{\rm D}$ are also provided.}
\end{table}
}

The presence of two $U(1)$ gauge groups, however, allows for the existence of a renormalisable and gauge-invariant kinetic mixing term already in the unbroken EW and dark symmetry phases, such that the Lagrangian of the $U(1)_Y\times U(1)_{\rm YD}$ sector is
\begin{equation}
 -\Lag_{\rm KM} = {1\over4} B_{\mu\nu} B^{\mu\nu} + {1\over4} B_{D\mu\nu} B_D^{\mu\nu} + {\varepsilon\over2} B_{\mu\nu} B_D^{\mu\nu}\;,
\end{equation}
where $B_{D\mu\nu}$ is the field tensor of $U(1)_{\rm YD}$ and $\varepsilon$ is the kinetic mixing parameter. The diagonalisation of the kinetic terms can be obtained through the rotation~\cite{Feldman:2007wj}:
\begin{equation}
 \left(\begin{array}{c} B^\mu \\ B^{0\mu}_{D0} \end{array}\right) = \left(\begin{array}{cc} {1\over\sqrt{1-\varepsilon^2}} & 0 \\ -{\varepsilon^2\over\sqrt{1-\varepsilon^2}} & 1 \end{array}\right) \left(\begin{array}{cc} \cos\theta_k & -\sin\theta_k \\ \sin\theta_k & \cos\theta_k \end{array}\right) \left(\begin{array}{c} B_1^\mu \\ B_2^\mu \end{array}\right)
\end{equation}

The kinetic-mixing term induces a modification in the mass mixing matrix of the fully neutral gauge bosons. Upon diagonalisation, two massless eigenstates are obtained, corresponding to the SM photon and to a massless dark photon, and two massive eigenstates, corresponding to the $Z$ boson and to a massive $Z^\prime$ boson. The full expressions of the mass mixing matrix and of the mass eigenstates can be found in \cref{app:darkEW}. Expanding the mass eigenstates of $Z$ and $Z^\prime$ for small $\varepsilon$, the lowest order terms assume a simple form:
\begin{eqnarray}
 M_Z^2 &=& {v^2\over4}\left[g^2+g^{\prime 2} \left(1 + {(g^2+g^{\prime 2}) v^2-g_D^2 v_D^2\over(g^2+g^{\prime 2}) v^2-(g_D^2+g_D^{\prime 2}) v_D^2} \varepsilon^2\right) \right] + \mathcal O(\varepsilon^4)\;, \\
 M_{Z^\prime}^2 &=& {v_D^2\over4}\left[g_D^2+g_D^{\prime 2}\left(1+{g^2 v^2 - (g_D^2+g_D^{\prime 2}) v_D^2 \over (g^2+g^{\prime 2}) v^2-(g_D^2+g_D^{\prime 2}) v_D^2} \varepsilon^2\right) \right] + \mathcal O(\varepsilon^4)\;,
\end{eqnarray}
which in the $\varepsilon\to0$ limit (no kinetic mixing) reduce to the SM value and eq.\eqref{eq:ZPmass2}, respectively.
Of course, analogously to the FPVDM model with the global $U(1)$ symmetry, after spontaneous breaking of EW and dark symmetries, kinetic and mass mixing terms arise at loop level as illustrated in \cref{sec:Dgaugemasses}, involving the four electrically and D-charge neutral gauge bosons.
The implications of this scenario and the derivation of its experimental bounds are beyond the scope of this analysis and are reserved for future developments.

\subsection{A composite origin}\label{sec:composite}
In the case of composite models the discrete symmetries allowing the stability of the DM particle depend on the model building details of the composite sector. However, this does not mean that the DM candidate and the corresponding discrete symmetries are an arbitrary choice. The composite effective chiral Lagrangian is invariant under a parity changing the signs of all the pseudo Nambu-Goldstone Bosons (pNGBs), as they appear in bilinear terms in the Lagrangian. Furthermore,  these models contain by construction explicit symmetry breaking terms, so more scrutiny is needed to understand if a pNGB can be stable due to a residual parity and therefore be used as a particle describing DM. The origin of the non-invariance with respect to parity (and also charge conjugation) is due to the choice of the vacuum while the strong techni-sector at the origin of these models is instead parity invariant as it is VL with respect to the composite gauge dynamics and the SM gauge group. Once possible parities acting on the pNGBs are identified, these models require a careful check of their invariance, including the Wess-Zumino-Witten terms. In explicit realisations studied in the literature, e.g.,  in \cite{Ma:2015gra,Wu:2017iji}, a stable pNGB multiplet allowing the description of DM can indeed be found. 

\section{A case study: top portal with no mixing between $h$ and $H$}
\label{sec:onlyTnoHHD}

This section is dedicated to a specific realisation of the model. It is assumed that only one VL partner exists, and interacts exclusively with the SM top quark. Moreover it is further assumed that the Higgs bosons $h$ and $H$ do not mix, {i.e.}, $\theta_S=0$. These choices significantly simplify the expressions of the Lagrangian parameters, which read:
\begin{eqnarray}
v   &=& {2 m_W \over g},~\mu^2 = {m_h^2\over2},~\lambda = {g^2 m_h^2\over8m_W^2},\\
v_D &=& {2 m_{V_D}\over g_D},~\mu_D^2 = {m_H^2\over2},~\lambda_D = {g_D^2 m_H^2\over8m_{V_D}^2},~\lambda_{\Phi_H\Phi_D}=0, \\
y_t &=& {g\;m_t\;m_T \over \sqrt 2 m_{t_D} m_W}=y_t^{\rm SM} {m_T\over m_{t_D}} , \quad y_t^\prime = {g_D\sqrt{(m_T^2 - m_{t_D}^2)(m_{t_D}^2 - m_t^2)}\over \sqrt2 m_{t_D} m_{V_D}}\;,\label{eq:TPVDMYuk}
\end{eqnarray}
where the $\Z_2$-even(-odd) partner of the top quark has been labelled $T$($t_D$), the SM Higgs sector is left unaffected by the new scalar, and $\Phi_D$ has a potential completely analogous to the Higgs potential. The hierarchy between the masses in the fermion sector is the same as that discussed in \cref{sec:fermionmasses}, {i.e.}, $m_t < m_{t_D} \leq m_T$, but $H$ can have any mass allowed by experimental bounds, including,
in principle,
being lighter than the SM Higgs boson.

The new physics parameter space for this model is five-dimensional:
\begin{equation}
g_D,m_{V_D}, m_H, m_T, m_{t_D}\;.
\label{eq:pars_case}
\end{equation}
In the following, we will denote this scenario as TPVDM -- a specific case  of top portal in the FPVDM  framework.
We chose this realisation as a case study since, on the one hand, it is minimal whilst, on the other hand, it allows us to explore a scenario where a non-Abelian dark sector is not connected to the SM via a Higgs portal at tree level. Furthermore, connecting the dark sector only with the SM top quark allows for an exploration of several interesting 
collider physics
signatures, whilst reducing the impact of constraints from direct detection.

Many other realisations are also very attractive. For example, the dark sector could be connected to SM leptons. The collider constraints on new VL leptons would then be milder, making the scenarios potentially less restricted, but the impact on the cosmological observables would not qualitatively change.\footnote{This is true except when the mass difference between DM and VL fermion  mediator is small.
In that case DM co-annihilation will be less intense
in comparison with strong co-annihilation with the $t_D$ quark.}
These kind of realisations are potentially interesting for a study of anomalies in the lepton sector (for example in connection with the muon anomalous magnetic moment) and will be developed in future studies.

As anticipated in \cref{sec:Dgaugemasses}, the mass splitting between $m_{V_D}$  and $m_{V'}$, $\Delta m_V= m_{V_D} - m_{V'}$, plays an important role for DM  phenomenology.
First of all, we have found that $\Delta m_V>0$ in the whole parameter space of the model, 
with the approximate expressions for 
$\Delta m_V$ given by \cref{eq:simple_mass_splitting_1,eq:simple_mass_splitting_2}.
Since $m_{V_D} > m_{V'}$, the 
$V_D V_D^* \to V'V'$ process for DM annihilation will
{\it always} take place for any point in the parameter space to contribute crucially to the list of processes 
affecting the relic density and to extend the viable parameter space compatible with constraints imposed by the relic density. The $V_D V_D^* \to V'V'$ process  also contributes to the DM indirect detection signals.

\begin{figure}[htb!]
	\centering
	\includegraphics[width=.4\textwidth]{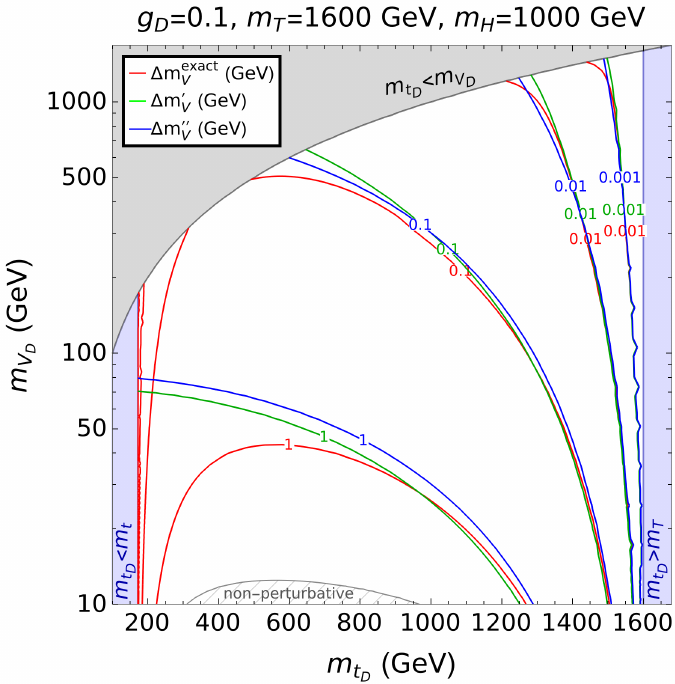}
	\ \ \ \
	\includegraphics[width=.4\textwidth]{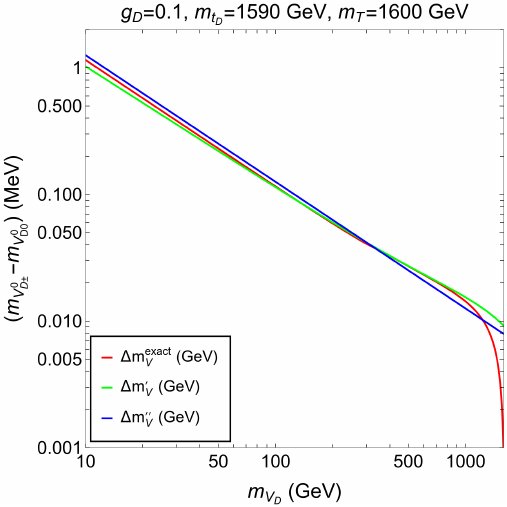}
	\caption{Values of the mass splitting $\Delta m_V=m_{V_D}-m_{V'}$ in the $(m_{t_D},m_{V_D})$ plane for a specific choice of $g_D$, $m_T$ and $m_H$ (left panel) and as a function of $m_{V_D}$ for a specific value of $m_{t_D}$  (right panel). The red, green and blue curves correspond to results from exact expression, approximated formulae \cref{eq:simple_mass_splitting_1} and \cref{eq:simple_mass_splitting_2}, respectively. The region where one-loop corrections to the masses of $V_D$ or $V'$ become larger than 50\%, so that a perturbative treatment is questionable, is also highlighted.\label{fig:dmv}}
\end{figure}

Numerically, the value of $\Delta m_V$ varies over a very wide range, since it scales as $g_D^2$ and it is proportional to $m_T^2-m_{t_D}^2$. One should also note that $\Delta m_V$ does not depend on $m_H$. 
In \cref{fig:dmv} (left) we present the iso-contours for  $\Delta m_V$ in the $\{m_{t_D}, m_{V_D}\}$ plane for $g_D=0.1$ and $m_T=1600$~GeV, whilst in \cref{fig:dmv} (right) we show how $\Delta m_V$ evolves as function of $m_{V_D}$ for the specific value of $m_{t_D}=1590$~GeV, all other parameters being the same.
The value of $m_T$ is chosen to be safely above the current upper limit on VL top partners at the LHC~\cite{ATLAS:2018ziw}.
For our particular choice of $g_D$ and $m_T$, $\Delta m_V$ can be as large as 1~GeV, while its minimal value reaches zero for a vanishing value of $m_T-m_{t_D}$.
In both frames we present a comparison of the exact one-loop result for $\Delta m_V$ and its approximations given by \cref{eq:simple_mass_splitting_1,eq:simple_mass_splitting_2}.
It is possible to see from \cref{fig:dmv} (right) that the approximate formulae are very accurate for a small $m_T-m_{t_D}$ splitting, but break down for $m_{V_D}$ close to the $m_t+m_{\psi_D}$ threshold, where the one-loop corrections are highly non-linear in the expansion parameters used in
approximate expressions for $\Delta m_V$.
Moreover, for small values of $m_{V_D}$, the one-loop mass corrections can be large, making 
the evaluation of $\Delta m_V$ perturbatively unstable. Therefore, we indicate by the hatched area the region where one-loop corrections to the masses of $V_D$ and/or $V'$ become larger than 50\% of the corresponding bare masses.

\begin{figure}[htb]
	\centering
	\begin{minipage}{.15\textwidth}
		\includegraphics[width=\textwidth]{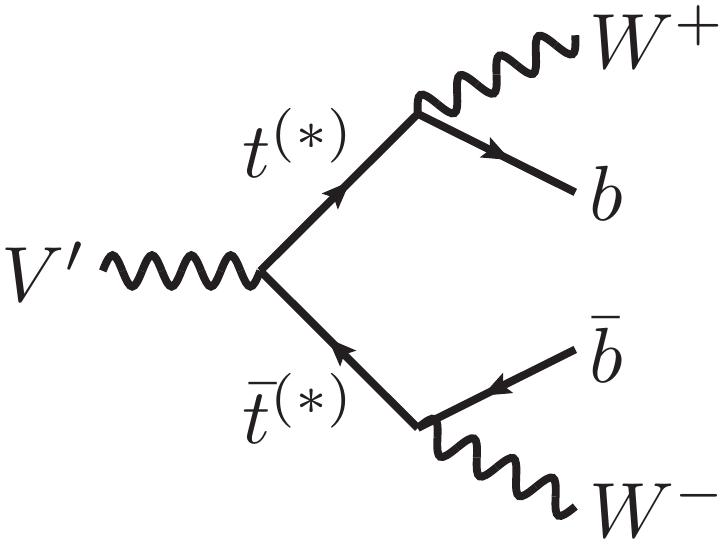}\\
		\hskip -10pt\includegraphics[width=.85\textwidth]{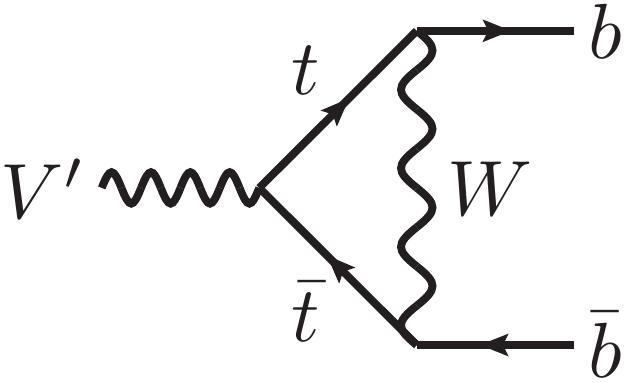}
	\end{minipage}
	\begin{minipage}{.4\textwidth}
		\includegraphics[width=\textwidth]{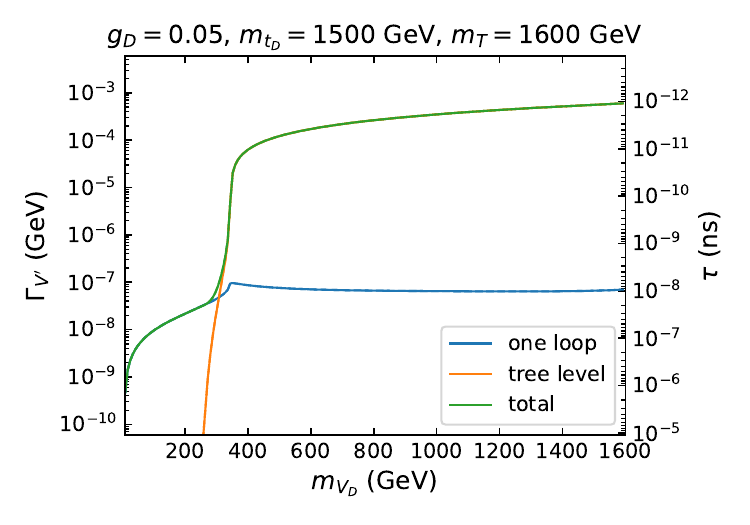}
	\end{minipage}
	\begin{minipage}{.4\textwidth}
		\includegraphics[width=\textwidth]{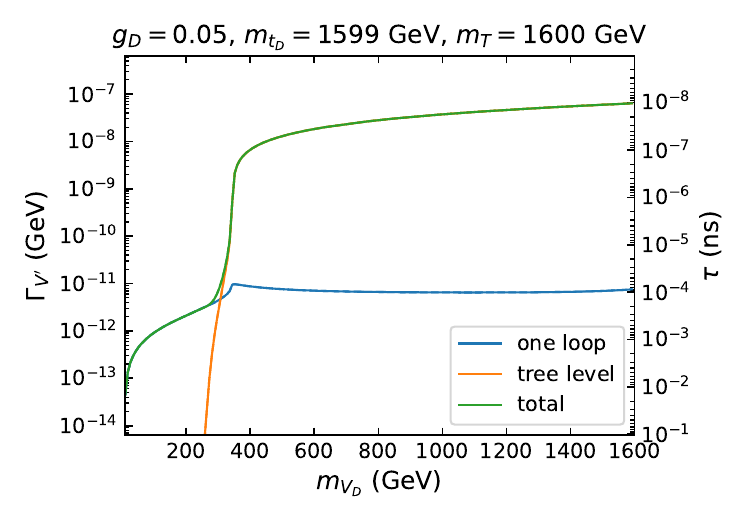}
	\end{minipage}
	\caption{\label{fig:V0decay} Left: Tree-level and one-loop diagrams for $V'$ decay. Center and right: decay width and lifetime of $V'$ at tree and one-loop level for $g_D=0.05$, $m_T=1600\GeV$ and different values of $m_{t_D}$.}
\end{figure}

The lifetime of $V'$ does not directly depend on $\Delta m_V$.
However, the $\Z_2$-even $\SUD$ gauge boson can also be long lived, if the DM is light enough. The only tree-level interaction of $V^{\prime}$ with SM particles is with top quarks, due to its mixing with $T$. 
If the mass of $V^{\prime}$ drops below the $t\bar t$ threshold, it can only decay directly to a three-body or four-body final state with $W$ bosons and $b$ quarks via the off-shell top quarks, or decay to a $b\bar b$ final state at one-loop, see the Feynman diagrams in ~\cref{fig:V0decay} (left). 
The latter, although 
only present at the one-loop level,
becomes dominant due to the reduced  phase space
for the four-body final state.
This is shown in \cref{fig:V0decay} (centre and right).
These loop-induced diagrams prevent $V'$ 
from having a sufficiently long lifetime
to spoil Big Bang Nucleo-synthesis (BBN). However, when the $g_D$ coupling is small, the $t_D$ mass approaches the decoupling limit ($m_{t_D}=m_T$) and the DM is light, $V'$ becomes long lived at colliders. Therefore, it could provide a signal for searches of long-lived neutral bosons decaying into $b\bar b$ pairs.

As mentioned in \cref{sec:scalarsector}, even if  TPVDM scenario does not contain a tree-level mixing, a loop-induced mixing between $h$ and $H$ still occurs, via SM top and the $\Z_2$-even top ($T$)  loops.
This contribution is eventually  suppressed.
A scenario with tree-level scalar mixing
is more constrained and can exhibit the following signatures:
1) the heavy scalar $H$ can decay also to any final state accessible to the Higgs boson, and therefore the model predicts further signatures at collider; 2) if the mass of the DM is small enough, the Higgs boson 
will decay into the DM itself or the $\Z_2$-even gauge boson $V^\prime$, affecting its width and branching ratios. From the cosmological point of view,
additional interactions from the tree-level
scalar mixing will affect  the relic density, and direct and indirect detection observables.

\begin{figure}[htb]
\centering
\begin{minipage}{.26\textwidth}
\includegraphics[width=\textwidth]{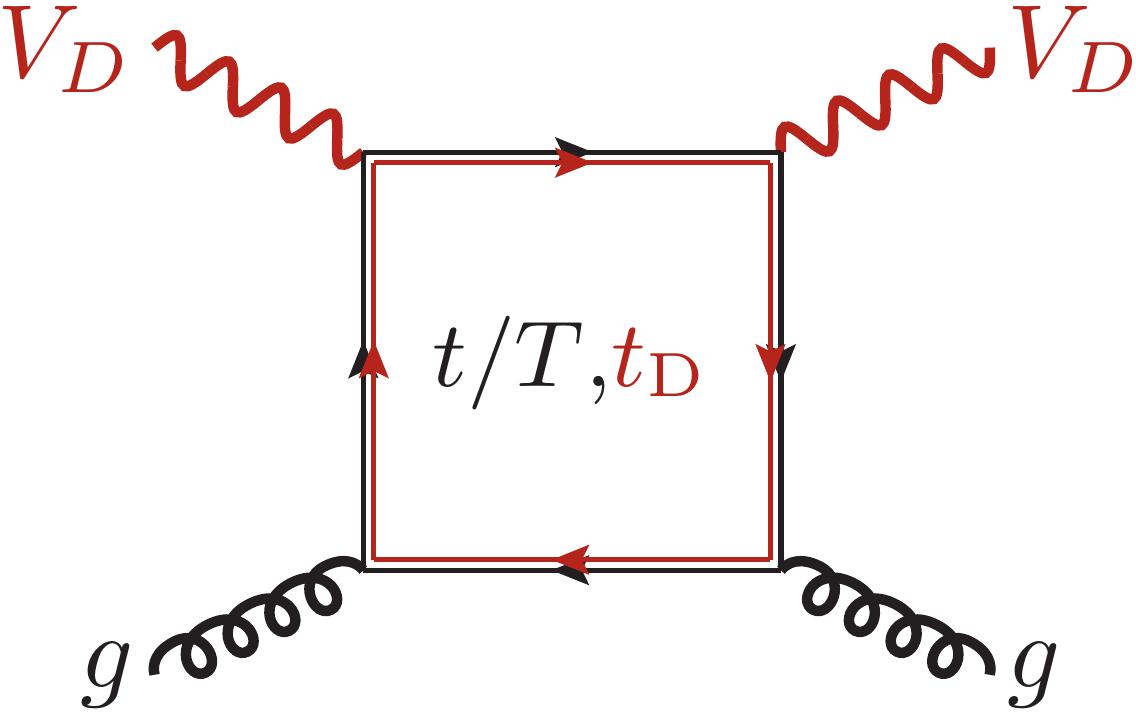}\\
(a)
\end{minipage}\hfill
\begin{minipage}{.22\textwidth}
\includegraphics[width=\textwidth]{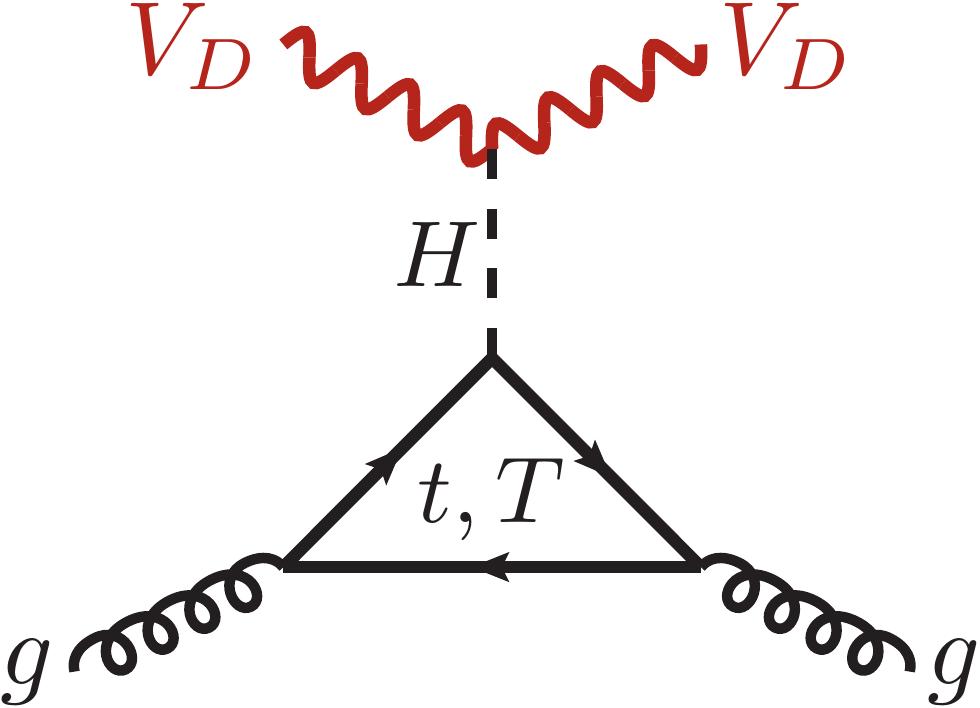}\\
(b)
\end{minipage}\hfill
\begin{minipage}{.15\textwidth}
\includegraphics[width=\textwidth]{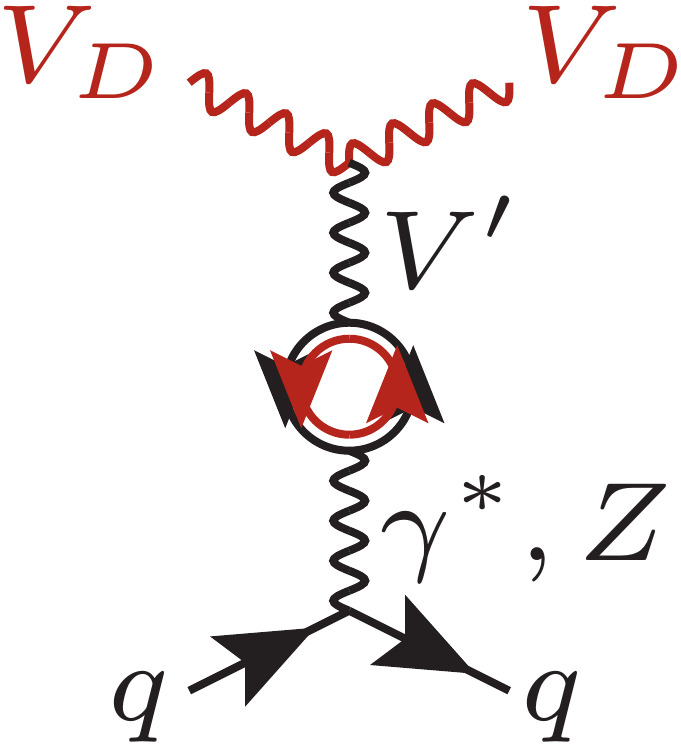}\\
(c)
\end{minipage}\hfill
\begin{minipage}{.215\textwidth}
\includegraphics[width=\textwidth]{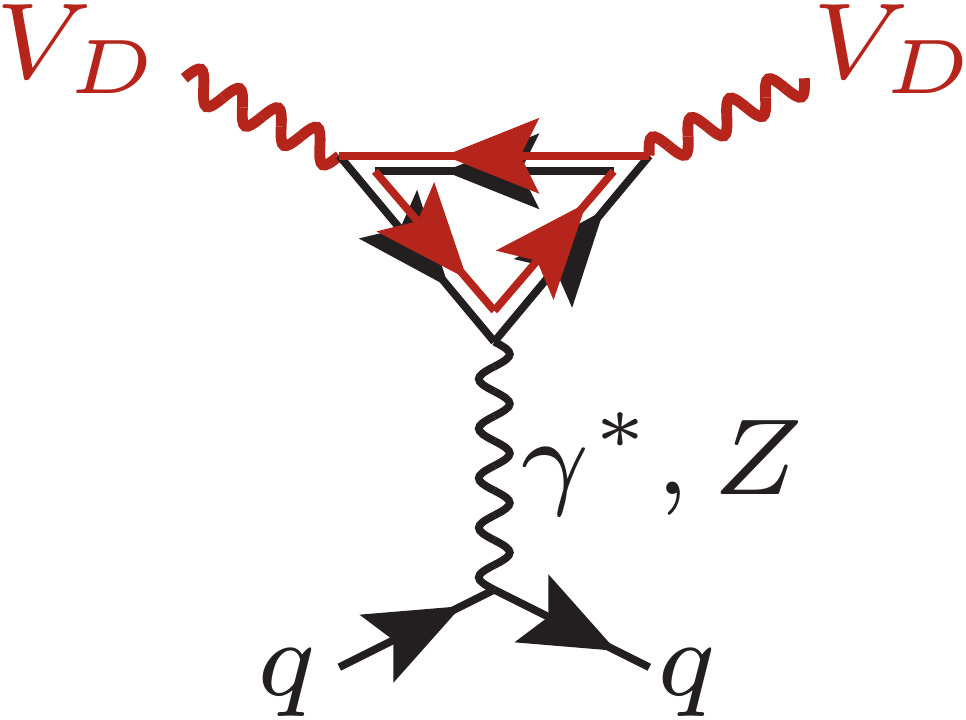}\\
(d)
\end{minipage}\hfill
\caption{\label{fig:DDtopology} Representative diagrams for direct detection processes. $H$ is the $\Z_2$-even scalar in the dark sector. $\Z_2$-odd particles are highlighted in red.}
\end{figure}

Since there is no  $h$-$H$ mixing in TPVDM scenario, DM scattering off the nuclei is induced only at loop-level. The Feynman diagrams for DM-gluon
interactions with quark box and triangle topologies are shown in \cref{fig:DDtopology}(a) and (b), while the DM-quark diagrams generated by the loop-induced $V'-\gamma/Z$ kinetic mixing and triangle diagrams are shown in \cref{fig:DDtopology}(c) and (d), respectively. The detailed evaluation of the triangle loop of fermions connected to gauge boson propagators is given in \cref{app:triangle}. As it will become clear in \cref{sec:combined}, the KM and triangle contributions play a crucial role in constraining the parameter space of the model through Direct Detection (DD) limits on DM. 

In the following sections, this model is tested against multiple observables from cosmology, direct DM detection experiments and LHC searches. For this analysis we implemented the Lagrangian of the model in the  
{\sc LanHEP}~\cite{Semenov2009}  and 
{\sc FeynRules}~\cite{Alloul:2013bka} packages whilst
model files have been generated in {\sc CalcHEP}~\cite{Belyaev:2012qa}, {\sc FeynArts}~\cite{Hahn:2000kx} and {\sc UFO}~\cite{Degrande:2011ua} formats.\footnote{The model implementations 
are available in the {\sc HEPMDB}~\cite{hepmdb} repository in {\sc CalcHEP} (\url{https://hepmdb.soton.ac.uk/hepmdb:0322.0335})
and {\sc UFO} (\url{https://hepmdb.soton.ac.uk/hepmdb:0322.0336})
formats.}
We used {\sc micrOMEGAs v5.2.7}~\cite{Belanger:2020gnr} for calculating DM observables and for setting the corresponding limits (see \cref{sec:cosmo}) as well as for the evaluation of some LHC processes. The model implementation in UFO format has been used in {\sc MG5\_aMC}~\cite{Alwall:2014hca} for the determination of the complete set of  LHC constraints (see \cref{sec:collider}). The {\sc FeynArts} model files from 
{\sc LanHEP}
were used to generated one-loop corrections to masses of $SU(2)_{\rm D}$ gauge bosons by {\sc FeynCalc}~\cite{Shtabovenko:2020gxv}, {\sc FeynHelpers}~\cite{Shtabovenko:2016whf} and {\sc Package-X}~\cite{Patel:2016fam}. A simplified version of the model has been implemented to calculate cross-sections at one-loop level in {\sc MG5\_aMC} and {\sc FormCalc9.8}~\cite{Hahn:2016ebn}.

\subsection{Constraints from DM relic density\label{sec:cosmo}}

There are many non-collider experiments dedicated to searching for signals of DM, both in space and on  Earth which play a very important role in limiting the DM parameter space and in the identification of viable  DM models.
These experiments are devoted to the precise determination of the DM relic density as well as to DD and ID of DM. 
In particular, the  PLANCK experiment has measured the relic density with a precision better than 1\%~\cite{Planck:2018vyg}:
\begin{equation}
\Omega^{\text{Planck}}_{\text{DM}}h^2=0.12\pm 0.0012 \,.
\label{eq:planck}
\end{equation}
In our analysis, we will select points that satisfy this constraint, bearing in mind that points which predict a relic density below the PLANCK constraint could still be allowed if new sources of DM exist besides $V_D$.

For DM DD we use the limits from  XENON1T~\cite{XENON:2018voc}. The XENON1T experiment provides the most stringent upper limit (compared to LUX (2017) and Panda-X (2017), see fig. 5 in the reference \cite{XENON:2018voc}). XENON1T provides the limit on  DM-nucleon's cross-section vs DM mass at $90\%$ C.L. together with the detector's efficiency as a function of nuclear recoil energy. We have evaluated the DM-nucleon scattering cross section and converted it into the number of events by taking in account  the efficiency of the XENON1T detector. This allowed us to find the corresponding p-value for the signal. The calculation was performed using a modified version of {\sc micrOMEGAs} package which allowed us to correctly 
evaluate DM DD rates from the loop-induced  $\gamma(Z)$-$V_D$-$V_D$ interactions. 
 We have scaled the number of registered events if the corresponding relic density is less than the measured value as follows:
\begin{eqnarray}
	\hat{N}_{\text{event}}&=&
	\begin{cases}
		\left(\frac{\Omega_{\text{DM}}}{\Omega^{\text{Planck}}_{\text{DM}}}\right)N_{\text{event}}, & \text{if}\ \Omega_{\text{DM}} h^2<0.12 \\
		N_{\text{event}}, & \text{otherwise}
	\end{cases}\;,
\end{eqnarray}
and have defined the $p$-value, $hat{p}$, as
\begin{eqnarray}
	\hat{p} &=& \exp(-\hat{N}_{\text{event}}) \;.
\end{eqnarray}
The exclusion of parameter space is imposed on the points where $\hat{p}<0.1$, which corresponds to the exclusion limit at $90\%$ C.L.

ID DM searches   are being  performed by many experiments, including Fermi-LAT \cite{Fermi-LAT:2017opo}, IceCube~\cite{IceCube:2016zyt},  ANTARES \cite{ANTARES:2020leh}, etc. However, these experiments rely on the DM local density and velocity distribution as well as the  propagation of the particles in the galactic plane. Therefore, the respective predictions are affected by various uncertainties of an astronomical nature. 
To be independent of these uncertainties, in this study we use the Cosmic Microwave Background (CMB)  limit on DM ID based on PLANCK data. We consider the product of the DM-self annihilation or the DM decay into SM particles. By studying the effect of energy injection from DM annihilation products (electrons, positrons, gamma-ray, neutrinos and anti-protons) on the galactic medium  which is sensitive to the CMB anisotropies, the upper limit on the energy injection measured by PLANCK is:
\begin{equation}
	P_{\text{ann}}<3.2\times 10^{-28}\,\, \frac{\rm{cm^3}}{\rm{s\,GeV}}\quad \text{at 95\% C.L.,}
\end{equation}
with
\begin{equation}
	P_{\text{ann}}=\sum_{j}\frac{f^{\text{eff}}_{j}\braket{\sigma v}_{j}}{M_{\text{DM}}}\left(\frac{\Omega_{\text{DM}}}{\Omega^{\text{Planck}}_{\text{DM}}}\right)^2 ,
	\label{eq:Pann_def}
\end{equation}
where $\braket{\sigma v}_{j}$ is the thermally averaged partial annihilation cross-section for the $j$ channel whilst  $f_{j}^{\text{eff}}$ is the energy fraction of DM annihilation transferring to the plasma for the $j$th channel. 
To construct the quantity $P_{\text{ann}}$, we use {\sc micrOMEGAs} to calculate $\braket{\sigma v}_{j}$ for all possible channels and neglect those that contribute to the total annihilation cross-section less than $0.1\%$. The effective fraction of energy $f^{\text{eff}}_{j}$ was thoroughly studied and provided for almost all DM annihilation processes into two SM particles in the final state in \cite{Slatyer:2015jla,Leane:2018kjk}. For non-SM particles in the final state of $2\to2$ processes, for example $V_D,V_D\to V^{\prime},V^{\prime}/V^{\prime},H/H,H$, we make the approximation $f^{\text{eff}}_{\text{non-SM}}\sim f^{\text{eff}}_{q\bar{q}}$. This approximation is reasonable because each $V_D/H$ eventually decays into 3 pairs of quarks anti-quarks and the energy fractions stored in each quark anti-quark pair ($u,d,s,c,b,t$) are not significantly different. The annihilation cross-section in eq.~(\ref{eq:Pann_def}) is rescaled by $(\Omega_{\text{DM}}/\Omega^{\text{Planck}}_{\text{DM}})^2$ due to the two DM particles in the initial state.

Finally,
we have  checked that the model does not spoil the predictions from BBN. When the lifetime of $V^{\prime}$ is too long, such that it decays during or after BBN, it would spoile the observed neutron to proton density ratio. 
For $m_{V^{\prime}} \lesssim 2 m_W$, the dominant decay to $b\bar{b}$ via the loop-induced process discussed above makes $V'$ lifetime much shorter than the value excluded by BBN. So, in this respect, BBN does not exclude any region of the parameter space of our model that is allowed by relic density constraints.

\subsection{Collider constraints}\label{sec:collider}

In the scenario under consideration the top quark is the only SM particle which interacts with the dark sector. Processes involving top quarks in propagators or final states are therefore affected by new physics contributions.
The model contains a complex vector DM candidate but two different kind of mediators: the VL and $\Z_2$-odd top partner $t_D$ and the two $\Z_2$-even bosons $H$ and {$V^{\prime}$}, which however can only be produced at the LHC via interactions with the top quark or its $\Z_2$-even partner $t^\prime$.

A list of relevant signatures for the scenario are provided in Table~\ref{tab:LHCprocesses}. A mono-jet signature can only arise at loop level, while the $t\bar t+E_T^{\rm miss}$ and $t\bar t t \bar t$ one can receive both tree- and loop-level contributions, which might be of similar size depending on the regions of parameter and phase space.
\begin{table}[h]
\centering
\begin{tabular}{c|c}
\toprule
Process & Representative diagrams\\
\midrule
mono-jet (only loop) & 
\begin{minipage}{100pt} \includegraphics[width=\textwidth]{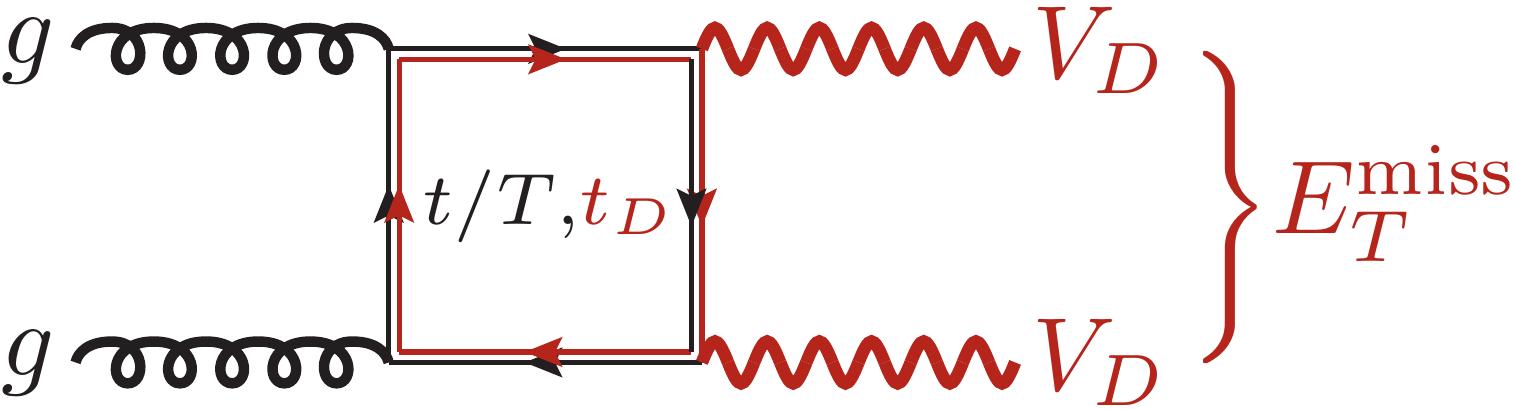}\end{minipage} 
\begin{minipage}{100pt} \includegraphics[width=\textwidth]{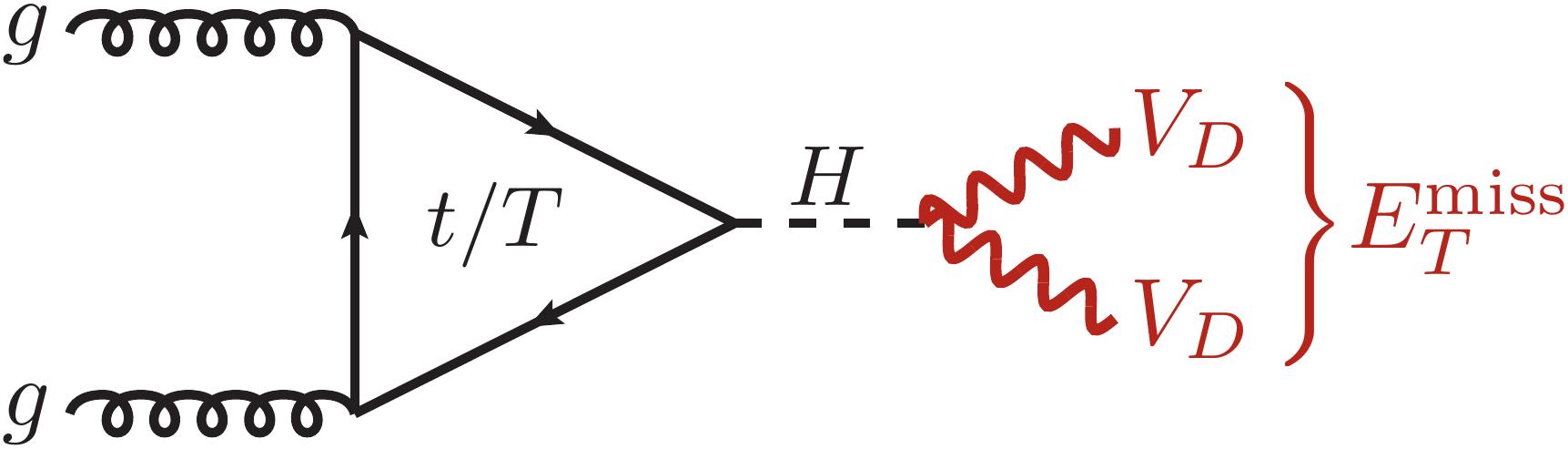}\end{minipage} + jet from ISR or from loop\\
\midrule
$t \bar t + E_T^{\rm miss}$ & 
\begin{minipage}{100pt} \includegraphics[width=\textwidth]{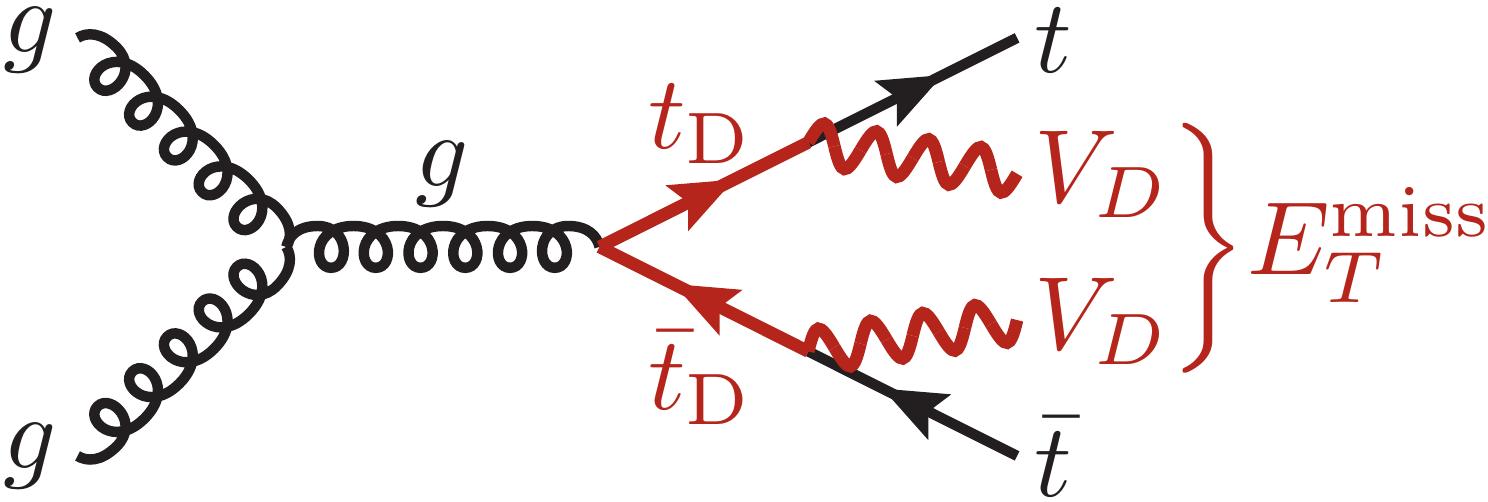}\end{minipage} 
\begin{minipage}{100pt} \includegraphics[width=\textwidth]{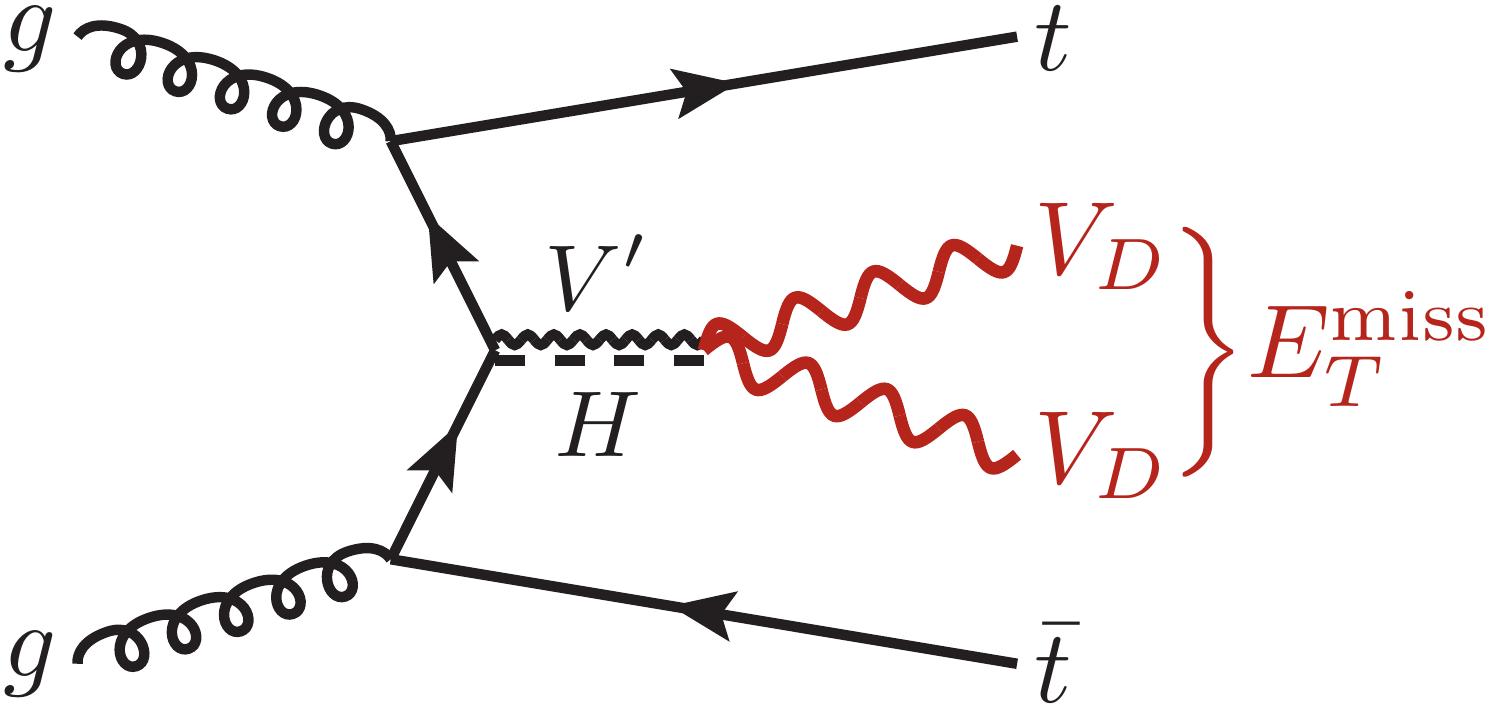}\end{minipage}\\
\midrule
$t \bar t t \bar t$ & 
\begin{minipage}{80pt} \includegraphics[width=\textwidth]{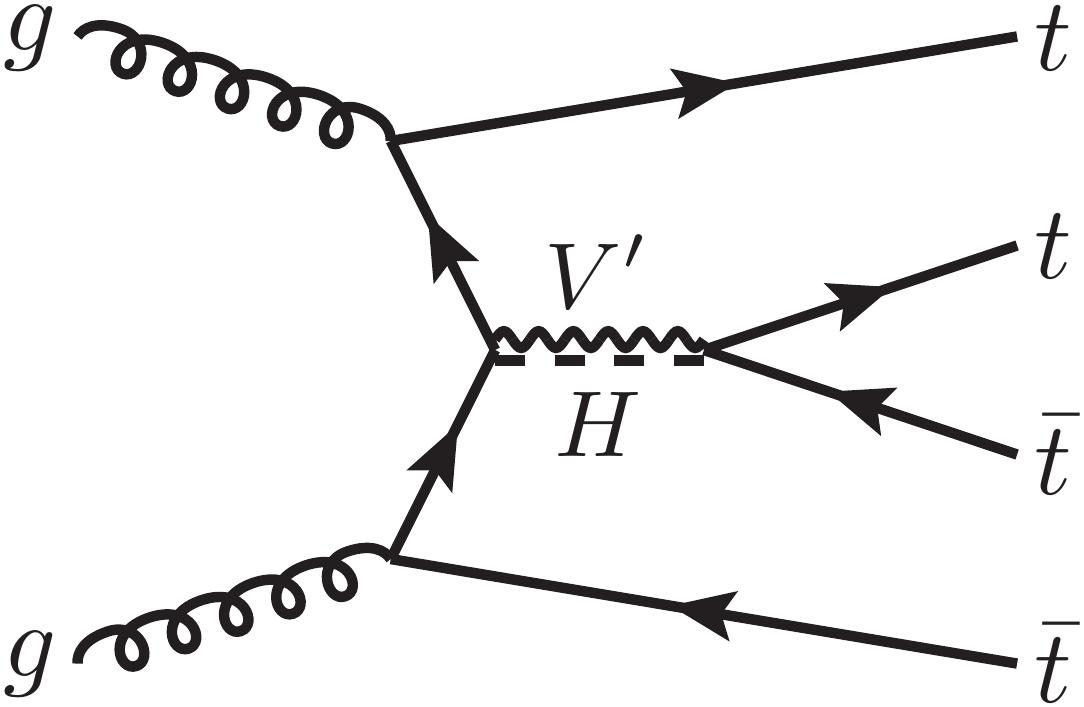}\end{minipage} 
\\
\midrule
$h V^\prime$ and $V^\prime V^\prime$ (only loop) & 
\begin{minipage}{90pt} \includegraphics[width=\textwidth]{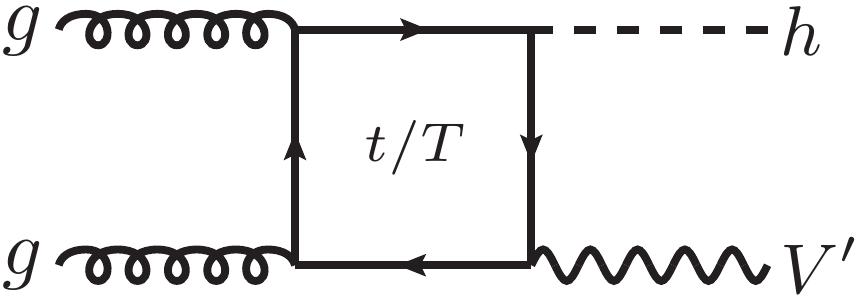}\end{minipage}\hskip 10pt 
\begin{minipage}{90pt} \includegraphics[width=\textwidth]{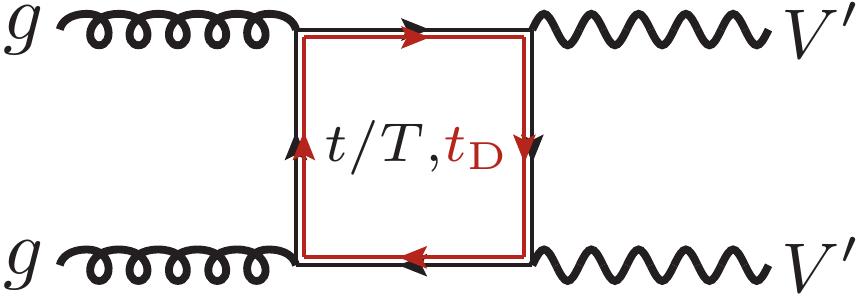}\end{minipage}
\hskip 10pt 
\begin{minipage}{90pt} \includegraphics[width=\textwidth]{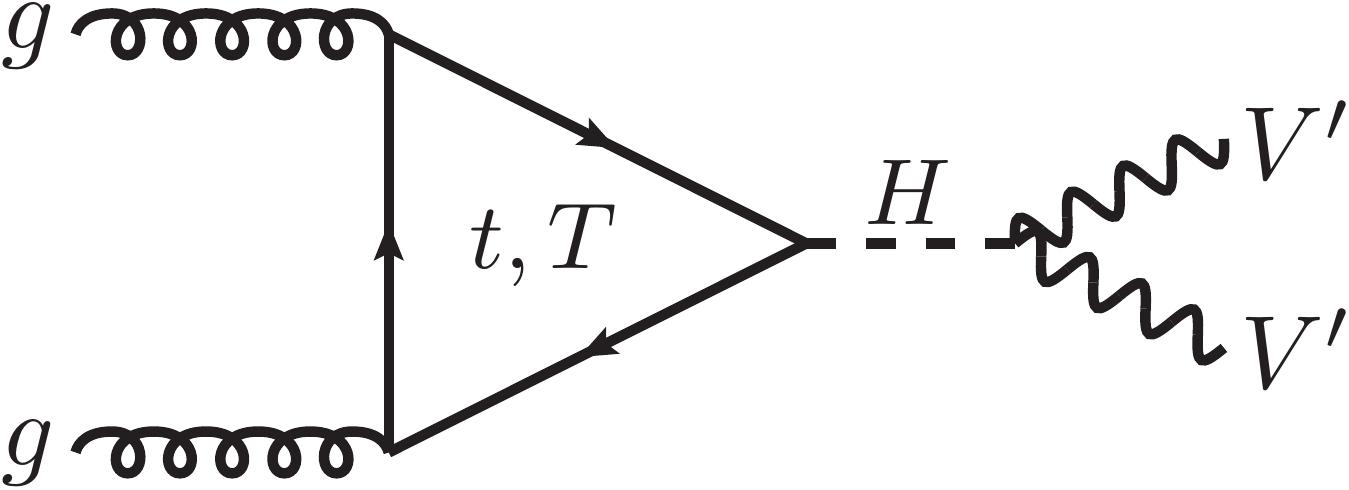}\end{minipage}
\\
\bottomrule
\end{tabular}
\caption{\label{tab:LHCprocesses} List of relevant processes at the LHC. $\Z_2$-odd particles are highlighted in red. Due to its purely VL nature, $t_D$ cannot interact with the scalars.}
\end{table}
Given the preliminary and explorative nature of this analysis, in the following we perform a recast of current LHC searches only for the tree-level processes to obtain constraints on the parameter space of the model. 

The simulations are performed at Leading Order (LO) with {\sc  MG5\_aMC}~\cite{Alwall:2014hca} in the 4-flavour scheme using the NNPDF3.0 LO set~\cite{NNPDF:2014otw} through the {\sc LHAPDF 6} library~\cite{Buckley:2014ana} (LHA index 262400). No resonant propagation of new particles is imposed, to allow for the inclusion of interference and off-shellness effects when relevant. For the $t \bar t + E_T^{\rm miss}$ signature, in the region of a small mass gap between $t_D$ and $V_D$, where $m_{t_D}-m_{V_D}<m_t$, simulations are performed for the $2\to6$ process $pp\to W^+bW^-\bar bV_D V_D$. The recast is done through the {\sc MadAnalysis 5} framework and the searches considered for the recast are different depending on the process:
\begin{itemize}
 \item for the $t \bar t + E_T^{\rm miss}$ processes we used a CMS search for top squark pair production decaying to DM, in final states with opposite sign leptons and missing transverse energy $E_T^{\rm miss}$~\cite{CMS:2017jrd}, recast in~\cite{BQM0T3_2021}. 
 \item for the $t \bar t t \bar t$ processes we used a CMS search for four top quarks in final states with either a pair of same-sign leptons or at least three leptons, in addition to multiple jets~\cite{CMS:2019rvj}, recast in \cite{DVN/OFAE1G_2020}.
\end{itemize}
In both cases, the searches target the very same final states predicted by our model, and are therefore ideal for determining constraints from collider. 

The model also predicts a 
signal from pair production of the $\Z_2$-even partners of the SM top-quark, $T\bar{T}$, which is constrained by ATLAS and CMS searches and only needs to be rescaled for different branching ratios. However, the $T$-quark primarily decays into $Wb/Zt/ht$ final state
with a $50\%/25\%/25\%$ branching ratio pattern, and the contribution of decays to new states is very small in the whole parameter space. Therefore, current LHC bounds leave the region of parameter space with $m_T \gtrsim 1.5$~TeV unconstrained~\cite{ATLAS:2021ibc,CMS:2022fck}.
Bounds from single $T$ production are more model-dependent, but less tight, as the production cross-section is driven by the $T-t$ mixing which is small.

The loop-level diagrams can be relevant especially when the particles which decay to the final states are produced 
at resonance
: in this case the loop suppression can be compensated by the lower multiplicity in the phase space. For the $h V^{\prime}$ and $V^{\prime}V^{\prime}$ processes we have only computed cross-sections using a simplified version of the model suitable for one-loop calculations in {\sc  MG5\_aMC}, to estimate if they can be tested against data from current searches.

\subsection{Combined bounds \label{sec:combined}}

\subsubsection{Full parameter scan}

\begin{figure}[htb!]
	\includegraphics[width=0.5\textwidth]{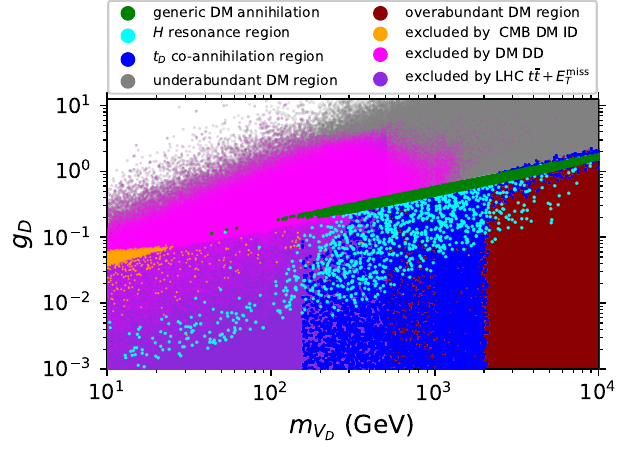}%
	\includegraphics[width=0.5\linewidth]{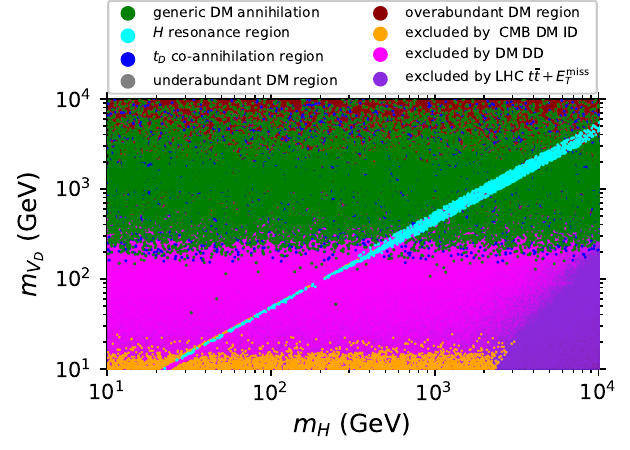}
	\\
	\hspace*{0.25\textwidth} 
	(a) \hfill (b)
	\hspace*{0.25\textwidth} 
	\\
	\vskip 0.5cm
	\includegraphics[width=0.5\linewidth]{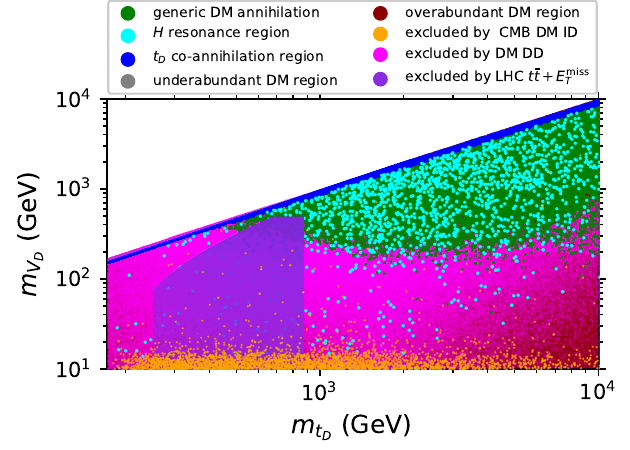}%
	\includegraphics[width=0.5\linewidth]{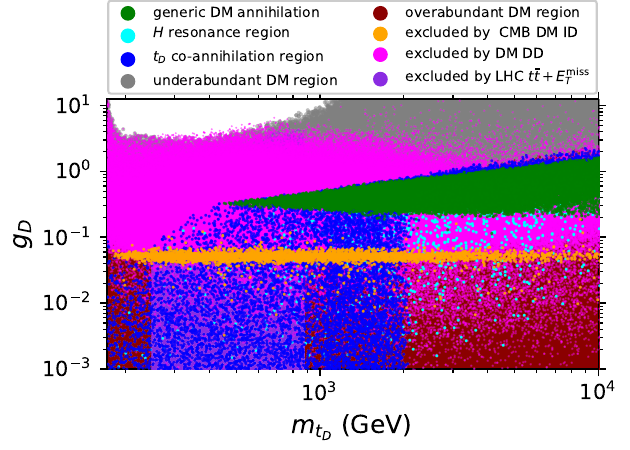}
	\\
	\hspace*{0.25\textwidth} 
	(c) \hfill (d)
	\hspace*{0.25\textwidth} 
	\caption{(New plots) Excluded and allowed region of the parameter space of the model from the full five-dimensional scan  of the parameter space projected into $(m_{V_D}, g_D)$, {$(m_H,m_{V_D})$},  $(m_{t_D},m_{V_D})$ and $(m_{t_D},g_D)$	planes. The white areas represent: top-left corner of panel (a) and bottom-right corner of panel (c) -- non-perturbative region of the parameter space; upper part of panel (c) -- kinematically inaccessible $m_{V_D}> m_{t_D}$ region.
		\label{fig:scatterplot}}
\end{figure}

We explore the viable parameter space of our model as well as   the effect of the  cosmological and collider constraints by performing a comprehensive scan over the 5D parameter space
in the following ranges:
\begin{eqnarray}
	\left\{
	\begin{array}{l}
		10^{-3}<g_D<4\pi \\
		10\GeV<m_{V_D}<m_{t_D}\\
		1.5 \TeV <m_T \\
		m_t<m_{t_D}\leq m_T<10\TeV\\
		10\GeV<m_H<20\TeV
	\end{array}
	\right. \ \ .
\end{eqnarray} 
In~\cref{fig:scatterplot} we present the results of this scan showing projections into various planes:  $(m_{V_D}, g_D)$ 
(a),  $(m_{H},m_{V_D})$ (b),  $(m_{t_D},m_{V_D})$ (c)  and $(m_{t_D},g_D)$ (d).

\begin{figure}[htb]
\centering
\begin{minipage}{.2\textwidth}
\includegraphics[width=.8\textwidth]{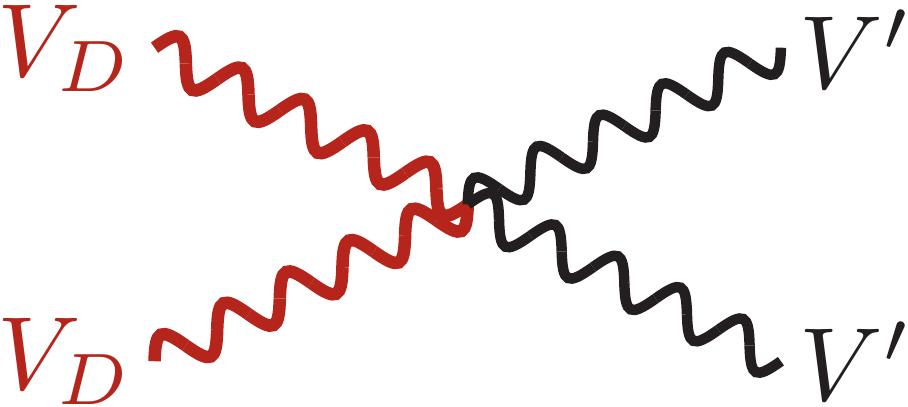}\\
(a)
\end{minipage}\hfill
\begin{minipage}{.2\textwidth}
\includegraphics[width=.8\textwidth]{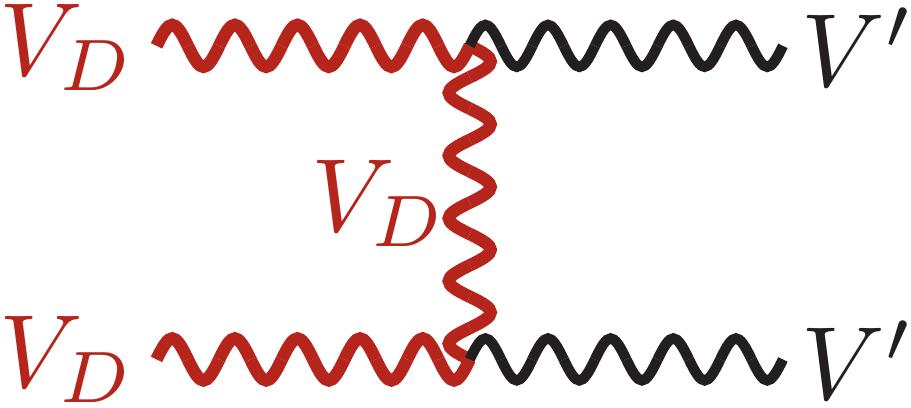}
\\
(b)
\end{minipage}\hfill
\begin{minipage}{.18\textwidth}
\includegraphics[width=.8\textwidth]{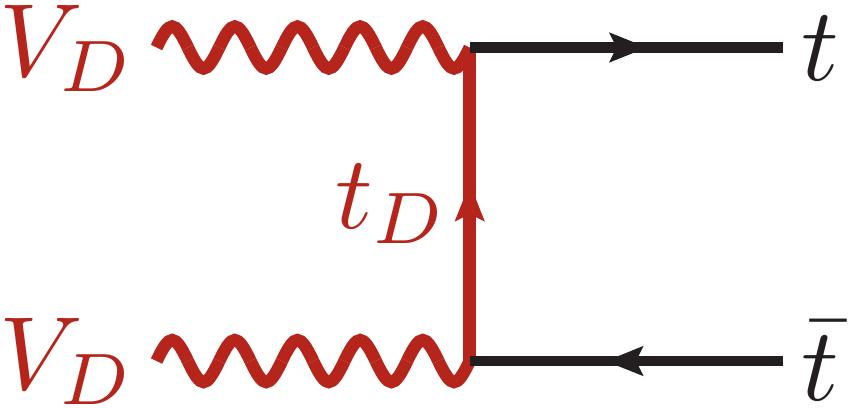}
\\
(c)
\end{minipage}\hfill
\begin{minipage}{.18\textwidth}
\includegraphics[width=\textwidth]{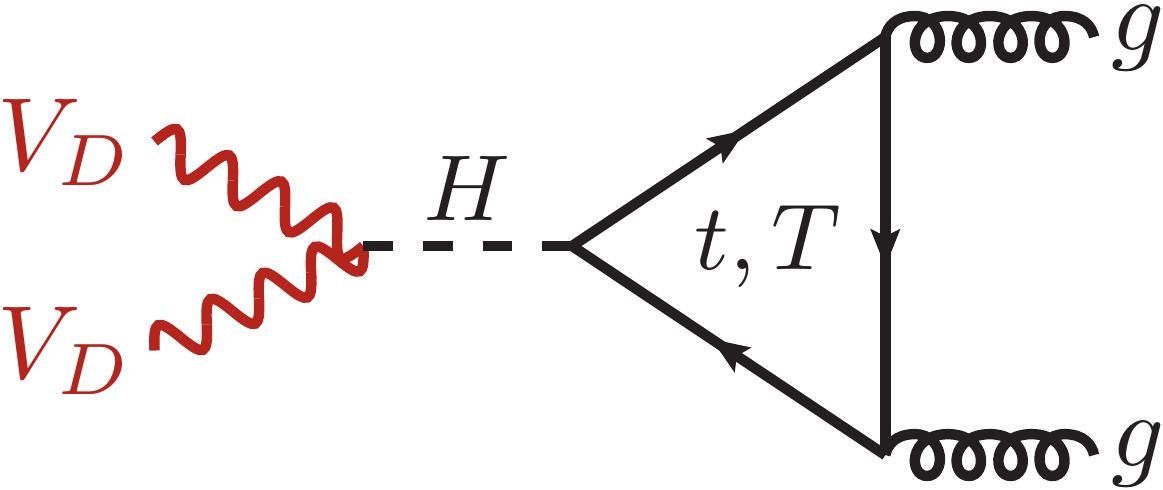}
\\
(d)
\end{minipage}\hfill
\begin{minipage}{.2\textwidth}
\includegraphics[width=\textwidth]{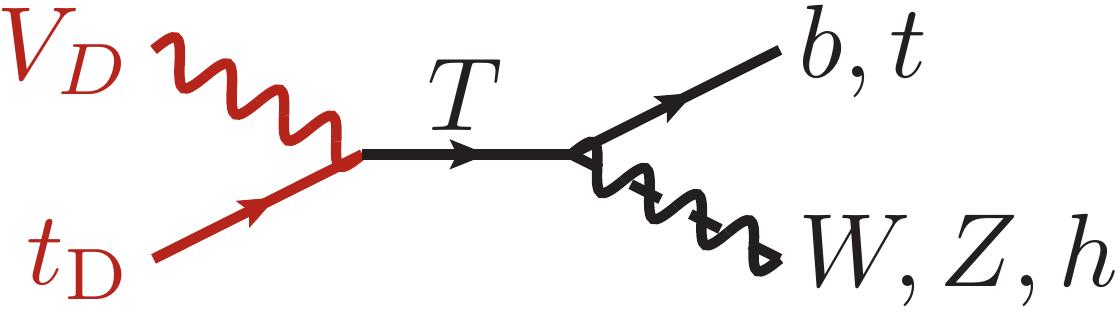}
\\
(e)
\end{minipage}
\caption{\label{fig:relicresonances} Representative contributions to relic density. From left to right: 4-leg; t-channel DM annihilation; DM annihilation via resonant $H$ (the $\Z_2$-even scalar in the dark sector); DM-mediator co-annihilation. $\Z_2$-odd particles are highlighted in red.}
\end{figure}

The allowed parameter space is indicated by the green, cyan and blue regions, corresponding to generic DM annihilation (via $V_D V_D \to V' V'$ and  $t$-channel $V_D V_D \to t\bar{t}$
processes), resonant $(H)$ annihilation and $\text{DM}-t_D$ co-annihilation regions respectively.
The representative Feynman diagrams for these channels are shown in \cref{fig:relicresonances}.

In these regions the relic density constraint from PLANCK is satisfied
to
within $5\%$. The  grey colour indicates the under-abundant DM relic density region. From  \cref{fig:scatterplot}(a) 
one can see that
the generic  DM annihilation (diagrams (a)--(c) of \cref{fig:relicresonances}) determines
 a narrow  strip in the $(g_D,m_{V_D}) $  plane indicating the  correlation between $g_D$ and $m_{V_D}$
 required
to arrange the right amount of DM.
For values of $g_D$ below this band these processes cannot provide large enough cross-section for DM annihilation and this leads to the excluded over-abundant DM region indicated  by the red colour.
One can clearly see this region
in all panels of~\cref{fig:scatterplot}  for  large DM masses.
However, there are additional processes  which provide an  effective DM annihilation   low DM relic density
respectively, consistent with PLANCK data. One of them is $V_D V_D \to H$ resonant annihilation, a representative diagram of which is shown in  \cref{fig:relicresonances}(d).
This process allows one to extend the viable  parameter space into the lower region of $g_D$ (by up to two orders of magnitude) indicated by the cyan colour.
This  can be clearly seen  in~\cref{fig:scatterplot}(b),
which presents the cyan $H$ resonant band
which goes across the  whole parameter space in the $(m_{H},m_{V_D})$ plane.

Another process, the DM-$t_D$ co-annihilation channel 
(see representative diagram in  \cref{fig:relicresonances}(e)),  
provides viable parameter space even for  lower values of $g_D$ for $m_{V_D}>m_t$ and $m_{t_D}$ values below 2 TeV. The respective region is indicated by the blue colour, which can be clearly seen
especially in  $(m_{t_D},m_{V_D})$ as a narrow resonance band.
At the same time, when  $m_{V_D}$ is above $2\TeV$, neither DM-$t_D$  co-annihilation nor  $H$-resonant annihilation  are effective enough to provide low enough relic density for $g_D$ values below the generic DM annihilation region.
Therefore, the region with low $g_D$ and large $m_{V_D}$ is excluded due to the  over-abundant relic density indicated by the red colour. 

Furthermore, notice that the regions with low $m_{V_D}$ and large $g_D$ values are partly excluded by DD and/or ID experiments as indicated by magenta and orange points, respectively. The region of DM masses which can be tested and excluded by the LHC is presented by the violet region.
This parameter space, which can be seen in all panels of~\cref{fig:scatterplot}, is related to  constraints on the $t\bar t + E_T^{\rm miss}$  signal at the LHC coming from $t_D\bar{t}_D$  pair production. For masses of $t_D$ below about 900 GeV  this signal would be observed if there is enough phase space for $t_D \to V_D t$ decay. This process is important in setting one of the main collider constraints on the model under study.

The four projections presented in~\cref{fig:scatterplot} reveal the non-trivial shapes of the allowed and excluded regions over the 5D  parameter space of the model.
For example, the orange colour, which presents the DM ID exclusion region, takes place for $m_{V_D}<20$ GeV (\cref{fig:scatterplot}(a,b,c)), $g_D\lesssim 0.06$ (\cref{fig:scatterplot}(a,d)) and $m_H \lesssim 3$~TeV(\cref{fig:scatterplot}(b)).
In \cref{fig:scatterplot} (b), one can see that  DM ID exclusion takes place  (besides the low $m_{V_D}$ region discussed above) and also along the very middle of the cyan band, where $m_{V_D} = m_H/2$.
Indeed, in this case, DM effectively annihilates through the $H$ state into $t\bar{t}$, $V'V'$ or $gg$, distorting precise CMB data, which therefore also limits the model parameter space. This region cannot be clearly seen in other panels, where it is presented just by randomly scattered points.

\subsubsection{Benchmark analysis}

In order to assess the relative role of the different constraints in identifying the allowed region of parameter space of our model we identify different benchmarks, characterised by fixed values for the masses of the $\Z_2$-even top partner, $m_T=1600\GeV$, and of the new scalar, $m_H=1000\GeV$, as well as  different values of the new gauge coupling $g_D=\{0.05,0.1,0.3,0.5\}$. These choices have the following rationale: 1) the gauge coupling can either assume a small value for which constraints from over-abundant relic density only allow tiny regions of the parameter space or a larger value for which such constraints become weaker; 2) the $\Z_2$-even partner of the top ($T$) is heavy enough to evade current LHC bounds based on pair production and considering decays into SM final states; 3) the mass of the $H$  state is large enough for it to decay into a top-quark pair. 
This affects the relative contribution of the diagrams mediated by $H$ in \cref{tab:LHCprocesses}.

\begin{figure}[htb!]
\centering
\includegraphics[width=.48\textwidth]{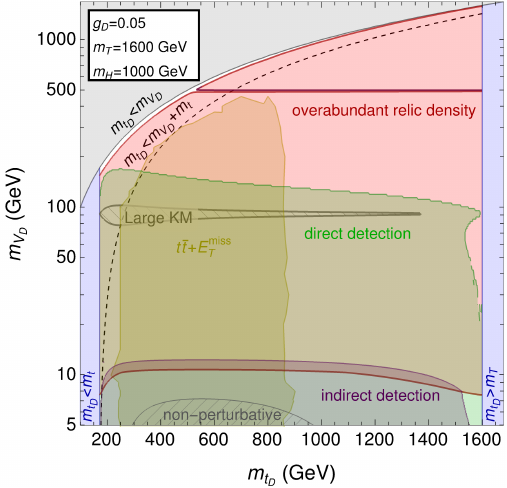}
\includegraphics[width=.48\textwidth]{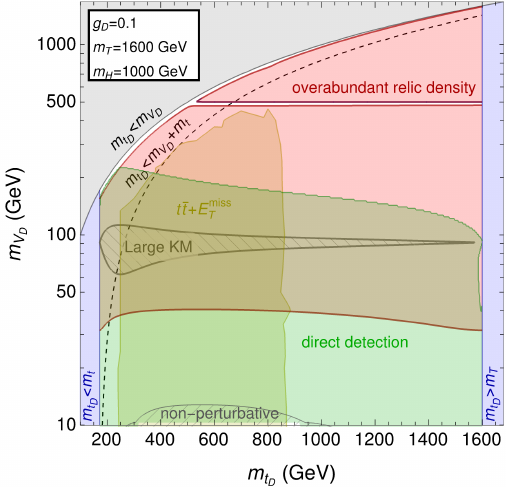}
\includegraphics[width=.48\textwidth]{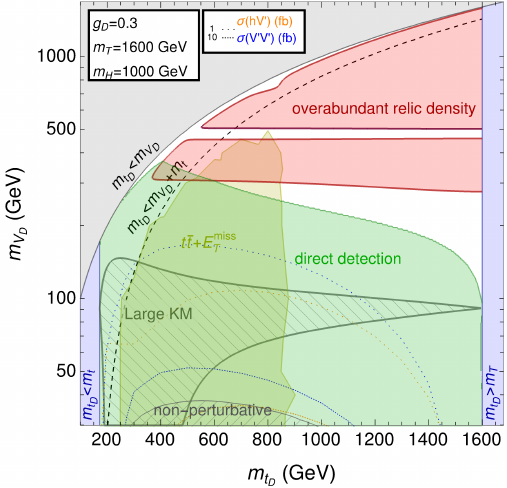}
\includegraphics[width=.48\textwidth]{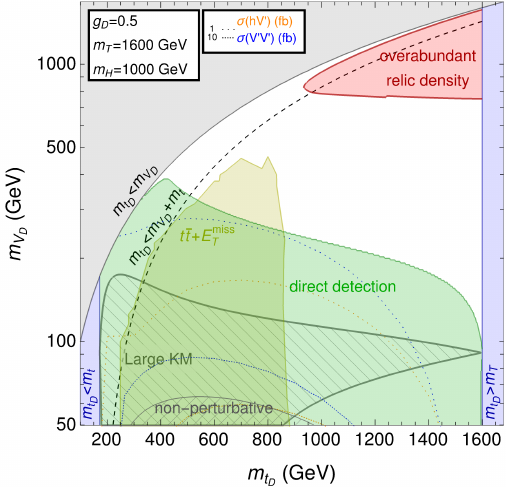}
\caption{\label{fig:BPresults} Combination of constraints from LHC, relic density ID and DD in the $\{m_{t_D},m_{V_D}\}$ plane for $m_T=1600\GeV$, $m_H=1000\GeV$ and different values of $g_D$. The coloured regions are excluded. The measured relic density value is reconstructed on the borders of the excluded region. When constraints from ID are absent, cross-sections for {$h V^{\prime}$} and {$V^{\prime} V^{\prime}$} production processes are shown. The non perturbative region corresponds to corrections to the gauge boson masses larger than 50\%. An estimate of the region of large KM is shown as a hatched area where at least one of the adimensional KM parameters $\{\epsilon_{AV},\epsilon_{ZV},\theta_{ZV}\}$ becomes larger than 10\%.}
\end{figure}

\begin{figure}[htb!]
\centering
\includegraphics[width=.48\textwidth]{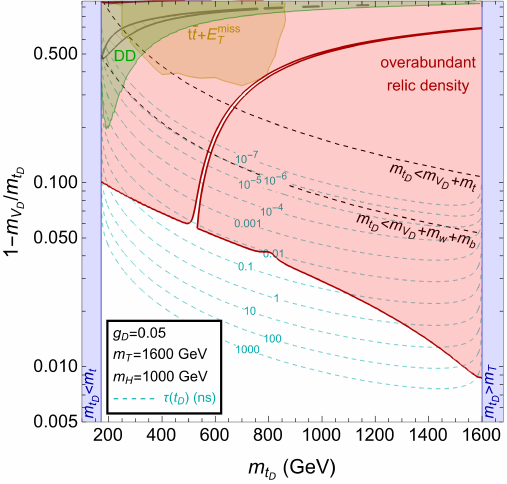}
\includegraphics[width=.48\textwidth]{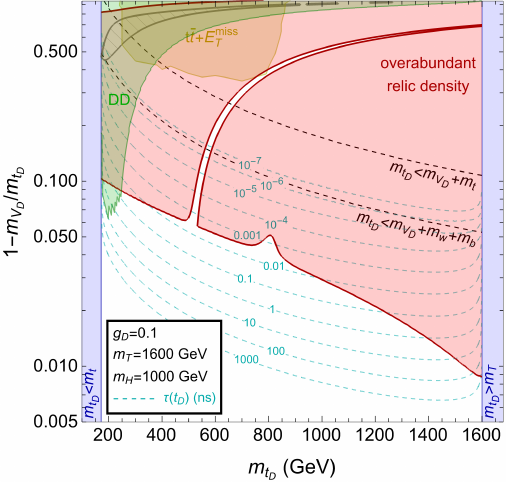}
\includegraphics[width=.48\textwidth]{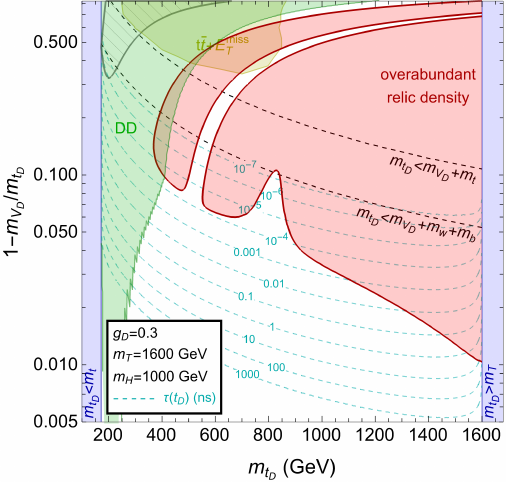}
\includegraphics[width=.48\textwidth]{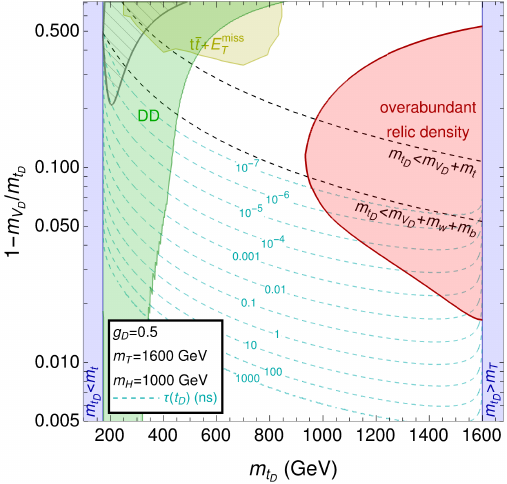}
\caption{\label{fig:BPresultssmallgap} Same as \cref{fig:BPresults} in the plane $\{m_{t_D},1-{m_{V_D}\over m_{t_D}}\}$, to highlight the region where the DM and $t_D$ have a small mass splitting. Contours corresponding to the lifetime of $t_D$ (in a region where it can be long-lived) are also shown.}
\end{figure}

The complementarity of cosmological and collider constraints can be represented in the $\{m_{t_D},m_{V_D}\}$ or $\{m_{t_D},1-{m_{V_D}\over m_{t_D}}\}$ planes. The former, shown in \cref{fig:BPresults}, allows us to highlight the low $m_{V_D}$ region while the latter, shown in \cref{fig:BPresultssmallgap}, emphasises the small mass gap region between $t_D$ and the DM particle.


The interplay between cosmological and collider bound is largely driven by the relative roles of relic density and DD bounds as function of the gauge coupling value, while indirect detection plays a role only for small coupling values. 

For smaller values of the gauge coupling, $g_D=0.05$ and  $g_D=0.1$, the measured amount of relic density is reconstructed only for light DM masses, $m_{V_D}\sim\mathcal{O}(10)\GeV$, and in a narrow region where the mass splitting between $t_D$ and the DM is small, less than $\sim10\%$ of $m_{t_D}$. 
In the co-annihilation region, where the mass gap between $V_D$ and $t_D$ is small, as well as 
in the $H$-resonant region around $m_{V_D}=m_H/2$, where $H$ is produced  near resonance, the relic density is drastically reduced, becoming under-abundant.


The small bell-shaped area visible in the middle of each panel of \cref{fig:BPresultssmallgap} with $g_D<0.5$ corresponds to the process in which $T$ is produced resonantly and decays into SM final states $Wb$, $Zt$ or $ht$, (see \cref{fig:relicresonances}).
If the gauge coupling becomes large enough, it 
eventually
becomes  impossible to reconstruct the measured value of the relic density  and the entire allowed parameter space of the model corresponds to an under-abundant relic density. In this case, the theory would not be able to explain the whole observed DM content of the universe and other sources of DM would be needed.

In the small $m_{V_D}$ region, strong constraints from ID limit the allowed parameter space to $m_{t_D}$ values approaching $m_T$, {i.e.}, the region where the mixing between $T$ and $t$ becomes small. ID constraints however disappear for increasing values of $g_D$, corresponding to a reduction of relic density values, owing to the scaling reported in \cref{eq:Pann_def}. 

However, the constraints from DD always exclude the region with small $m_{V_D}$ regardless of the gauge coupling value. 
The contribution of DM-gluon topologies is limited to the region with either minimal or maximal mixing in the fermion sector, corresponding to dominant contributions of the topologies (a) or (b) of \cref{fig:DDtopology}, respectively. These contributions destructively interfere for generic mixing otherwise, reducing the impact of this process in driving the DD bounds.
But the main contributions to DM DD is driven by the 
topologies with kinetic mixing induced by gauge boson self-energies, see \cref{fig:DDtopology}(c), and by loop-induced effective couplings $V_D$-$V_D$-$Z/\gamma$ which lead to DM-quark interactions through multipole moments, see \cref{fig:DDtopology}(d). The evaluation of the amplitudes for triangle diagrams leading to $V_D$-$V_D$-$Z/\gamma$ multipole interactions is given in detail in \cref{app:triangle}.\footnote{The role of multipole contributions in DM DD has also been studied in \cite{Hisano:2020qkq}. In our study, however, we took into account also KM topologies and the interference between them.}
For DM masses below about $\sim$400 GeV, the kinetic mixing with Z-boson plays dominant role for DM DD constraints. In the hatched region of \cref{fig:BPresults}, one can see that the KM contribution becomes strongest when the DM mass is comparable to the mass of $Z$ boson ({i.e.}, dominated by the mass mixing between $Z$ and $V^\prime$). As the gauge coupling increases, the effect of KM becomes strong also when the DM mass is small and the ratio between $t_D$ and $T$ is small (compatibly with the behaviour of the KM functions in \cref{fig:KMZVplot}. 
On the other hand, for heavier DM and sufficiently large $y_t^\prime$ coupling, triangle diagrams, defining multipole DM interactions with the photon can play a dominant role. Therefore, taking into account of the complete set of Feynman diagrams and their interference is an important element for the consistent and correct estimation of DM DD rates and constraints in the FPVDM framework.\\

The LHC bound comes exclusively from the $t\bar t + E_T^{\rm miss}$ signature, dominated by the pair production of $t_D$ states. The bound is almost independent of the mass of $t_D$ and constrains the region $250\GeV \lesssim m_{t_D} \lesssim 850\GeV$, independently of $g_D$, until the mass difference between $t_D$ and the DM becomes small: in this case the missing energy component of the events decreases and the sensitivity of the relevant CMS search reduces, allowing the small mass gap region. Effects coming from the width of $t_D$ are negligible, as the $t_D$ is narrow in the whole parameter space for each choice of $g_D$. The 4-top-quark  search does not show any sensitivity over the whole parameter space, regardless of the value of $g_D$. The loop processes of {$hV^{\prime}$} associate production and {$V^{\prime}V^{\prime}$} pair production are not testable at current luminosities, as their cross-sections are always well below $\sigma\gtrsim\mathcal{O}(10\fb)$ in the region where the relic density is reproduced. Higher luminosities and/or higher energies would be needed to be sensitive to such final states.

A very interesting feature of this scenario emerges for small values of $g_{D}$ in the small region where the DM and $t_D$ have a small mass gap: the decay width of $t_D$ becomes significantly small, such that $t_D$ becomes long-lived (its lifetime in the small mass gap region is shown in \cref{fig:BPresultssmallgap}) and can be probed by dedicated searches at the LHC or future colliders.
Different $T$ or $H$ masses would not modify this qualitative picture. 
	
One should also note that the model predicts
that the $tth$ Yukawa coupling, $y_t$ is always
bigger than the SM one (see eq. \ref{eq:TPVDMYuk}). This happens due to the the 
presence of a non-zero $y'$ coupling -- the key point of the model, which provides the portal between 
the SM and dark sectors. The current direct constraints on $y_t$ are quite weak (of the order of 50\%) from $pp\to ttH$ production at the LHC.
We have checked that imposing even 10\% constraint on 
 $y_t$, e.g., requiring $\delta y_t/y_t<0.1$
does not qualitatively change 
our results. On the other hand, the $y_t$ constraint will play a very important role at future 
$e^+e^-$ colliders, which will measure $y_t$
to within an accuracy of one
percent.
The importance of  such a constraint as future colliders is the subject of a separate study.\\

As a general conclusion, the combination of cosmology and LHC bounds always favours the region with a small mass splitting between $t_D$ and the DM. Other regions can be accessed depending on the value of other model parameters.
This specific realisation of the model is in any case an example dictated by its simple features.
Including mixing in the scalar sector, further VL partners or further interactions of the same VL representation would enlarge the possible signatures and change the complementarity between different observables in constraining the model, potentially opening up further new interesting signatures.

\section{Conclusions}\label{sec:conclusion}

To summarise, in this paper we have defined a new class of FPVDM scenarios based on an additional $\SUD$ dark gauge group connected to the SM symmetry structure through a VL fermion mediator. As such, this scenario does not require a Higgs portal mediating the interactions between the dark sector and the standard one.
Spontaneous breaking of the $\SUD$ symmetry provides the mass to the triplet of the corresponding gauge bosons. Two of these, which transform under a $U(1)$ global symmetry differently from  the SM particle, are the DM candidates. This symmetry, which contains a discrete $\Z_2$ subset and provides stability to the DM particles, can naturally be interpreted either in terms of of a dark EW sector or
in terms 
of a possible composite nature of the dark sector.

This general framework allows for multiple realisations, depending on the specific properties of the VL partner and the actual form of the scalar potential. 
As a simple example, we have studied the case of a VL top-quark partner and no mixing between the SM Higgs doublet and the new scalar sector,
which we have therefore called a `top portal' (or TPVDM). We have explored the phenomenology of such a minimal scenario and have provided bounds from both collider (chiefly, the LHC) and astroparticle 
(relic density, DD and ID) observables sensitive
to the presence of DM, specifically discussing the role of the new states and interactions. 
In doing so, we have found that LHC and non-collider search experiments have significant complementary power to decode the scenario under study provided that several interesting signatures are observed.
The signals could include direct or indirect evidence of the simultaneous presence of VL, $t_D$ and $T$ quarks and/or the new $H$ and/or $V_D$ and $V'$ bosons from $SU(2)_D$ in both open (i.e, real) and closed (i.e., virtual) production of such new physics states.

In fact, the specific BSM scenario introduced here presents one with the unique possibility of a multi-prong approach to a variety of distinctive signatures which would serve the purpose of enabling one to delineate all its key features. While we defer the detailed quantitative treatment of this approach to future publications, we highlight here what would be the
salient features of it.
The presence of a VL top companion $T$ and its dark counterpart $t_D$ subject to QCD interactions opens the obvious possibility of establishing their evidence at the LHC, through strong production processes. Furthermore, the additional Higgs and gauge states, as they couple to each other, would offer complementary evidence of such an extended dark sector -- particularly of its symmetry breaking pattern. 
Besides the generic mono-jet signature from $V_D$ pair production (first row of diagrams in table~\ref{tab:LHCprocesses}), which is hard to use to measure the model parameters, even the DM mass itself, there are several important complementary signals.
Among these, there is associated production of $V_D$ pairs with a $t\bar t$ pair, yielding $t\bar t + E_T^{\rm miss}$, providing certain sensitivity to the presence of $V^{\prime}$ and $H$ propagation. This can be achieved via the study of the momentum recoiling against the top-antitop system in the transverse plane (second graph in the second row of diagrams in table~\ref{tab:LHCprocesses}).
Indeed, the same final state may also make manifest the presence of the dark state $t_D$ in a specific form, when it becomes a LLP exhibiting a displaced vertex, in which a charged track
(or invisible neutral dark hadron)  decays into the DM itself plus SM hadrons and/or leptons. A measurement of the (proper) decay length of this signature could offer one the chance of extracting the value of the $t_D$ width and  this information 
could be used
to decode related model parameters.
Furthermore, the presence of $V^{\prime}$ and $H$ states would be even clearer in $4t$ final states
(diagram in the third row in table~\ref{tab:LHCprocesses}), especially when the transitions $V^{\prime}\to t\bar t$ and $H\to t\bar t$ are resonant. All such processes are potentially accessible at Run 3 of the LHC already. Furthermore, when the High Luminosity LHC (HL-LHC) option of the CERN machine becomes available, also $hV^{\prime}$ and $V^{\prime}V^{\prime}$ production and decay would be accessible (fourth row of diagrams in table~\ref{tab:LHCprocesses}).
Finally, it is worth mentioning that, if the $V'$ mass is below the $t\bar{t}$ threshold, it can be long-lived and dominantly decay to $b\bar{b}$ pairs through loop-induced diagrams. In this case, $hV'$ or $V^\prime V^\prime$ production would provide new striking signatures such as associate Higgs boson production together with a displaced $b\bar{b}$ resonance or pairs of $b\bar{b}$ displaced resonances, respectively. 

However, this strongly depends on the value of the $\Z_2$-odd VL mass and on the specific model realisation (i.e., which fermionic partner is present), as, for example, in the TPVDM direct-detection constraints limit the region with low DM mass (and therefore low $V^\prime$ mass). In evaluating such constraints, we have computed triangle-loop induced DM-DM-$Z/\gamma$ amplitudes  which define multipole DM Z-boson/photon interactions and lead to an important constraints from DM direct detection experiments.
We provide the respective detailed generic formulas which can be used for analogous models.

The minimal realisation of a FPVDM scenario adopted here has already significant potential to explain astrophysical DM phenomena as well as to exhibit smoking gun signals at the LHC. However, non-minimal FPVDM 
models, whose structure depends upon the concrete realisation of the mediator (Higgs and/or flavour sectors) would imply an even richer set of predictions and could well be used to explain currently observed data anomalies. For example, if the VL fermion interacts with the leptonic sector of the SM, it might explain the muon $(g-2)$~\cite{Muong-2:2021ojo} 
or $W$ mass~\cite{CDF:2022hxs} anomalies,
while at the same time provide novel physics cases for future $e^+e^-$ colliders~\cite{Aicheler:2012bya,Baer:2013cma,An:2018dwb,FCC:2018evy}. Finally, allowing for mixing in the scalar sector, further VL partners and/or additional interactions of the same VL representation, would open up a long list of possibilities for future studies, both theoretical and experimental. This would allow one to also explore the complementarity between collider and non-collider observables in such scenarios in ever greater depth than can be afforded by the minimal realisation tackled here.

\begin{acknowledgements}

We would like to thank Alexander Pukhov for help with the  micrOMEGAs modification for the correct evaluation of the DD due to the $\gamma(Z)$-$V_D$-$V_D$ loop-induced interactions, and Rogerio Rosenfled for pointing to the potential constraint from the  modified Yukawa coupling, $y_t$. The authors would like to thank referee for pointing out crucial aspects about the construction and testing of our model.
AB and SM acknowledge support from the STFC Consolidated Grant ST/L000296/1 and are partially
financed through the NExT Institute.
AB also acknowledge support from Soton-FAPESP grant.
LP's work is supported by the Knut and Alice Wallenberg foundation under the SHIFT project, grant KAW 2017.0100. AD is grateful to the LABEX Lyon Institute of Origins (ANR-10-LABX-0066)  for its financial support within the program ``Investissements d'Avenir". AD acknowledges partial support from the National Research Foundation in South Africa. NT is supported by the scholarship from the Development and Promotion of Science
and Technology Talents Project (DPST). All authors
acknowledge the use of the IRIDIS High-Performance Computing Facility and associated
support services at the University of Southampton in completing this work.

\end{acknowledgements}

\appendix

\newpage
\section{Mass splitting at one loop}\label{app:masssplitting}
At tree level, the neutral and charged components of $SU(2)_D$ gauge triplet are degenerate in mass as one can see in \cref{eq:VPmass}. Nevertheless, the radiative correction at one-loop level breaks their mass degeneracy. 
The difference between $m_{V_D}$ and $m_{V^{\prime}}$ takes place due to the $T-t$ mixing and the different $\Z_2$ parities of the members of the $SU(2)_D$ fermion doublet, which results in distinct particles circling in the loops. 
In the limit $m_{T}\to m_{t_D}$ there is no mixing between the $T-t$ quarks, and the radiative corrections give zero contribution to masses of new vector bosons.

Loops involving the two scalars $h$ and $H$ are non-zero in case of mixing in the scalar sector. However, the contribution of such loops is identical for $V_D$ and $V^\prime$ and therefore they will not be considered in the calculation of mass differences.

In \cref{fig:vector_SE} all possible self-energy diagrams with fermions circulating in the loops for $V_D$ and $V^{\prime}$, contributing to a two-point function at one loop, are shown.
 The self-energy amplitude of a vector boson can be decomposed into two components:
\begin{equation}
	i\Pi^{\mu\nu}_{V}(p^2)=\left(g^{\mu\nu}-\frac{p^\mu p^\nu}{p^2}\right)i\Pi^{T}_{V}(p^2)+\left(\frac{p^\mu p^\nu}{p^2}\right)i\Pi^{L}_{V}(p^2)\;,
\end{equation}
where $\Pi^{T}_{V}$ and $\Pi^{L}_{V}$ are the transverse and longitudinal amplitudes, respectively. Here we use a symbol $V$ to indicate either $V_D$ or $V^{\prime}$. 
To get the transverse and longitudinal components of the self-energy amplitude, we extract each part by using the following operators.

\begin{eqnarray}
	\Pi^{T}_{V}(p^2)&=& \frac{1}{3-2\epsilon}\left(g_{\mu\nu} - \frac{p_{\mu}p_{\nu}}{p^2}\right)\Pi_V^{\mu\nu}(p^2),\nonumber\\
	\Pi^{L}_{V}(p^2)&=&\frac{p_\mu p_\nu}{p^2} \Pi_V^{\mu\nu}(p^2).
\end{eqnarray}
We work in $d$-dimensions, 
$D = g_{\mu\nu}g^{\mu\nu}=4-2\epsilon$.
The physical mass, $m_{VD}$, is defined
as the  position of the propagator's pole and is given by 
\begin{equation}
m_{VD}^2 \equiv 	(m_{V}^{\text{pole}})^2=m^2_{V}+\text{Re}(\Pi^{T}_{V})\;,
	\label{eq:pole_mass_sq}
\end{equation}
where $m_{V}$ is the (divergent) bare mass, which is the same for both $V_D$ and $V^{\prime}$, and  $\text{Re}(\Pi^{T}_{V})$ stands for the real part of $\Pi^{T}_{V}$ . 
We use the physical (one-loop corrected) mass of DM ($V_D$) as an input parameter of the model.The mass of $V^\prime$ is given by
\begin{eqnarray}
	m_{V^\prime}^2 &=& m_{V_D}^2 - \Pi^{T}_{V^\prime} + \Pi^{T}_{V_D},\nonumber\\
	m_{V^\prime}&=& m_{V_D}\sqrt{1-\frac{\left(\Pi^{T}_{V^\prime} - \Pi^{T}_{V_D}\right)}{m^2_{V_D}}}\nonumber\\
	&=& m_{V_D}-\frac{\left(\Pi^{T}_{V^\prime} - \Pi^{T}_{V_D}\right)}{2m_{V_D}} + ...
	\label{eq:pole mass of Vprime}
\end{eqnarray}

 After truncating of the  expansion up to  the first order in $\Pi^{T}$,	the $V_D - V'$ mass splitting at one-loop reads:
\begin{equation}
	\Delta m_V=m_{V_D}-m_{V^{\prime}}=\frac{1}{2} \left(\frac{\Pi^{T}_{V_D}-\Pi^{T}_{V^{\prime}}}{m_{V_D}}\right)
	\label{eq:deltaMV}\;.
\end{equation}

The transverse component of the self-energy function $\Pi^T$ of gauge bosons with  fermion $F_1$ and $F_2$ in  the loop is given by
\begin{eqnarray}
	\Pi^{T}_{F_1,F_2}(p^2)&=&\frac{1}{16\pi^2}\left[2(v^2_{12}+a^2_{12})(A_0(m_1^2)+A_0(m_2^2)) -8(v^2_{12}+a^2_{12})B_{00}(p^2,m_1^2,m_2^2) \right.\nonumber\\
	&&\left. + 2(v^2_{12}(m_1-m_2)^2 + a^2_{12}(m_1^2+m_2^2) -p^2(v^2_{12}+a^2_{12}))B_0(p^2,m_1^2,m_2^2)\right],
	\label{eq:general self-energy function of gauge boson}
\end{eqnarray}
where the $v_{12}$ and $a_{12}$ are the vector and axial-vector couplings of $F_1F_2V$ vertices, respectively. The $A_0, B_0$ and $B_{00}$ are the standard one- and two-point Veltman-Passarino functions. The one-loop function for $V^\prime$ and $V_D$ are defined as

\begin{eqnarray}
	\Pi_{V_D}^T &=& \Pi^T_{t,t_D} + \Pi^T_{T,t_D},\nonumber\\
	\Pi_{V^\prime}^T &=& \Pi^T_{t,t} + \Pi^T_{T,T} + \Pi^T_{t,T} + \Pi^T_{t_D,t_D},
\end{eqnarray}
where $\Pi^T_{F_1,F_2}$ is the transverse component of self-energy function in which the fermions $F_1$ and $F_2$ are circulating. In this case, they are top quark and VL partners of top quark.

We have evaluated  eq.\eqref{eq:deltaMV} by using eq. \eqref{eq:general self-energy function of gauge boson}, 
the expressions for  couplings from \cref{tab:couplings of VVA/VVZ vertex}, and then set the square incoming momentum and the renormalisation scale equal to the mass of DM, $\mu^2=p^2=m^2_{V_D}$,
which leads to the following simple expression

\begin{equation}
	\Delta m_V^{\prime}=\frac{1}{640 \pi^2 m_{V_D}}g_D^2 m_T^2 \epsilon_1^2\left[(20+3\epsilon_3-15\epsilon_2+20\epsilon_2\epsilon_3)+10(3\epsilon_2-\epsilon_3-2\epsilon_2\epsilon_3)\log\epsilon_3\right]\;.
\end{equation}
where
\begin{equation}
	\epsilon_1=\frac{m_T^2-m_{t_D}^2}{m_T^2},\quad \epsilon_2=\frac{m_t^2}{m_T^2},\quad \epsilon_3=\frac{m_V^2}{m_T^2}\;.
\end{equation}
This formula was derived 
in the approximation $\epsilon_1,\epsilon_2,\epsilon_3\ll 1$. 
Keeping only the leading term of $\epsilon_1$ provides the following very simple expression for the $V_D - V'$ mass split: 
\begin{equation}
	\Delta m^{\prime\prime}_{V}=\frac{1}{32 \pi^2 m_{V_D}}g_D^2 m_T^2 \epsilon_1^2=\frac{1}{32 \pi^2 m_{V_D}}g_D^2m_T^2\left(\frac{m_T^2-m_{t_D}^2}{m_T^2}\right)^2\;.
\end{equation}

\section{Kinetic mixing functions}\label{app:KMfunctions}
{
The functions describing the $Z-V$ kinetic and mass mixings in \cref{eq:ZVKM} are given by

\begin{eqnarray}
 F_{qT1+qL}^{ZV}(r_f,r_{\psi_D})
 &=& \frac{2(r_f^2-1)(r_{\psi_D}^2-1)}{3}\bigg[\frac{3r_f^6r_{\psi_D}^6 - 5r_f^6r_{\psi_D}^4 -21r_f^4r_{\psi_D}^4 +22r_f^4r_{\psi_D}^2 -21r_f^2r_{\psi_D}^2 -5r_f^2 +3}{(r_f^2r_{\psi_D}^2-1)^4}\bigg.\nonumber\\
 &&+\bigg.6\frac{\left(r_f^8 r_{\psi_D}^6 -3r_f^6r_{\psi_D}^4+12r_f^4r_{\psi_D}^4-3r_f^4r_{\psi_{D}}^2+r_f^2\right)}{(r_f^2r_{\psi_D}^2-1)^5}\log{(r_fr_{\psi_D})}\bigg] \\
 F_{qT2}^{ZV}(r_f,r_{\psi_D}) &=& 8 \bigg[ \log{\left(\frac{r_f}{r_{\psi_D}} \right)} + \frac{(r_{\psi_D}^2-r_f^2)\log{(r_fr_{\psi_D})}}{r_f^2 r_{\psi_D}^2 -1} \bigg]\\
 F_m^{ZV}(r_f,r_{\psi_D}) &=& (r_f^2-1)(r_{\psi_D}^2 - 1)\left[\frac{1-4r_{\psi_D}^2+r_f^2r_{\psi_D}^2}{r_{\psi_D}^2(r_f^2r_{\psi_D}^2-1)^2} + \frac{4(r_f^2r_{\psi_D}^2-r_f^2 +1)\log{(r_f r_{\psi_D})} }{(r_f^2 r_{\psi_D}^2-1)^3}\right]
\end{eqnarray}

}

\section{Mixing structure in the gauge sector for the dark EW sector}\label{app:darkEW}
Defining $\mathcal V^0_{D0\mu}=(B_\mu,W^3_\mu,B^0_{D0\mu},V^0_{D0\mu})^T$ and using analogous notation as eq.\eqref{eq:gaugeLag} for the fully neutral gauge boson Lagrangian term after EW and dark symmetry breaking, 
\begin{equation}
\Lag_{\mathcal V^0_{D0}}^{\rm kin}|_{v,v_D} \supset (\mathcal V^0_{D0})^T \mathcal M_{\mathcal V^0_{D0}}^2 \mathcal V^0_{D0}\;,
\end{equation}
the entries of the mass mixing matrix in the gauge sector are: 
\begin{eqnarray}
\mathcal M^2_{\mathcal V^0_{D0}}|_{11} &=& \frac{\left(g^{\prime 2} v^2+g_D^{\prime 2} v_D^2 \epsilon^2\right) \cos^2\theta_k - g_D^{\prime 2} \epsilon  \sqrt{1-\epsilon^2} \sin2\theta_k v_D^2 + g_D^{\prime 2} \left(1-\epsilon^2\right) \sin^2\theta_k v_D^2}{8 \left(1-\epsilon^2\right)} \\
\mathcal M^2_{\mathcal V^0_{D0}}|_{12} &=& \mathcal M^2_{\mathcal V^0_{D0}}|_{21} = -\frac{g g^\prime v^2 \cos\theta_k}{8 \sqrt{1-\epsilon^2}} \\
\mathcal M^2_{\mathcal V^0_{D0}}|_{13} &=& \mathcal M^2_{\mathcal V^0_{D0}}|_{31} = \frac{g_D^{\prime 2} v_D^2 \left(\left(1-2 \epsilon^2\right) \sin2\theta_k-2 \epsilon  \sqrt{1-\epsilon^2} \cos2\theta_k\right)-g^{\prime 2} \sin2\theta_k v^2}{16 \left(1-\epsilon^2\right)} \\
\mathcal M^2_{\mathcal V^0_{D0}}|_{14} &=& \mathcal M^2_{\mathcal V^0_{D0}}|_{41}  =\frac{1}{8} g_D g_D^\prime v_D^2 \left(\frac{\epsilon  \cos\theta_k}{\sqrt{1-\epsilon^2}}-\sin\theta_k\right) \\
\mathcal M^2_{\mathcal V^0_{D0}}|_{22} &=& \frac{g^2 v^2}{8}  \\
\mathcal M^2_{\mathcal V^0_{D0}}|_{23} &=& \mathcal M^2_{\mathcal V^0_{D0}}|_{32} = \frac{g g^\prime v^2 \sin\theta_k}{8 \sqrt{1-\epsilon^2}} \\
\mathcal M^2_{\mathcal V^0_{D0}}|_{24} &=& \mathcal M^2_{\mathcal V^0_{D0}}|_{42} = 0 \\
\mathcal M^2_{\mathcal V^0_{D0}}|_{33} &=& \frac{g_D^{\prime 2} \left(1-\epsilon^2\right) \cos^2\theta_k v_D^2+ g_D^{\prime 2} \epsilon  \sqrt{1-\epsilon^2} \sin2\theta_k v_D^2+\left(g^{\prime 2} v^2+g_D^{\prime 2} v_D^2 \epsilon^2\right) \sin^2\theta_k}{8 \left(1-\epsilon^2\right)} \\
\mathcal M^2_{\mathcal V^0_{D0}}|_{34} &=& \mathcal M^2_{\mathcal V^0_{D0}}|_{43}  =-\frac{1}{8} g_D g_D^\prime v_D^2 \left(\cos\theta_k+\frac{\epsilon  \sin\theta_k}{\sqrt{1-\epsilon^2}}\right) \\
\mathcal M^2_{\mathcal V^0_{D0}}|_{44} &=& \frac{g_D^2 v_D^2}{8}\;.
\end{eqnarray}
The mass eigenstates corresponding to the eigenvalues of the mixing matrix are $\gamma$, $\gamma_D$, $Z$ and $Z^\prime$. Their masses do not depend on the rotation angle $\theta_k$ and read:
\begin{eqnarray}
 m_\gamma&=&m_{\gamma_D}=0 \\
 M_{Z,Z^\prime}^2 &=& {1\over8} \left[ g^2 v^2 + g_D^2 v_D^2 + {1\over1-\epsilon^2} \left( g^{\prime2} v^2 + g_D^{\prime2} v_D^2 \mp \sqrt{ \mathcal K_0 + \mathcal K_2 \epsilon^2+\mathcal K_4 \epsilon^4 }\right)\right]
\end{eqnarray}
where the $\mathcal K$ functions are defined as:
\begin{eqnarray}
 \mathcal K_0 &=& \left(\left(g^2+g^{\prime2}\right)v^2 - \left(g_D^{\prime2}+g_D^2\right)v_D^2 \right)^2 \\
 \mathcal K_2 &=& - 2 \left[ g^2 (g^2+g^{\prime2})v^4 + g_D^2 (g_D^2+g_D^{\prime2})v_D^4 - \left(g^2(2g_D^2+g_D^{\prime2})+g^{\prime2}\left(g_D^2+2g_D^{\prime2}\right)\right)v^2 v_D^2\right]\\
 \mathcal K_4 &=& (g^2 v^2 - g_D^2 v_D^2)^2
\end{eqnarray}
and the sign in front of the square root is chosen to reconstruct the SM value of the $Z$ mass for $\epsilon\to0$ and $\left(g^2+g^{\prime2}\right)v^2>\left(g_D^{\prime2}+g_D^2\right)v_D^2$.

\section{Contributions from fermion triangle digrams to direct detection of DM}\label{app:triangle}

\begin{figure}[htb!]
\centering
\begin{minipage}{0.35\textwidth}
\includegraphics[width=\textwidth]{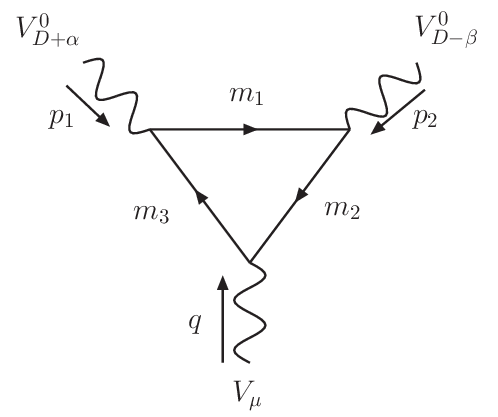}\\
(a)
\end{minipage}
\begin{minipage}{0.35\textwidth}
\includegraphics[width=\textwidth]{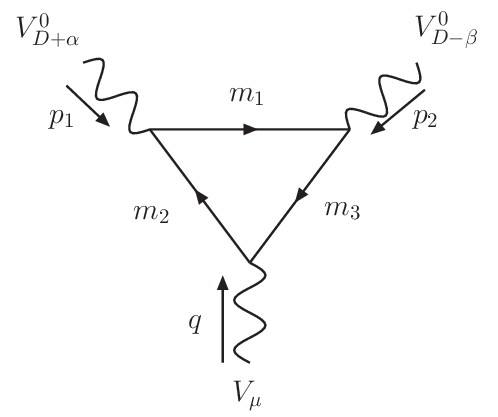}\\
(b)
\end{minipage}
\caption{\label{fig:triangle diagrams} The generic triangle diagrams for $V^0_{D+ \alpha}V^0_{D- \beta}V_{\mu}$ where $V_\mu$ stands for either photon and Z-boson. The ingoing momenta for $V^0_{D+},V^0_{D-}$ are $p_1,p_2$ and $q$, respectively. 
The vector and axial coupling constants for the vertex between a vector boson and fermions with masses $m_i$ and $m_j$  are $v_{ij}$, $a_{ij}$ respectively.}
\end{figure}

The  computed direct detection limit at one loop level is  based on the interaction between DM and Standard model particles through the box and triangle (scalar propagating) diagrams in \cref{fig:DDtopology} (c) and (d). Furthermore, there are two additional vertices, $V^0_{D+}V^0_{D-}\gamma$ and $V^0_{D+}V^0_{D-}Z$, which can also contribute to the direct detection limit, depicted in \cref{fig:triangle diagrams}. 
	
The most general (CP conserving) effective Lagrangian  \cite{Hagiwara:1986vm,Hisano:2020qkq} for on-shell DM $V^0_{D\pm}$ interacting with neutral vector bosons $\gamma/Z$ is given by
	\begin{eqnarray}
		\mathcal{L}^{eff}_{V^0_{D+}V^0_{D-}V}&=& \lambda_1^V \left[(\partial_\mu V^0_{D+\nu} - \partial_\nu V^0_{D+\mu})V^{0\mu}_{D-} V^\nu - (\partial_\mu V^0_{D-\nu} - \partial_\nu V^0_{D-\mu})V^{0\mu}_{D+} V^\nu\right]\nonumber\\
		&& -  \lambda_2^V V^0_{D+\mu} V^0_{D-\nu}(\partial^\mu V^\nu - \partial^\nu V^\nu) \nonumber\\
		&& + \frac{\lambda_3^V}{M^2_{V^{\pm}}}(\partial_\lambda V^0_{D+\mu} - \partial_\mu V^0_{D+\lambda})(\partial^\mu V^0_{D-\nu} - \partial_\nu V^0_{D-\mu})(\partial^\nu V^\lambda - \partial^\lambda V^\nu) \nonumber\\
		&& -i\lambda_5^V \epsilon^{\mu\nu\rho\sigma}(V^0_{D+\mu} \partial_\rho V^0_{D-\nu} -  V^0_{D-\nu} \partial_\rho V^0_{D+\mu})V_\sigma \nonumber\\
		&& -i\lambda_6^V \epsilon^{\mu\nu\rho\sigma}(V^0_{D+\mu} \partial_\rho V^0_{D-\nu} - V^0_{D-\nu} \partial_\rho V^0_{D+\mu})\partial^\lambda\partial_\sigma(\partial^2)^{-1}V_\lambda,
		\label{eq:effective Lagrangian}
	\end{eqnarray}
where $V$ can be either $\gamma$ or $Z$. The DM vector particles $V^0_{D^\pm}$ are taken to be on-shell with mass $M_{V^\pm}$. Furthermore, since in the direct detection  process, the momentum transferred between DM particles and SM particles is much smaller than the masses of the DM particles, we may therefore use the approximation of setting $q^2$ to zero.

This effective Lagrangian (in momentum space)  can be expressed in terms of the vertex function which is a function of all incoming momenta 
\begin{equation}
\mathcal{L}^{eff}_{V^0_{D+}V^0_{D-}V} =  -i V_V^{\alpha\beta\mu}(p_1,p_2,q)V^0_{D+\alpha}(p_1) V^0_{D-\beta}(p_2) V_\mu(q),
\end{equation} 
where the vertex function reads
\begin{eqnarray}
	V_V^{\alpha\beta\mu}(p_1,p_2,q)&=& f^V_1(p_1 - p_2)^\mu g^{\alpha\beta} + \frac{f^V_2}{M_{V^\pm}^2}(p_1 - p_2)^\mu q^\alpha q^\beta + f^V_3 (q^\alpha g^{\mu\beta} - q^\beta g^{\mu\alpha})\nonumber\\
	&& + if^V_5 \epsilon^{\alpha\beta\mu\rho}(p_1 - p_2)_\rho + i\frac{f^V_8}{M_{V^\pm}^2}(p_1 - p_2)_\rho q_\sigma (\epsilon^{\mu\alpha\rho\sigma}q^\beta - \epsilon^{\mu\beta\rho\sigma}q^\beta).
\end{eqnarray}
The CP-conserving\footnote{The more general vertex function found in appendix A of \cite{Hagiwara:1986vm} includes additional CP-violating form-factors  $f_4, f_6, f_7$ and $f_9$. However, these are irrelevant for direct detection of DM in this model, and are therefore omitted.} form factors, $f_i$, are related to the couplings $\lambda_i$  of 	(\ref{eq:effective Lagrangian}) by 
\begin{eqnarray}
	f_1^V &=& \lambda_1^V + \frac{q^2}{2M_{V^\pm}^2}\lambda_3^V,\\
	f_2^V &=& -\lambda_3^V,\\
	f_3^V &=& \lambda_1^V + \lambda_2^V + \frac{1}{2}\lambda_3^V, \\
	f_5^V &=& \lambda_6^V - \lambda_5^V, \\
	f_8^V &=&-\frac{M_{V^\pm}^2}{q^2}\lambda_6^V.
\end{eqnarray}

We explicitly calculate the form-factors of the $V^0_{D+}V^0_{D-}V$ vertex according to the diagrams in \cref{fig:triangle diagrams} where $p_1,p_2$ and $q$ are the momenta of $V^0_{D+},V^0_{D-}$ and $V_\mu$, respectively.
The vector and axial coupling constants for the vertex between a vector particle and fermions $i$  annd $j$ (with masses  $m_i$ and $m_j$) are denoted by $v_{ij}, \, a_{ij}$.

For the prototype graphs shown in  \cref{fig:triangle diagrams}, we find the following expressions for the form-factors:
 \begin{align}
f_1^{V}&= \frac{(v_{23}v_{12}v_{13}+v_{23}a_{12}a_{13}+a_{23}v_{12}a_{13}+a_{23}a_{12}v_{13})}{4\pi^2}\left[1-(m_2^2+m_3^2)\bar{C}_1-(m_3^2-m_2^2)\Delta C_1 + 8\bar{C}_{001}\right]\nonumber\\
& + \frac{(v_{23}v_{12}v_{13}-v_{23}a_{12}a_{13}+a_{23}v_{12}a_{13}-a_{23}a_{12}v_{13})}{4\pi^2}m_1m_2\left[\bar{C}_0+2\bar{C}_1\right]\nonumber\\
& + \frac{(v_{23}v_{12}v_{13}-v_{23}a_{12}a_{13}-a_{23}v_{12}a_{13}+a_{23}a_{12}v_{13})}{4\pi^2}m_1m_3\left[\bar{C}_0+2\bar{C}_1\right]\nonumber\\
& - \frac{(v_{23}v_{12}v_{13}+v_{23}a_{12}a_{13}-a_{23}v_{12}a_{13}-a_{23}a_{12}v_{13})}{4\pi^2}m_2m_3\left[2\bar{C}_1\right]\;,\\
f_2^{V}&= - \frac{(v_{23}v_{12}v_{13}+v_{23}a_{12}a_{13}+a_{23}v_{12}a_{13}+a_{23}a_{12}v_{13})}{4\pi^2}M_{V^\pm}^2\left[8\bar{C}_{112}+4\bar{C}_{12}\right]\;,\\
f_3^{V}&= \frac{(v_{23}v_{12}v_{13}+v_{23}a_{12}a_{13}+a_{23}v_{12}a_{13}+a_{23}a_{12}v_{13})}{4\pi^2}\left[-2-2m_1^2(\bar{C}_0+2\bar{C}_1) + (m_3^2+m_2^2)\bar{C}_1 \right.\nonumber\\
& + \left. (m_3^2-m_2^2)\Delta C_1 + 8 \bar{C}_{00}+8 \bar{C}_{001}\right]\nonumber\\
& -\frac{(v_{23}v_{12}v_{13}-v_{23}a_{12}a_{13}+a_{23}v_{12}a_{13}-a_{23}a_{12}v_{13})}{4\pi^2}m_1m_2\left[\bar{C}_0 + 2\Delta C_1\right]\nonumber\\
& + \frac{(v_{23}v_{12}v_{13}-v_{23}a_{12}a_{13}-a_{23}v_{12}a_{13}+a_{23}a_{12}v_{13})}{4\pi^2}m_1m_3\left[2\Delta C_1 - \bar{C}_0\right] \nonumber\\
& + \frac{(v_{23}v_{12}v_{13}+v_{23}a_{12}a_{13}-a_{23}v_{12}a_{13}-a_{23}a_{12}v_{13})}{4\pi^2}m_2m_3\left[2\bar{C}_1\right]\;,\\
f_5^{V}&=- \frac{(a_{23}a_{12}a_{13}+a_{23}v_{12}v_{13}+v_{23}a_{12}v_{13}+v_{23}v_{12}a_{13})}{4\pi^2}\left[(m_2^2+m_3^2)\bar{C}_1 + (m_3^2-m_2^2)\Delta C_1\right]\nonumber\\
& -\frac{(a_{23}a_{12}a_{13}-a_{23}v_{12}v_{13}+v_{23}a_{12}v_{13}-v_{23}v_{12}a_{13})}{4\pi^2}m_1m_2\left[\bar{C}_0 + 2\bar{C}_1\right]\nonumber\\
& -\frac{(a_{23}a_{12}a_{13}-a_{23}v_{12}v_{13}-v_{23}a_{12}v_{13}+v_{23}v_{12}a_{13})}{4\pi^2}m_1m_3\left[\bar{C}_0+2\bar{C}_1\right]\nonumber\\
& -\frac{(a_{23}a_{12}a_{13}+a_{23}v_{12}v_{13}-v_{23}a_{12}v_{13}-v_{23}v_{12}a_{13})}{4\pi^2}m_2m_3\left[2\bar{C}_1\right]\;,\\
f_8^{V}&=\frac{(a_{23}a_{12}a_{13}+a_{23}v_{12}v_{13}+v_{23}a_{12}v_{13}+v_{23}v_{12}a_{13})}{4\pi^2}M_{V^\pm}^2[2\bar{C}_{12}]\;.
\end{align}
The average and difference of three-point Passarino-Veltman C-functions are defined as
	\begin{eqnarray}
		\bar{C}_{\{i\}} \equiv \frac{1}{2}\left(C^{(a)}_{\{i\}}+C^{(b)}_{\{i\}}\right),\nonumber\\
		\Delta{C}_{\{i\}} \equiv \frac{1}{2}\left(C^{(a)}_{\{i\}}-C^{(b)}_{\{i\}}\right).
	\end{eqnarray}
where $C_{\{i\}}^{r},  \, (r=a,b)$ are given in terms of to one-loop triangle Feynman integrals

\begin{align}
\frac{1}{i\pi^2}\int d^4k\,\, \frac{1}{D^{(r)}} &= C_0^{(r)}\;,\nonumber\\
\frac{1}{i\pi^2}\int d^4k\,\, \frac{k^\mu}{D^{(r)}} &= -p_1^\mu C_1^{(r)} + p_2^\mu C_2^{(r)}\;,\nonumber\\
\frac{1}{i\pi^2}\int d^4k\,\, \frac{k^\mu k^\nu}{D^{(r)}} &= g^{\mu\nu}C_{00}^{(r)}+p_1^\mu p_1^\nu C_{11}^{(r)} + p_2^\mu p_2^\nu C_{22}^{(r)} - \left(p_1^\mu p_2^\nu + p_1^\nu p_2^\mu \right)C_{12}^{(r)}\;,\nonumber\\
\frac{1}{i\pi^2}\int d^4k\,\, \frac{k^\mu k^\nu k^\rho}{D^{(r)}} &=- \left(p_1^\mu g^{\nu\rho} + p_1^\nu g^{\mu\rho} + p_1^\rho g^{\mu\nu}\right)C_{001}^{(r)} + \left(p_2^\mu g^{\nu\rho} + p_2^\nu g^{\mu\rho} + p_2^\rho g^{\mu\nu}\right)C_{002}^{(r)}\nonumber\\
& +\left(p_1^\mu p_1^\nu p_2^\rho  + p_1^\mu p_1^\rho p_2^\nu + p_1^\nu p_1^\rho p_2^\mu    \right)C_{112}^{(r)} - \left(p_1^\mu p_2^\nu p_2^\rho + p_1^\nu p_2^\mu p_2^\rho + p_1^\rho p_2^\mu p_2^\nu \right)C_{122}^{(r)} \nonumber\\
& - p_1^\mu p_1^\nu p_1^\rho C_{111}^{(r)} + p_2^\mu p_2^\nu p_2^\rho C_{222}^{(r)}.
\end{align} 	
		The denominators of the Feynman integrals are
		\begin{eqnarray}
			 D^{(a)}  & = &
		 (k^2-m_1^2)\left((k-p_1)^2 - m_3^2\right)\left((k+p_2)^2 - m_2^2\right).
              \nonumber \\
               D^{(b)}  & = &
              (k^2-m_1^2)\left((k-p_1)^2 - m_2^2\right)\left((k+p_2)^2 - m_3^2\right).
              \end{eqnarray}

          Both the graphs in
          \cref{fig:triangle diagrams}, have the fermion direction in the clockwise (CW)
          direction. Diagrams for which the fermion line is in the counter-clockwise direction
          (CCW) give contributions to the the form-factors, which are related to the clockwise
          form-factor contributions by
          
          \begin{eqnarray}
          	f_1^{\text{CCW}} &=& -f_1^{\text{CW}},\nonumber\\
          	f_2^{\text{CCW}} &=& -f_2^{\text{CW}},\nonumber\\
          	f_3^{\text{CCW}} &=& -f_3^{\text{CW}},\nonumber\\
          	f_5^{\text{CCW}} &=& f_5^{\text{CW}},\nonumber\\
          	f_8^{\text{CCW}} &=& f_8^{\text{CW}}.
          \end{eqnarray}

	\begin{figure}[htb]
		\centering
		\begin{subfigure}[b]{0.3\textwidth}
			\centering
			\includegraphics[width=\textwidth]{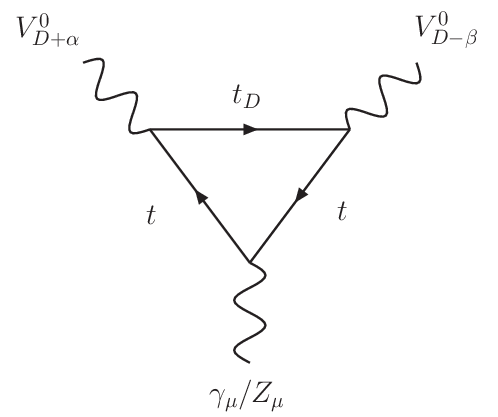}
			\caption{}
			\label{fig:triangle 1}
		\end{subfigure}
		\begin{subfigure}[b]{0.3\textwidth}
			\centering
			\includegraphics[width=\textwidth]{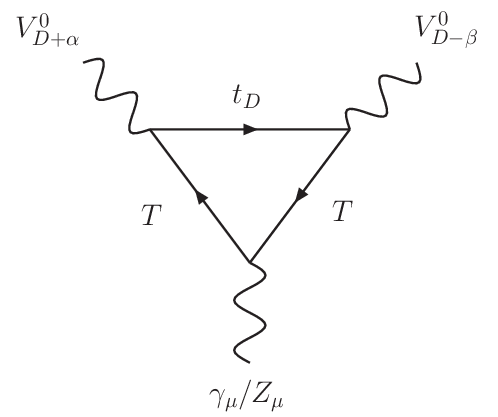}
			\caption{}
			\label{fig:triangle 2}
		\end{subfigure}
		\begin{subfigure}[b]{0.3\textwidth}
			\centering
			\includegraphics[width=\textwidth]{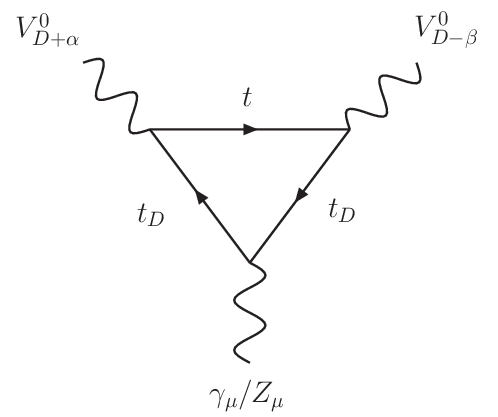}
			\caption{}
			\label{fig:triangle 3}
		\end{subfigure}
		\begin{subfigure}[b]{0.3\textwidth}
			\centering
			\includegraphics[width=\textwidth]{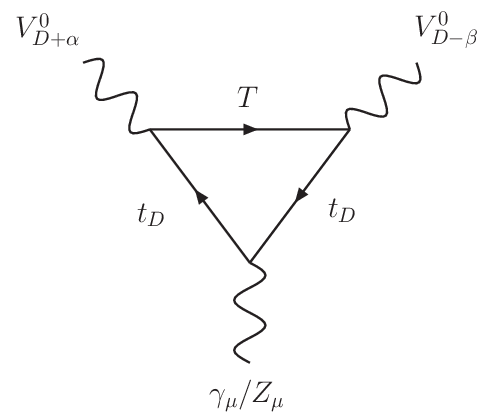}
			\caption{}
			\label{fig:triangle 4}
		\end{subfigure}
		\begin{subfigure}[b]{0.3\textwidth}
			\centering
			\includegraphics[width=\textwidth]{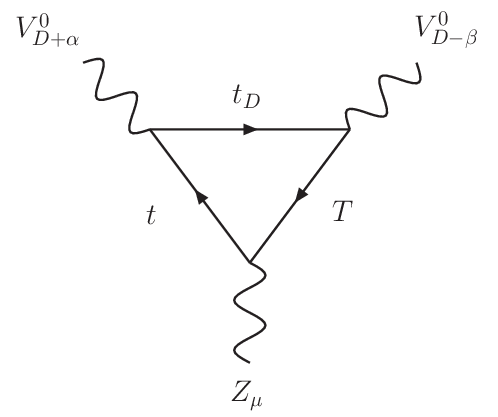}
			\caption{}
			\label{fig:triangle 5}
		\end{subfigure}
		\begin{subfigure}[b]{0.3\textwidth}
			\centering
			\includegraphics[width=\textwidth]{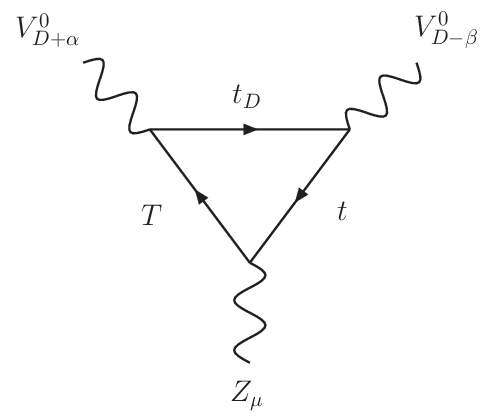}
			\caption{}
			\label{fig:triangle 6}
		\end{subfigure}
		\caption{\label{fig:triangle} The complete set of Feynman graphs for $V^0_{D+} V^0_{D-}\gamma$ and $V^0_{D+} V^0_{D-} Z$ form factor calculations. The diagram \cref{fig:triangle 1,fig:triangle 2,fig:triangle 3,fig:triangle 4} contribute to the $V^0_{D+} V^0_{D-}\gamma$ vertex and \cref{fig:triangle 1,fig:triangle 2,fig:triangle 3,fig:triangle 4,fig:triangle 5,fig:triangle 6} to the $V^0_{D+} V^0_{D-} Z$ vertex.} 
	\end{figure}

\begin{table}[htbp]
	\setlength{\tabcolsep}{10pt} 
	\renewcommand{\arraystretch}{1.5} 
	\centering
	\begin{tabular}{ |c|c|c| } 
		\hline
		\textbf{Vertices} & \textbf{Vector couplings $(v_{ij})$} & \textbf{Axial couplings $(a_{ij})$}\\
		\hline
		$\gamma \overline{t} t$ & $\frac{2e}{3}$ & $0$ \\ 
		$\gamma \overline{t}_D t_D$ & $\frac{2e}{3}$ & $0$ \\
		$\gamma \overline{T} T$ & $\frac{2e}{3}$ & $0$ \\
		$Z\overline{t}t$ & $\frac{e}{s_W c_W} \left(\frac{\cos^2{\theta_{fL}}}{4} - \frac{2 s_W^2}{3}\right)$ & $\frac{e}{s_W c_W}\frac{\cos^2{\theta_{fL}}}{4}$\\ 
		$Z\overline{t}_D t_D$ & $-\frac{2e s_W^2}{3 s_W c_W}$ & $0$ \\ 
		$Z \overline{T} T$ & $\frac{e}{s_W c_W}\left(\frac{\sin^2{\theta_{fL}}}{4}-\frac{2 s_W^2}{3}\right)$ & $\frac{e}{s_W c_W}\frac{\sin^2{\theta_{fL}}}{4}$ \\
		$Z \overline{t} T$ & $\frac{e \sin{\theta_{fL}}\cos{\theta_{fL}}}{4s_W c_W}$ & $\frac{e \sin{\theta_{fL}}\cos{\theta_{fL}}}{4s_W c_W}$ \\
		$V^0_{D+} \overline{t}_D t$ & $-\frac{\sqrt{2}g_D}{4}\left(\sin{\theta_{fL}}+\sin{\theta_{fR}}\right)$ & $-\frac{\sqrt{2}g_D}{4}\left(\sin{\theta_{fL}}-\sin{\theta_{fR}}\right)$ \\
		$V^0_{D+} \overline{t}_D T$ & $\frac{\sqrt{2}g_D}{4}\left(\cos{\theta_{fL}}+\cos{\theta_{fR}}\right)$ & $\frac{\sqrt{2}g_D}{4}\left(\cos{\theta_{fL}}-\cos{\theta_{fR}}\right)$ \\
		$V^0_{D0} \overline{t} t$ & $-\frac{g_D}{4}\left(\sin^2{t_R}+\sin^2{t_L}\right)$ & $\frac{g_D}{4}\left(\sin^2{t_R}-\sin^2{t_L}\right)$ \\
		$V^0_{D0} \overline{T} T$ & $-\frac{g_D}{4}\left(\cos^2{t_R}+\cos^2{t_L}\right)$ & $\frac{g_D}{4}\left(\cos^2{t_R}-\cos^2{t_L}\right)$ \\
		$V^0_{D0} \overline{t} T$ & $\frac{g_D}{4}\left(\sin{t_R}\cos{t_R}+\sin{t_L}\cos{t_L}\right)$ & $-\frac{g_D}{4}\left(\sin{t_R}\cos{t_R}-\sin{t_L}\cos{t_L}\right)$ \\
		$V^0_{D0} \overline{t}_D t_D$ & $\frac{g_D}{2}$ & $0$ \\
		\hline
	\end{tabular}
	\caption{The vector and axial part of coupling in the form of $v\gamma^\mu-a\gamma^\mu \gamma_5$ where $v$ is the vector part, $a$ the axial part and $\mu$ is the Lorentz index of a vector field. Here we suppress the $SU(2)_D$ charge of $V^+/V^-$ and write them as $V$.}
	\label{tab:couplings of VVA/VVZ vertex}
\end{table}

	For the direct detection calculation, we need to evaluate the triangle integrals that correspond to the Feynman diagrams shown in \cref{fig:triangle}. The vertex $V^0_{D+} V^0_{D-}\gamma$ receives the contributions from \cref{fig:triangle 1,fig:triangle 2,fig:triangle 3,fig:triangle 4}, while The vertex $V^0_{D+} V^0_{D-} Z$ from \cref{fig:triangle 1,fig:triangle 2,fig:triangle 3,fig:triangle 4,fig:triangle 5,fig:triangle 6}. The complete set vertex couplings required to evaluate these triangle graphs is provided in \cref{tab:couplings of VVA/VVZ vertex}.

For the numerical evaluation of triangle loops, we have created our own code written in {\sc C} and {\sc python} for computing the necessary Passarino-Veltman (PV) functions, as {\sc LoopTools} \cite{HAHN1999153} does not provide stable and reliable results for small momentum of $\gamma/Z$.\footnote{These codes are available together with the model files in the {\sc HEPMDB}~\cite{hepmdb} repository at the following link \url{https://hepmdb.soton.ac.uk/hepmdb:0322.0335}}.

\clearpage
\newpage
\bibliographystyle{apsrev4-1}
\bibliography{bibliography}

\end{document}